%% file: arxiv.tex
\documentclass[11pt]{article}
\usepackage{fullpage}
\usepackage{amsmath,amsthm,amssymb}
\usepackage{graphicx,float,psfrag,epsfig,amssymb}
\usepackage[top=1in, bottom=1.25in, left=1in, right=1in]{geometry}
\usepackage[usenames,dvipsnames,svgnames,table]{xcolor}
\definecolor{darkgreen}{rgb}{0.0,0,0.9}
\usepackage[pagebackref,letterpaper=true,colorlinks=true,pdfpagemode=none,citecolor=OliveGreen,linkcolor=BrickRed,urlcolor=BrickRed,pdfstartview=FitH]{hyperref}
\usepackage{relsize}
\usepackage{color}
\usepackage{pict2e}
\usepackage{caption}

\usepackage{nameref}
\usepackage{makecell}
\usepackage[font={small}]{caption} 
\usepackage{subcaption}
\usepackage{mathtools}
\usepackage{enumitem}

\let\chapter\section
\usepackage[ruled,vlined]{algorithm2e}
\usepackage[noend]{algorithmic}
\usepackage{diagbox}

\DeclareMathAlphabet{\mathpzc}{OT1}{pzc}{m}{it}

\newtheorem{propo}{Proposition}[section]
\newtheorem{lemma}[propo]{Lemma}

\newtheorem{assumption}[propo]{Assumption}

\newtheorem{defi}[propo]{Definition}
\newtheorem{coro}[propo]{Corollary}
\newtheorem{thm}[propo]{Theorem}
\newtheorem{rmk}[propo]{Remark}

\usepackage{natbib}

\usepackage{hyperref}       % hyperlinks
\usepackage{url}   

\addtocontents{toc}{\protect\setcounter{tocdepth}{2}}
\newcommand{\indep}{\perp \!\!\! \perp}
\newcommand{\notindep} {\not\!\perp\!\!\!\perp}
 \input{notation}

\def\cD{\mathcal{D}}

\def\bX{\mathbf{X}}
\def\bY{\mathbf{Y}}
\def\bZ{\mathbf{Z}}

\def\hP{\widehat{P}}
\def\hX{\widehat{X}}
\def\normal{\mathsf{N}}
\def\tv{\mathsf{TV}}

\def\tbX{\widetilde{\bX}}
\def\tcG{\widetilde{\cG}}

\def \myg {n_g}
\def \mys{s}
\def \bx{\boldsymbol{x}}
\def \bs{\boldsymbol{s}}
\def \bR{\boldsymbol{R}}
\def\tbx{\widetilde{\bx}}
\def \by{\boldsymbol{y}}
\def \bz{\boldsymbol{z}}
\def \bSigma{\boldsymbol{\Sigma}}
\def \bV{\boldsymbol{V}}
\def \br{\boldsymbol{r}}
\def \hbX{\widehat{\bX}}
\def \bU{\boldsymbol{U}}
\def \bI{\boldsymbol{I}}
\title{\bf Pearson Chi-squared Conditional Randomization Test}

\author{ 
Adel Javanmard\thanks{Data Sciences and Operations Department, University
of Southern California} \thanks{A.~Javanmard was supported in part by the Sloan Research Fellowship in
mathematics, an Adobe Data Science Faculty Research Award, an Amazon
Faculty Research Award, NSF Award 2311024, and a grant from Institute for
Outlier Research in Business (iORB) at USC Marshall School of Business.
Part of this work was done when A. Javanmard was a visiting scientist at
the Simons Institute for the Theory of Computing.} \and 
Mohammad Mehrabi\thanks{Department of Operations, Information \&
Technology, Stanford Graduate School of Business,Stanford University} \thanks{The names of the authors are in alphabetical order. }
}
\begin{document}

\maketitle
\begin{abstract}

Conditional independence (CI) testing arises naturally in many scientific problems and applications domains. The goal of this problem is to investigate the conditional independence between a response variable $Y$ and another variable $X$, while controlling for the effect of a high-dimensional confounding variable $Z$. In this paper, we introduce a novel test, called `Pearson Chi-squared Conditional Randomization' (PCR) test,  which uses the distributional information on covariates $X,Z$ and constructs randomizations to test conditional independence.  
PCR leverages the i.i.d-ness property of the observations to obtain high-resolution p-values with a very small number of conditional randomizations.

We also provide a power analysis of the PCR test, which captures the effect of various parameters of the test, the sample size and the distance of the alternative from the set of null distributions, measured in terms of a notion called `conditional relative density'.  In addition, we propose two extensions of the PCR test, with important practical implications: $(i)$ parameter-free PCR, which uses Bonferroni's correction to decide on a tuning parameter in the test; $(ii)$ robust PCR, which avoids inflations in the size of the test when there is slight error in estimating the conditional law $P_{X|Z}$.

\end{abstract}
\medskip


\section{Introduction}

Understanding the statistical relationship between random variables is a cornerstone of many scientific experiments. Various measures of dependency were developed in the statistics literature to capture the association between random variables, such as the mutual information and information theoretic coefficients \citep{reshef2011detecting}, the kernel-based measures \citep{pfister2018kernel, zhang2018large}, the correlation coefficients that are based on sample ranks 
\citep{drton2018high, deb2021multivariate, weihs2018symmetric}, and the dependency metrics that are based on copulas \citep{zhang2019bet, shih2021copula}; We refer to the survey by \citet{josse2013measures} for other dependency measures. 

Inferential tasks in data science and statistics often require a more thorough analysis of the associations between random variables. In particular, a desired analysis must control for the presence of  confounding factors.  This happens when (an often unmeasured) factor $Z$ affects both of the variables of interest (say $X$ and $Y$), and hence  can lead to misleading conclusions about the association of the variables. For example, in genome-wide association studies (GWAS), researchers are interested in finding loci that are causal for the trait. However, spurious association can arise due to ancestry-induced correlations between causal and non-causal loci, or when ancestry is correlated with both the genotype and the trait~\citep{campbell2005demonstrating,bhaskar2017novel}.
 
 Conditional independence (CI) testing controls for the effect of such confounding factors. To further highlight the significance of the CI problem, it is worth noting that many important problems in statistics can indeed be cast as a CI testing problem, with examples ranging from the classic concepts of sufficient and ancillary statistics \citep{dawid1979conditional}, to the well-known concepts in graphical models \citep{koller2009probabilistic, friedman2004inferring, dobra2004sparse}, and the causal discovery problems \citep{pearl2000models, zhang2012kernel, peters2017elements}, where at the heart of all these settings, one can find a CI testing problem. 

In the recent work of \citet{shah2020hardness}, it is argued that the CI testing is provably a hard problem without assumptions being placed on the distribution of variables. Concretely, \citet{shah2020hardness} shows that no uniformly valid test\footnote{A test that controls the type I error at a predetermined significance level $\alpha$ for all absolutely continuous (with respect to the Lebesgue measure)  random variables $(X,Z,Y)$ that are conditionally independent} can have nontrivial power (power  exceeding $\alpha$) against any alternative hypothesis (a triple $(X,Z,Y)$ that are not conditionally independent).  By and large, this impossibility result can be perceived as a consequence of an interesting phenomenon that happens in the CI testing problem: while the space of the null distributions are separated from the alternatives, in fact the convex hull of the null space is a dense set in the alternative space with respect to the total variation metric \citep{shah2020hardness}.

The discouraging result of \citet{shah2020hardness} highlights the crucial role of the assumptions on the distribution of $(X,Z,Y)$ in the CI testing problem. This is a noteworthy observation that such assumptions may make the null space smaller, so the aforementioned no-free-lunch theorem can not be applied anymore. 
%In fact, focusing merely on a null subset, instead of the entire set of conditionally independent triples, sheds light on the existence of statistical tests with valid type I  error control and non-trivial statistical power in the CI testing problem. 
During the past few years, several methods have been developed for CI testing under different setups, such as  \citet{neykov2020minimax} for one-dimensional variables satisfying certain smoothness assumptions, and \citet{canonne2018testing} for discrete variables. Also there exists  quite a large body of work on model-specific methods, where a parametric model is assumed between the response and the covariates (assumptions on the law $\cL(Y|X,Z)$) \citep{liang2018bayesian, crawford2018bayesian, belloni2014inference}. There is also other concurrent work which goes beyond  testing for the conditional independence and aims at measuring the strength of dependency when the CI hypothesis does not hold; e.g.,  \citep{zhang2020floodgate, azadkia2019simple,newey2018cross,huang2022kernel}.

%\aj{Simultaneously, there has also been great effort to go beyond CI testing and try to capture the conditional dependency  power \citep{zhang2020floodgate, azadkia2019simple}.}
%e.g. \citep{shah2020hardness} assumes that the conditional regression models $\E[Y|Z],\E[Y|X]$ can be estimated perfectly.  

Another complementary line that has been pursued in the past few years is the model-X perspective \citep{candes2018panning}.  In this framework, contrary to the classic setup no assumption is made on the conditional law $\cL({Y|X,Z})$, rather it shifts the focus on $(X,Z)$ and requires an extensive knowledge on the law $\cL({X,Z})$. To emphasize the importance of the model-X setup, one should note that a set of CI tests that have been developed for a certain family of distribution $\cL({Y|X,Z})$ leads to type I error inflation under model misspecification. On the other hand, in many settings, you may have access to abundant unlabeled data which allows for good approximation of $\cL({X,Z})$. For example, in genetic studies \citep{peters2016comprehensive, cong2013multiplex} the joint distribution of covariates can be well approximated. In particular, \citet{wen2010using} proposed an estimator to approximate the covariance matrix of covariates for the genome-wide association study (GWAS), in which genetic distance information is used.

In this paper, we will focus on CI testing in the model-X setup. In this setting, we would like to examine the independence of a covariate $X\in \reals$ and a response value $Y\in \reals$,  while controlling for the effect of a potentially high-dimensional confounding covariate vector $Z\in \reals^q$. This is formalized via a hypothesis testing problem:
{
\begin{equation}\label{eq: CI hypothesis}
 H_0: X\indep Y|Z\,,\quad\quad H_A: X\notindep Y|Z\,.
\end{equation}
}
%The model-X framework implies that an ample amount of information on covariates is available, while the dependency rules $Y|XZ$ or $Y|Z$ remain unknown. In short, 
In the model-X CI testing problem, we are given access to the conditional law $P_{X|Z}$  along with $n$ i.i.d. observations $(X_i,Z_i,Y_i)$ as data, while the conditional laws $Y|X,Z$ or $Y|Z$ are unknown.  A large body of proposed CI tests in the model-X setup, such as the conditional randomization test (CRT) \citep{candes2018panning},  and the holdout-randomization test (HRT) \citep{tansey2018holdout} are based on constructing counterfeit data sets using the law $P_{X|Z}$, and scoring them by a certain score function $T$. In our work, we follow a similar strategy and propose novel schemes for scoring counterfeits that work at sample level as well as novel test statistics. Our test leverages the i.i.d-ness property of the observations to obtain high-resolution p-values with a very small number of conditional randomizations. We begin by thoroughly explaining the motivation behind our proposal as well our contributions, and then discuss the related work on model-X CI tests.

\subsection{Motivation and summary of contributions}

In model-X conditional independence testing, the Conditional Randomization Test (CRT) has proven to be highly effective, demonstrating strong statistical power when paired with a well-chosen score function and a large number of randomizations. However, in practice, the number of randomizations is often limited by data availability, problem-specific constraints, and the need to minimize computational overhead. Additionally, simpler classes of score functions are often employed to maintain interpretability and further reduce computational costs.

These constraints—limited randomizations and simpler score functions—can present challenges for the CRT and its variants. Specifically, a small number of randomizations reduces the resolution of p-values in the CRT family, making it more difficult to reject the null hypothesis and leading to lower statistical power. Additionally, when simpler score functions are used, and the score function is only moderately sensitive to the alternative hypothesis, certain difficult scenarios can be problematic for the CRT family, further diminishing its statistical power.

In this work, we introduce a novel conditional test, the \textbf{P}earson {C}hi-squared \textbf{C}onditional \textbf{R}andomization (PCR) test.  Similar to the CRT, the PCR test uses randomization to construct multiple counterfeits of the data and rank the original data among the counterfeits according to a score function. The score function can be based on arbitrary (potentially complex) predictive models. Unlike the CRT whose score function takes in the \emph{entire dataset}, the PCR test works with score function that applies to \emph{subgroups} of data, where by changing the size of groups, it can go from sample level to the entire dataset level. 

At its core, the PCR test utilizes the Pearson Chi-squared test, allowing for the flexible scoring of data subsets by leveraging the i.i.d-ness property of the samples. This approach enables the generation of high-resolution p-values with only a small number of conditional randomizations. This novel perspective—scoring data subsets and aggregating them through the Pearson $\chi^2$-test—provides several advantageous properties, as outlined below.

\subsubsection{Few randomizations: high-resolution p-values and speed-up}
For a data set $(\bX,\bZ,\bY)$ consisting of $n$ independent samples with $\bX\in \reals^{n\times d_x}$, $\bZ\in\reals^{n\times d_z}$ an $\bY\in\reals^{n\times d_y}$, the CRT constructs $M$ counterfeits $(\widetilde{\bX}_1,\bZ,\bY),...,(\widetilde{\bX}_M,\bZ,\bY)$ where $\widetilde{\bX}_j$ is sampled independently from the conditional law $P_{\bX|\bZ}(\cdot|\bZ)$. (By independence of samples, this means that the entries $\tX_{j,\ell}$ are drawn independently from the law $X|Z$, for $\ell \in\{1,2,\dotsc, n\}$.) Then, for score function $T$ define the normalized rank: 
 {
 \begin{equation}\label{eq: p-crt}
p = \frac{1}{M+1}\Big(1+\sum\limits_{j=1}^{M}\ind{\{T(\bX,\bZ,\bY)\geq T(\widetilde{\bX}_j,\bZ,\bY)\} } \Big)\,.
\end{equation}  
}
Given that $\widetilde{\bX}|\bZ,\bY\sim\cL(\bX|\bZ)$, under the null hypothesis we have 
{ 
$
 T(\bX,\bZ,\bY) \stackrel{d}{=} T(\tbX,\bZ,\bY) | \bZ,\bY\,,
 $
 }
% \vskip{-1cm}
so the original and counterfeit scores are i.i.d. and so exchangeable, and therefore the normalized rank $p$ follows a uniform distribution, provided that the number of counterfeits is sufficiently large.

The CRT interprets extreme values of the normalized rank (close to 0 or 1) as evidence against the null hypothesis. Specifically, it rejects the null hypothesis when the p-value is smaller than $\alpha$ or, in the case of two-sided tests (e.g., two-sided CRT \citep{wang2020power}), based on the two extreme $\alpha/2$ tails. When working with $M$ counterfeits, the smallest possible value for the normalized rank is $1/(M+1)$, which requires selecting a sufficiently large $M$ to perform tests at a small significance level $\alpha$. Achieving high-resolution p-values in CRT family tests, therefore, requires a large number of randomizations, which can become computationally prohibitive.

In contrast, the PCR test can be conducted with very few randomizations while still producing high-resolution p-values. We explore the speed-up factor of PCR in Section~\ref{sec:Computational}. Specifically, we demonstrate that in certain problem settings, using the same score functions, PCR can achieve higher statistical power with only \emph{one-fifth} the number of randomizations compared to CRT. We consider both regression-based and covariance-based score functions. It is important to note that for many model-X tests, such as CRT, dCRT, and the conditional permutation test (CPT) \citep{berrett2020conditional}, computational time increases linearly with the number of randomizations.

\subsubsection{High statistical power with simple score functions}

As PCR is formulated at a finer granularity level and operates directly with data points--or small groups, it facilitates a more comprehensive examination of deviations from the $[0,1]$ range compared to the final p-value of CRT test statistics. This finer approach potentially allows PCR to achieve considerable statistical power relative to a \emph{broader range of alternatives}. Specifically, in Section~\ref{sec: compare-crt-alternatives} we demonstrate that under certain conditional independence (CI) testing setups—particularly when simpler score functions are used—the CRT family can become powerless, even with an infinite sample size. In contrast, PCR, using the \textit{same} score functions, maintains robust power. We show that this occurs with both \textit{model-agnostic} score functions (e.g., marginal covariance) and those derived from \textit{fitted models} (e.g., LASSO).

Using moderately simple score functions, such as the coefficients of a fitted LASSO model, as described in the original CRT paper \citep{candes2018panning}, is very common in practice. This approach is motivated by the desire for interpretability, reduced computational overhead, and the uncertainty about how much the model must be enriched to detect rejections, especially given the possibility that the null hypothesis may hold true. In the model-X setup, where the assessment relies on residual values, designing score functions based on complex model fitting can lead to overfitting both the original and randomized datasets, making the relative comparison meaningless.

\subsubsection{Standard and Non-Standard Setups: High Flexibility}

The PCR test is fundamentally based on the Pearson Chi-squared test statistic, which has a rich history in various multinomial testing problems, including uniformity testing and tolerance testing. Our approach reduces the model-X CI testing problem to a multinomial testing framework, allowing us to leverage the extensive literature and techniques available for standard multinomial testing settings.

The PCR test not only offers flexibility for multinomial testing setups, but it can also incorporate recent advances in CRT frameworks to improve computational efficiency and robustness. Specifically, when the groups used in the PCR test statistic are of moderately large size, more complex functions—such as fitted LASSO coefficients or fitted neural network loss—can be employed as score functions. This flexibility allows many extensions designed to enhance the computational efficiency or robustness of the CRT, such as the Holdout Randomization Test (HRT)~\citep{tansey2018holdout}, the Distilled CRT~\citep{liu2020fast}, or the Conditional Permutation Test (CPT)~\citep{berrett2020conditional}, to be applied in scoring groups of data points, thereby improving the overall performance of the PCR test.

In non-standard settings, such as those involving covariate shifts—where different populations are pooled together or when data collection involves adaptivity—the corresponding multinomial testing problem is highly flexible and easy to modify. For example, \citet{xu2024covariate} recently utilized our framework to propose the Covariate Shift Corrected Pearson Chi-Squared Randomization (csPCR) test for conditional independence testing in model-X under covariate shift. They achieved this by applying importance weights and leveraging the data-point granularity of PCR test statistics.\footnote{The csPCR procedure by \citet{xu2024covariate} was developed after the release of this work and is based on our PCR framework.}

The rest of the paper presents the following contributions:

\begin{enumerate}
%\item Section~\ref{sec: crt-failure}: We discuss some alternatives which while stands far from the null distribution (i..e, the response value $Y$ and the covariate $X$ are highly dependent), yet the CRT is provably powerless in rejecting them, even with arbitrarily large sample size and number of counterfeits. Let us stress that this is for a ``fixed'' common score function, namely the marginal covariance score $T(\bX,\bZ,\bY) = n^{-1} \bX^\sT \bY$, and not for all score functions. The goal here is to discuss that under a broad set of alternatives, it is possible that the deviation from the null distribution (uniform) occurs at the central part and so missed by the CRT rejection rule.

\item Section~\ref{sec:PCR}: We present the PCR test statistic, and provide two rejection thresholds for it to control the size of the test under a target level 
$\alpha$. One threshold indicated by $\th_{L,\alpha}^{\mathsf{finite}}$ is guaranteed to control the size even in finite-sample regime, while the other threshold $\th_{L,\alpha}^{\mathsf{asymp}}$ controls the size for large enough sample size (asymptotic regime). Of course, the former turns out to be more conservative and in our numerical study we observe that for $n$ of order a few hundreds, the size of test is already controlled using the  threshold $\th_{L,\alpha}^{\mathsf{asymp}}$.    

\item Section~\ref{sec: pwr}: We provide a power analysis of the PCR test. Distance of alternative distributions to the set of null distributions is measured via a notion called `conditional relative density', which depends on both the joint law $\cL(X,Z,Y)$ as well as the score function. Our analysis reveals the role of different factors, such as sample size, number of counterfeits and number of labels which are the input parameters for the PCR test.  

\item Section~\ref{sec:pf}: As our power analysis reveals, the number of labels ($L$) used in the PCR test affects its power in a non-trivial way. Here, we suggest to run PCR test for different choices of $L$ and then use Bonferroni's correction to combine the resulting $p$-values into a valid $p$-value for the conditional independence hypothesis.

\item Section~\ref{sec:robust}: While in the model-X framework it is assumed that the conditional law $\cL(X|Z)$ is known, in practice one may need to estimate this distribution (e.g., from unlabeled data). In this section, we provide a more conservative version of the PCR test which is more robust to errors in estimating $\cL(X|Z)$, and avoids inflation in the type I error.
\item Section~\ref{sec: numerical}: 
We assess the performance of the PCR test  and its extensions using multiple synthetic datasets to measure its size and power. In addition, we apply our test to the Capital Bikeshare dataset. We then explore the potential benefits of our PCR test compared to other CRT-type procedures, highlighting its advantages in terms of power and computational efficiency through detailed numerical examples that consider various alternative hypotheses and different score function choices.
\end{enumerate}

\noindent\textbf{Notations.} Throughout the paper, we use the shorthands $[n]=\{1,2,...,n\}$ for an integer $n\geq1$,  also $a\wedge b=\min\{a,b\}$, and $a\vee b=\max\{a,b\}$ . We use the capital letters for random variables and the small letters for the specific values they may take. We use bold symbols for vectors and matrices. %For a function $f$ over $[0,1]$, let
%$\||f\||_{\infty}=\text{ess} \sup_{x\in [0,1]}^{} f(x)$, and $L^{\infty}([0,1]$ be the space of all measurable functions on $[0,1]$ with bounded $\|\cdot\|_\infty$ norm. 
%Further, let $\Phi(t)=\int\limits_{-\infty}^{t}e^{-x^2/2}dx/\sqrt{2\pi}$-- standard normal cdf,
% and $\Phi_{\chi^2_m}=P(m/2,x/2)$-- cdf of the Chi-squared distribution with $m$ degrees of freedom, with $P(.,.)$ being the regularized gamma distribution. 
 For random variables or vectors $U,V$, $\cL(U)$ represents the probability law (distribution) of $U$ and $\cL(U|V)$ represents the conditional distribution of $U$ given V. We write $U\stackrel{d}{=}V$ to indicate that $U$ and $V$ have the same distribution. 
 For an event $E$, we denotes its probability by $\prob(E)$.  We use $\stackrel{p}{\Rightarrow}$ to indicate convergence `in probability' and $\stackrel{d}{\Rightarrow}$ for convergence `in distribution'. 
 %Formally,  for a sequence of random variables $\{X_n\}_{n\in\naturals}$ and a random variable $X$, we write $X_n\stackrel{p}{\to}X$ if $\forall \eps>0$, we have $\lim_{n\to\infty}\prob(|X_n - X|> \eps) = 0$. 
 Throughout, $\phi(t) = e^{-t^2/2}/\sqrt{2\pi}$ is the Gaussian density and $\Phi(u) = \int_{-\infty}^u \phi(t)\de t$ is the Gaussian distribution. For positive sequences $a_n,b_n$ indexed by $n\ge 1$, we adopt the asymptotic notation $a_n \asymp b_n$ where there exists positive constants $c_1\le c_2$ and integer $N$ such that for $n\ge N$ we have $c_1 b_n \le a_n \le c_2 b_n$.

\subsection{Related literature on conditional randomization tests}
The Conditional Randomization Test (CRT) was originally proposed by \citet{candes2018panning} as a generic framework that exploits the distributional information $X|Z$ to control the type I error. A salient feature of CRT is that it is a valid test (controlling type I error) for any choice of score function $T$. This flexibility of the CRT allows for using any advanced black box predictive model, which plays a key role in achieving high statistical power for the CI testing problem. Of course, the specific choice of $T$ would impact the power of the test. Indeed, \citet{katsevich2020theoretical} prove that the most powerful model-X conditional independence test against any given point alternative is a CRT, and this is obtained by taking $T$ to be the corresponding likelihood score, which requires knowing the alternative distribution.  There are some common choices for the score function, such as marginal covariance \citep{wu2010screen,mcmurdie2014waste} or the absolute value of the Lasso coefficient for $\bX$ \citep{wu2010screen}, which do not require to know the alternative distribution.

 In \citet{wang2020power} the authors analyze the power of CRT in a high-dimensional linear regression setting for three different score functions: marginal covariance based scores,  the ordinary least square coefficient and the LASSO \citep{tibshirani1996regression}. Further, \citet{katsevich2020theoretical} shows that any valid CRT test $\phi^{\mathsf{CRT}}_T$ (for different score functions $T$) must also be valid conditionally on $Y,Z$, and this conditioning allows to reduce the composite null to a point null. Also as a result of Neyman-Pearson lemma it is argued that the CRT based on the likelihood score is the { most powerful \textit{conditionally} valid test} against a point alternative. { In addition, they leave this interesting question open that whether the CRT-based test is the most powerful test not only among \textit{conditionally} valid tests but also among \textit{marginally} valid tests. Given that there are, at the very least, marginally valid tests that do not meet the criteria of being conditionally valid.}  This work also considers MX(2) model under which only the first two moments of $X|Z$ are known (as compared to the vanilla CRT which requires the full knowledge of the law of $X|Z$), and proposes a MX(2) F-test building upon the generalized covariance measure statistics of~\citet{shah2020hardness}. In addition, this work derives the asymptotic power of the CRT against local semiparamteric alternatives of the form $H_1: \cL(Y|X,Z) = \normal(X^\sT\beta+g(Z), \sigma^2)$.

On the computational side, using  advanced black box predictive models in the CRT can be prohibitively daunting, due to the repetitive fittings of the score function on the resampled data. This issue is even exacerbated in multiple testing, where the CRT is used for the feature selection problem. In this approach, the CRT is run for each covariate separately to test its relevance to the response, conditioned on the other covariates. Such multiple usage of the CRT is computationally prohibitive in high-dimensional problems. Alternatively, one can use the model-X knockoff approach  proposed  by \citet{candes2018panning} to circumvent this issue, which of course assumes the knowledge of the covariates  joint distribution. Several recent works extended this procedure beyond the multivariate Gaussian distribution for a broader range of the covariates joint population, see \citet{sesia2017gene} for hidden Markov models,  and \citet{bates2020metropolized} which introduced the Metropolis knockoff sampling for cases where
the covariates are continuous and follow a graphical model. Despite the fact that the model-X knockoff procedure has alleviated the CRT computational burden, this benefit often comes at the cost of a lower statistical power \citep[Section 5.3]{candes2018panning}.  For high-dimensional linear models,~\citet{wang2020power} shows that the CRT provably dominates model-X knockoffs in the variable selection problem. More precisely, they show that under the high-dimensional linear setup, when the Benjamini–Hochberg (BH) procedure \citep{benjamini1995controlling}, or the adaptive p-value thresholding (AdaPT) procedure \citep{lei2018adapt} is applied on the CRT p-values, a higher statistical power is achieved in comparison to the model-X framework.

Several other methods have also been proposed recently to improve the heavy computational cost of CRT, such as the Holdout Randomization Test (HRT) \citep{tansey2018holdout} and the Conditional Randomization Test with Distillation (dCRT) \citep{liu2020fast}.  In \citet{berrett2020conditional} the authors  have proposed the Conditional Permutation Test (CPT)  to enhance the robustness of CRT with respect to approximation errors in the law of $X|Z$. { In addition,  to use CRT for variable selection with FDR control guarantee,  a natural choice is to apply the (BH) procedure \citep{benjamini1995controlling} on the $p$-values returned by the CRT. However, this can be challenging for problems with large number of predictors $p$, because at a significance level $\alpha$, in order to make at least one rejection the number of randomizations $M$ should be large enough such that $\frac{1}{M+1}\leq \frac{\alpha}{p}$. The reason is that the CRT $p$-values are inherently discrete and belong to the set $\{1/(M+1),2/(M+1),\dots,1\}$. For this end, \citet{li2021deploying} proposes sequential CRT that combines CRT $p$-values with Selective SeqStep+ procedure \citep{barber2015controlling} to address the variable selection problem. Our method can be seen as an alternate approach, where we leverage the i.i.d. property of  data samples to construct high-resolution $p$-values, using a small number of randomizations. }
%\vspace{-1.3cm}

\section{Pearson Chi-squared randomization (PCR) test}\label{sec:PCR}
Motivated by the issues of CRT discussed in the previous section, in this work we propose a novel test, called Pearson $\chi^2$ conditional   randomization (PCR) test.
%Con an arbitrary score function $T:\reals^{q+2} \rightarrow \reals$ that satisfies certain regularity conditions (will be provided later),  we propose the Pearson $\chi^2-$CI testing procedure. 
We start by describing the PCR test and its test statistic. We then characterize the null distribution of its statistic by which we propose two rejection thresholds, for finite and infinite sample regimes.
%We will characterize the asymptotic distribution of the Pearson $\chi^2$-CI test statistic, and provide two rejection thresholds that each controls the type I error rate in finite and asymptotic number of samples.  We start by describing the Pearson $\chi^2-$CI test statistic construction.

\subsection{PCR test statistic}

We construct the PCR test statistic in four main steps: 
\begin{description}

\item[{\bf Data grouping.}] We first split the entire data set $\cD=\{(X_j,Z_j,Y_j)\}_{j=1:n}$ into $\myg$ groups of equal size $\{\cG_i\}_{i=1:\myg}$. This means that $|\cG_i|=n/\myg$, for $i\in [n]$.  In this step, $\myg$ is an input value which is known upfront, and for simplicity we assume that $n$ is divisible by $\myg$ (otherwise remove the extra samples). Ideally, we want to have a moderately large value for $\myg$ as it will be used later as the number of samples for the uniformity testing problem in the multinomial model with the Pearson chi-squared test statistic.  

\item[{\bf Counterfeit sampling.}] This is a common step in model-X conditional independence testing methods, where for each group of data points $\cG_i$, for example  $\cG_i=\{(X_j, Z_j,Y_j), j=1,\dotsc, n/\myg\}$, several counterfeits of the form $\tcG_i=\{(\tX_j,Z_j,Y_j), j=1,\dotsc, n/\myg\}$ are constructed by sampling $\tX_j\sim \cL_{X|Z}(.|Z_j)$ while keeping $Y_j, Z_j$ intact.  As we will discuss a main distinction of our PCR test with other CRT approach is that the PCR test works with few number of counterfeits while, in CRT approach, one requires a large number of counterfeits (at least of order $1/\alpha$), given that the normalized rank statistic \eqref{eq: p-crt} is intrinsically discrete. 
 
%
% a significant distinction with other methods is the number of counterfeits required per each sample. In particular, the Pearson CI test works well with very few number of counterfeits per each sample, while in the CRTs, the final test statistic \eqref{eq: p-crt} is intrinsically discrete and requires a large number of counterfeits ($M$) to have valid statistical results.

\item[{\bf Score and label.}] Given a score function $T$, we first score each group $\tG_i$ and then label groups based on the relative ranking of the score of original groups among scores of its counterfeits. Specifically, we partition the range of possible ranks in to $L$ subsets, $S_1, \dotsc, S_L$, of equal size and assign label $\ell$ to groups whose score rank falls in $S_{\ell}$.  
Special cases of this idea (with $L=2$ labels and unbalanced groups) can be traced back in the conformal inference literature \citep{vovk2005algorithmic, lei2018distribution, lei2014distribution, romano2019conformalized}, where the sample quantile of non-conformity scores are compared to a certain threshold to construct prediction intervals.

\item[{\bf Uniformity testing in a multinomial model.}] Under the null hypothesis~\eqref{eq: CI hypothesis}, by using the exchangeability of data scores and their counterfeits scores, it is straightforward to see that each label occurs with equal frequency (with expected count of each label being $\myg/L$). In this step, we use the Pearson Chi-squared test statistic $U_{\myg,L}$ to test uniformity of label occurrences in a multinomial model with $\myg$ samples and $L$ labels. Note that, in general $L$ can scale with $\myg$, and as discussed in \citet{balakrishnan2019hypothesis}, the $\chi^2$ test can have bad power due to the fact that the variance of the $\chi^2$ statistics is dominated by small entries of the multinomial. A truncated version of $\chi^2$ statistic has been proposed by \citet{balakrishnan2019hypothesis} to mitigate this issue by limiting the contribution to the variance from each label. However, when testing for a uniform distribution, as in our case, the truncation becomes superfluous. This implies that in this case, the usual $\chi^2$ statistic inherits several appealing properties of the truncated $\chi^2$ statistic. In particular, \citet{balakrishnan2019hypothesis} showed that
truncated $\chi^2$ test is globally minimax optimal for the multinomial problem. It is worth noting that for the multinomial testing problem in high dimension ($L$ growing with $\myg$), the upper and lower bounds on the critical radius $\eps$ has been established in \citep{paninski2008coincidence, valiant2017automatic}. 
Concretely,    
%Similar phenomenon for uniformity testing in the high-dimensional setting $(\ell>n)$ in multinomial models is also noted in \citep{paninski2008coincidence, valiant2017automatic}, where 
it is shown that $O(\sqrt{L}/\eps^2)$ number of samples are sufficient and information-theoretically necessary for distinguishing uniform distributions from alternatives that are $\eps$ far in the $\ell_1$-ball, with success probability larger than $2/3$. 

\end{description}

A detailed description for construction of the PCR statistic is given in Algorithm~\ref{algorithm: model-xz}. {{It is worth mentioning that a common trait in randomization tests, in particular in Model-X setup, including CRT, HRT, distilled CRT, and our PCR test, is the inherent randomness in the procedure due to data splitting and draw of counterfeits. While these methods come with rigorous guarantee on type I error, the specific $p$-value may change depending on the random seed set for the procedure. A common approach to make this more stable (other than fixing the random seed) is to consider multiple runs of the procedure (often in a cross-validation scheme) and then combine the possibly dependent $p$-values using a multiplicity- corrected  method such as Bonferroni. }}

\subsection{Decision rule}\label{sec: dec-rule} We introduce two rejection thresholds for the hypothesis testing problem \eqref{eq: CI hypothesis} with the statistic $U_{\myg,L}$ given by~\eqref{eq:U_nell}. At significance level $\alpha$, the decision rule is based on the test statistic:

{
\begin{equation}\label{eq:decision rule}
\phi\left(\bX,\bZ,\bY\right)=\begin{cases} 1 & U_{\myg,L}\geq \th_{L,\alpha}\quad (\text{reject } H_0)\,,\\
	0&  \text{otherwise}\quad ~~(\text{accept }H_0)\,. \
\end{cases}
\end{equation}
}
For the threshold $\theta_{L,\alpha}$ we consider two proposals:
{
\begin{align}\label{eq:thresholds}
 \th_{L,\alpha}^{\mathsf{asym}}:= \chi^2_{L-1}(1-\alpha), 
 \quad \th_{L,\alpha}^{\mathsf{finite}}=L+\sqrt{\frac{2L}{\alpha}}\,,
 \end{align}
 }
 where $\chi^2_{L-1}(1-\alpha)$
denotes the $1-\alpha$ quantile of a $\chi^2$ distribution with $L-1$ degrees of freedom.
%\item $\th_{\ell,\alpha}^{\mathsf{finite}}=\ell+\sqrt{\frac{2\ell}{\alpha}}$
As we show in the next section, the size of PCR test is controlled asymptotically (as $n\to \infty$) with using $\th_{L,\alpha}^{\mathsf{asym}}$. In addition, by using $\th_{L,\alpha}^{\mathsf{finite}}$, we prove that the size is controlled at finite sample settings.

As clear form its description, and similar to the CRT, the PCR test looks for statistically significant deviations between the distribution of the rank of original scores and the uniform distribution. While CRT only examines the tails of the distributions, the PCR test examines the entire support by comparing the two distributions on $L$ bins (corresponding to labels) of equal size and is able to capture deviations occurring in the middle range as well as at tails.
{
{
\begin{algorithmic}[!t]
	\begin{algorithm}
	{ {
		\REQUIRE $n$ data points $(\bX_j,\bZ_j,\bY_j)\in {{\reals^{1\times d_x} \times \reals^{1\times d_z}\times \reals^{1\times d_y}}}$, a positive integer $\myg$ as the number of groups (let $\mys=n/\myg\in \integers$), a real-valued score function $T:\reals^{\mys \times d_x}\times\reals^{\mys\times d_z}\times\reals^{\mys \times d_y} \to \reals$, and integers $K,L\geq1$ (let $M=KL-1$).
		\ENSURE Test statistics $U_{\myg,L}$ for testing the conditional independence hypothesis \eqref{eq: CI hypothesis}. 
		        \begin{itemize} \item Split the data into $\myg$ groups $\{\cG_j=(\bX_j,\bZ_j,\bY_j)\}_{j=1:\myg}$ of equal size $s$, where $\cG_j\in  {{\reals^{\mys\times d_x} \times \reals^{\mys\times d_z}\times \reals^{\mys\times d_y}}}$.
		        %where $\bX_i\in \reals^s, \bY_i\in \reals^s,$ and $\bZ_i\in \reals^{s\times q}$. 
		        \end{itemize}
			\For{$j\in [\myg]$}{
			\begin{itemize}
			\item  Draw $M$ i.i.d. samples $\tbX_j^{(1)},..., \tbX_j^{(M)}$ from $\cL_{\bX|\bZ}(\cdot|\bZ_j)$.
			\item Construct $M$ counterfeit groups $\{\tcG_j^{(i)}=(\tbX_j^{(i)},\bZ_j,\bY_j)\}_{i=1:M}$.
			\item Use $T$ to score the initial group $\cG_j$ and its $M$ counterfeits $\tcG_j^{(1:M)}$. 
			%$(\tbX_j^{(1:M)},\bY_j,\bZ_j)$:
			\begin{eqnarray*}
			T_j &=&T(\cG_j)\,,\\
			\widetilde{T}_j^{(i)} &=& T(\tcG_j^{(i)}),\quad\text{ for } i \in[M]\,.
		    \end{eqnarray*}
	    \item Let $R_j$ denote the rank of $T_j$ among $\{T_j,\tT_j^{(1)},...,\tT_j^{(M)}\} $:
	    \[
	  R_j=  1+\sum\limits_{i=1}^{M}\ind{\left\{T_j \geq \tT_j^{(i)}\right \}}
	    \]
		\item Partition $[M+1] = S_1\cup \dotsc \cup S_L$ with $S_{\ell} := \{(\ell-1)K+1, \dotsc, \ell K\}$.
		Assign label $\ell_j\in \{ 1,2,...,L\}$ to group $\cG_j$ if $R_j\in S_{\ell_j}$.
		
				\end{itemize}
}

\For{$\ell\in \{1,2,\dotsc,L\}$}{
			\begin{itemize}
				\item Let $W_{\ell}$ be the number of groups with label $\ell$: \,	$W_{\ell}:=\Big|\big \{j\in \{1,2,...,\myg\}: \ell_j=\ell \big\}  \Big|\,.$
				
				\end{itemize}
		}
	\begin{itemize}[leftmargin = *]
\item
	Define the test statistic $U_{\myg,L}$ as follows
		\begin{align}\label{eq:U_nell}
		U_{\myg,L}=\frac{L}{\myg}\sum\limits_{\ell=1}^{L}\left(W_{\ell}-\frac{\myg}{L}\right)^2\,.
		\end{align}	
%\item At significance level $\alpha$, reject $H_0$ if $U_{n,\alpha}\geq \ell+\sqrt{\frac{2\ell}{\alpha}}$. 

%with $c_{m,1-\alpha}$ being the $\alpha$-th upper quantile of Chi-squared distribution with $m$ degrees of freedom. 
	\end{itemize}		
				\caption{PCR test statistic}\label{algorithm: model-xz}
		
		}}		
	\end{algorithm}
\end{algorithmic}

}} 
\subsection{Size of the PCR test}\label{sec:size}

 Under the null hypothesis, the original and counterfeit scores are coming from a similar population. Our next assumption on the continuity of random variables ensures that the different data points achieve distinct score values, with probability one. This symmetry on distinct values implies that each data point gets label $\ell \in[L]$ uniformly at random. In short, we change the problem of conditional independence testing into the uniformity testing problem on data points coming from a multinomial distribution. 
 %this is due to the aforementioned fact that under the conditional independence, data points are distributed uniformly on $\ell$ labels. 

\begin{assumption}\label{assum: mu_mapping}
	
	For a score function $T:\reals^{\mys \times d_x}\times\reals^{\mys\times d_z}\times\reals^{\mys \times d_y} \to \reals$, assume that the following conditional CDFs are continuous, for every pair $(\bz,\by)\in \reals^{\mys\times d_z}\times\reals^{\mys\times d_y}$:
	{
	\begin{eqnarray}
	F_{T|\bZ\bY}(t;\bz,\by)&:=&\prob_{{\bX|\bZ\bY}}\left(T(\bX,\bz,\by)\leq t|\bZ=\bz,\bY=\by\right)\,,\label{eq:FZY}\\
	F_{T|\bZ}(t;\bz,\by)&:=&\prob_{{\bX|\bZ}}\left(T(\bX,\bz,\by)\leq t|\bZ=\bz,\bY=\by\right)\,.\label{eq:FZ}
	\end{eqnarray}
	}
	Note that both $F_{T|\bZ\bY}$ and $F_{T|\bZ}$ are conditional on $\bY,\bZ$, and randomness is coming from $\bX$. The difference is that in $F_{T|\bZ\bY}$, we have $\bX\sim \cL(\bX|\bZ\bY)$, while in $F_{T|\bZ}$, we have $\bX\sim \cL(\bX|\bZ)$.  
\end{assumption}

It is worth noting that the above assumption, which is used to transform the conditional independence testing problem into a multinomial uniformity testing problem, is indeed a weak assumption. It is used to avoid ties when ranking the scores, and alternatively one can use a random tie-breaking decision rule and remove this assumption.  

In the next theorem, we show that by using $\th_{L,\alpha}^{\mathsf{asym}}$ in the decision rule \eqref{eq:decision rule} asymptotic control on type I error is guaranteed. It is an immediate consequence of characterizing the asymptotic distribution of $U_{\myg,L}$ statistic in Algorithm \ref{algorithm: model-xz}. Furthermore,  we show that deploying the rejection threshold  $\th_{L,\alpha}^{\mathsf{finite}}$ results in finite-sample control on the type I error.  
 %\begin{align*}
%&\lim\limits_{n\rightarrow \infty}^{}\prob\left( U_{n,\ell} \geq \th_{\ell,\alpha}^{\mathsf{asym}}| H_0 \text{ holds }\right) \leq \alpha\,,\\
% &~~\prob\left( U_{n,\ell} \geq \th_{\ell,\alpha}^{\mathsf{finite}}| H_0 \text{ holds }\right) \leq \alpha\,.
%\end{align*}
 \begin{thm}\label{thm: chi^2-CI-size}
 	Under the null hypothesis \eqref{eq: CI hypothesis} and Assumption \eqref{assum: mu_mapping} , the statistic $U_{\myg,L}$ constructed in Algorithm \ref{algorithm: model-xz} {{converges uniformly to the $\chi^2$ distribution with $L-1$ degrees of freedom, for $L\ge 2$. Concretely, let $V\sim \chi^2_{L-1}$. Then,
	\begin{align}\label{eq:uniform}
	\sup_{\eta\in\reals} \left|\prob(U_{\myg,L} \ge \eta) - \prob(V\ge \eta) \right| \le C\myg^{-1/2} (L-1)^{5/4} \,,
	\end{align}
	for an absolute positive constant $C$.  In addition, uniformly across $L,\alpha, \myg$},} we have  
	\[\prob\left( U_{\myg,L} \geq \th_{L,\alpha}^{\mathsf{finite}} \right) \leq \alpha\,, \quad  \text{with } \th_{L,\alpha}^{\mathsf{finite}}=L+\sqrt{\frac{2L}{\alpha}}\,.\]
\end{thm} 
We refer to Section \ref{proof:thm: chi^2-CI-size} for the proof of Theorem~\ref{thm: chi^2-CI-size}. It can be observed immediately that if $L = o(\myg^{2/5})$, the Type I error can still be controlled, since $\myg^{-1/2}(L-1)^{5/4} \to 0$ as $n_g \to \infty$.

Based on the above characterization of the null distribution, in finite sample and asymptotic regimes, we can construct the following $p$-values for the testing problem~\eqref{eq: CI hypothesis}:
%
% statistic $U_{n,\ell}$ and the decision rule~\eqref{eq:decision rule} with $\th^{\mathsf{asym}}_{n,\ell}$ and $ \th^{\mathsf{finite}}_{n,\ell}$ thresholds  we can construct the following $p-$values $P^{\mathsf{asym}}_{n,\ell}$ and $P^{\mathsf{finite}}_{n,\ell}$:
{
\begin{equation}\label{eq: p-val}
P^{\mathsf{finite}}_{\myg,L}=\begin{cases}
	1,& U_{\myg,L} \leq L\,,\\
	\min\left\{\dfrac{2L}{(U_{\myg,L}-L)^2},1 \right\},&\text{otherwise}\,.
\end{cases}
\end{equation}
\begin{equation}\label{eq: p-val-asymptotic}
P^{\mathsf{asym}}_{\myg,L} =1-{\sf F}_{L-1}(U_{\myg,L})\,,
\end{equation}
}
where ${\sf F}_k$ is the cdf of a chi-squared random variable with $k$ degrees of freedom.

Note that under the null hypothesis, $p$-value $P^{\mathsf{asym}}_{\myg,L}$ is asymptotically uniform, whereas $P^{\mathsf{finite}}_{\myg,L}$ is super-uniform for finite $\myg$.
{{Note that Theorem~\ref{thm: chi^2-CI-size}} gives us uniform control over the size of PCR test, as formalized below:
\begin{align}\label{eq:uniforms_pvals}
\prob\left( P^{\mathsf{finite}}_{\myg,L} \leq \alpha \right)\le \alpha\,, \quad \forall \myg, L \ge 1,  \quad
\underset{\myg\to\infty}{\lim\sup} \underset{P\in \mathcal{P}_0}{\sup} \prob\left( P^{\mathsf{asym}}_{\myg,L} \leq \alpha \right)\le \alpha\,, 
\end{align}
for all $\alpha\in[0,1]$, and $\mathcal{P}_0$ indicating the set of joint distributions on $(X,Z,Y)$ which satisfy the null hypothesis (conditional independence).
}

% Formally, for all $t\in [0,1]:$
%{\footnotesize
%\begin{align*}
%&\lim\limits_{\myg\rightarrow \infty}^{} \prob\left( P^{\mathsf{asym}}_{\myg,L} \leq t \right)=t\,, \\ 
%&\prob\left( P^{\mathsf{finite}}_{\myg,L} \leq t \right)\le t\,, \quad \forall \myg\ge 1\,.
%\end{align*}
%}
%i.e., $\lim\limits_{n\rightarrow \infty}^{} \prob\left( P^{\mathsf{asym}}_{n,\ell} \leq \alpha \right)=\alpha$, for all $\alph\in [0,1]$.  the p-value $P^{\mathsf{finite}}_{n,\ell}$ stochastically dominates the uniform distribution, and  it is more conservative. 
% Note that p-value $P$ constructed above is valid for all number of sample values $n$,  more precisely under the null hypothesis \eqref{eq: CI hypothesis} we have $\prob\left(P\leq \alpha\right)\leq \alpha$.  This is an immediate result of Theorem \ref{thm: chi^2-CI-size}. It worth noting that p-value $P$ is extracted from decision rule \eqref{eq:decision rule} with rejection threshold $\th_{\ell,\alpha}^{\mathsf{finite}}$. Using rejection threshold $\th_{\ell,\alpha}^{\mathsf{asym}}$ gives $p-$value $\tP=1-F_{\chi^2_{\ell-1}}(U_{n,\ell})$ which is  asymptoticly uniform, i.e $\lim\limits_{n\rightarrow \infty}^{}\prob(\tP\leq \alpha)=\alpha$ (here $F_{\chi^2_{k}}(.)$ is cdf of Chi-squared distribution with $k$ degrees of freedom.)\\\\

\section{A power analysis of the PCR test}\label{sec: pwr}
We next provide a power analysis of the PCR test. To this end, we need a notion of distance between a probability density function $p_{XZY}(x,z,y)$ and its corresponding conditional independence density $p_X(x)p_{Z|X}(z|x)p_{Y|Z}(y|z)$, where $p_{Y|Z}(y|z)$ is obtained by marginalizing out $X$, i.e.,
$p_{Y|Z}(y|z)= \int p_{Y|XZ}(y|x,z)p_{X|Z}(x|z) \de x$. As expected, the larger this distance, the easier to discern the conditional dependency. The metric that we use here to analyze the power of PCR test is a generalization of the notion of \emph{ordinal dominance curve} (ODC) \citep{hsieh1996nonparametric, bamber1975area}. For two densities $p$ and $q$ defined on the real line, the ODC is given by $F_p(F_q^{-1}(t))$, where $F_p, F_q$ respectively denote the cdfs corresponding to $p$ and $q$. In other words, the ODC is the population analogous of the PP plot. The derivative of the ODC (if exists) is given by $f_p(F_q^{-1}(t))/f_q(F^{-1}_q(t))$ and is called the \emph{relative density function} (\citep{thas2010comparing}, Section 2.4). 

We next define the conditional ODC and the conditional relative density function, along with two assumptions. Let us emphasize that the upcoming assumptions are made to facilitate the power analysis, and the validity of the PCR test (control on type I error) holds even without these assumptions. 
\begin{defi}\label{def: conditonal-odc}(Conditional ODC and relative density function).  For a score function $T$ defined in Assumption \ref{assum: mu_mapping}, recall the conditional cdfs $F_{T|\bZ\bY}(t;\bz,\by)$ and $F_{T|Z}(t;\bz,\by)$ given by equations~\eqref{eq:FZY} and \eqref{eq:FZ}. The conditional ODC is  $\sR_T:[0,1]\rightarrow [0,1]$ which is defined as
	{
	\[
	\sR_{T}(u)=\E_{(\bZ,\bY) \sim \cL(\bZ,\bY)}\left[ F_{T|\bZ\bY}\left( F_{T|\bZ}^{-1}\big(u; \bZ,\bY\big) ; \bZ,\bY\right)  \right]\,.
	\]
	}
	For Differentiable $\sR_{T}$, we 
	 call its derivative the conditional relative density function:
	 $\sr_{T}(u):=\frac{\partial}{\partial u}\sR_{T}(u)\,,$ for $u\in (0,1)$.
	 	
\end{defi}
{ We next assume that the conditional relative density function $r_T(\cdot)$ is bounded and Lipschitz. This assumption allows us to efficiently approximate the function using polynomials of degree $N$ with an approximation error of order $O\left(\frac{\log N}{N}\right)$, as detailed in \eqref{eq:unif-Bernstein-polynomials}.}

\begin{assumption}\label{assum: sr_continuity}
	Assume the conditional relative density function $\sr_{T}(u)$ is  $C$-Lipschitz continuous. This also implies that $\sr_{T}(u)$ is uniformly bounded, i.e., $\sup\limits_{u \in [0,1]} ^{} |\sr_{T}(u)| \leq B\,,$ for some positive constant $B$.
	\end{assumption}
	
Our next assumption is a sufficient condition to replace the order of the expectation and the derivative in the definition of $\sr_{T}(u)$ (see~\eqref{eq:interchange}). 
\begin{assumption}\label{assum: dR-L1-integrable}
We assume that
{
\[
\int_0^1 \E_{(\bZ,\bY) \sim \cL(\bZ,\bY)}\Big[ \Big|\frac{\partial}{\partial u}F_{T|\bZ\bY}\Big( F_{T|\bZ}^{-1}\big(u; \bZ,\bY\big) ; \bZ,\bY\Big)  \Big|  \Big] \de u <\infty \,.
\]
}
\end{assumption}

We are now ready to define a distance between the distribution of $(\bX,\bZ,\bY)$ and $(\tbX,\bZ,\bY)$ where $\tbX\sim\cL(\bX|\bZ)$, independently of $\bY$. Note that the two densities match under the null hypothesis~\eqref{eq: CI hypothesis}.

\begin{defi}\label{def: conditional-dependency-power}
	For a score function $T$ and its relative density function $\sr_{T}(.)$, define \textit{conditional dependency power}  as 
	$\Delta_T(\cL(\bX,\bZ,\bY))=\int\limits_{0}^{1}|\sr_{T}(u)-1|\de u\,.$
\end{defi}	
	We next state some properties of the measure $\Delta_T(\cL(\bX,\bZ,\bY))$. Recall that for two random variables $U, V$ with density functions $p, q$ (with respect to the Lebesgue's measure), the total variation distance is defined as $d_{\tv} = \frac{1}{2}\int_{-\infty}^\infty |p(t) - q(t)|\de t\,.$
	
	\begin{rmk}\label{rmk: Delta upper bound} The followings hold for the measure $\Delta_T(\cL(\bX,\bZ,\bY))$.
	\begin{itemize}
	\item[(a)] Under the null hypothesis \eqref{eq: CI hypothesis}, for any score function $T$ satisfying Assumption \ref{assum: mu_mapping} we have $\Delta_{T}(\cL(\bX,\bZ,\bY))= 0$.
	\item[(b)] The following upper bound holds in general:
	{
	\[\Delta_T(\cL(\bX,\bZ,\bY))\leq \E_{(\bZ,\bY) \sim \cL(\bZ,\bY)}\left[2 d_{\tv}\left(  (T(\tX,\bZ,\bY)|\bZ,\bY), (T(\bX,\bZ,\bY)|\bZ,\bY) \right)\right],		\]
	}
		with $\bX\sim \cL(\bX|\bZ,\bY)$ and $\tbX\sim \cL(\bX|\bZ)$.
	\end{itemize}
	\end{rmk}
We refer to Section~\ref{proof:rmk: Delta upper bound} for the proof of Remark~\ref{rmk: Delta upper bound}.
 As discussed earlier, the PCR test transforms the conditional independence problem into the problem of uniformity testing under a multinomial model. That said, in order to analyze the power of PCR test we focus on the later problem.
We use the results of~\citep{balakrishnan2019hypothesis} which characterize the power of truncated $\chi^2$-test for a high-dimensional multinomial model, in terms of the $\ell_1$ distance between the nominal probabilities and the uniform distribution over the categories. However, it is not clear how the nominal probabilities in the multinomial model are related to the distribution of $(\bX,\bZ,\bY)$ in the original conditional independence testing problem. Our next proposition answers this question and relates the $\ell_1$ distance between the nominal probabilities and the discrete uniform distribution, in the multinomial problem, to the measure $\Delta_T(\cL(\bX,\bZ,\bY))$ given in Definition~\ref{def: conditional-dependency-power}.

% and relate the  we need to capture the deviation of nominal probabilities from the uniformity in the multinomial model. The next proposition demonstrates a lower bound on $\ell_1$ distance of the transformed multinomial model from the uniform multinomial distribution. In fact, this lower bound is highly dependent on the conditional dependency power of the primary conditional independence problem, and implies that the higher the dependency power, the larger is the deviation.
%We first propose a closed-form relation on the probability that one triple $(X,Z,Y)$ gets label $s\in \{1,2,...,\ell\}$, then we will use this and establish the aforementioned  lower bound. 
\begin{propo}\label{propo: lambda_bound}
Under Assumption \ref{assum: dR-L1-integrable}, in Algorithm \ref{algorithm: model-xz}, each group $\cG_i\sim \cL(\bX,\bZ,\bY)$ admits label $t\in \{1,2,...,L\}$, independently from other data points with probability
{
	\begin{equation}\label{eq: tp_j}
		{p}_t= \sum\limits_{j=(t-1)K}^{tK-1} \binom{M}{j}\int\limits_{0}^{1}u^{j}\big(1-u\big)^{M-j}r_{T}(u)\de u\,,
	\end{equation} 
	}
where $\sr_T(.)$ is the conditional relative density function given by Definition \ref{def: conditonal-odc}. Under the null hypothesis \eqref{eq: CI hypothesis}, we have $p_{t}=\frac{1}{L}$. In addition, under Assumption \ref{assum: sr_continuity}, the partial sums of $\{p_t\}_{t=1}^\ell$  satisfies the following bounds:
	\begin{enumerate} [label=\roman*)]
		\item For every $\ell\in [L]$, we have 
		{
		\begin{equation}\label{eq:sum-p_ell-lower}
			\sum\limits_{t=1}^{\ell}p_t \geq \sR_{T}\left(\frac{\ell}{L}\right)\,,
		\end{equation}
		}
where $\sR_T(u)$ is the conditional dominance curve given by Definition \ref{def: conditonal-odc}.
		\item Let $D=C/2+2B$ with $B,C$ given according to Assumption \ref{assum: sr_continuity} and introduce $\nu_K:= 2\left(\frac{9D^2\log{K}}{\sqrt{K}}  \right)^{2/5}$. Then for $K$ sufficiently large such that $\nu_K<1$, we have 
		{
		\begin{equation}\label{eq:sum-p_ell-upper}
			\sum\limits_{t=1}^{\ell}p_t \leq \sR_{T}\left(\frac{\ell}{L}\right)+\nu_K
			\,.
		\end{equation}
		}
		\item 
		We have
		{
		\begin{equation}\label{eq: tmp10}
		\sum\limits_{\ell=1}^{L}\Big|p_\ell-\frac{1}{L}\Big|\geq 
		\left( \Delta_T(\cL(\bX,\bZ,\bY)) -L\nu_k-\frac{C}{L} \right)\,.	
		\end{equation}
		}
%\item 
%If the relative density function $r_T(u)$ satisfies a higher order smoothness condition, then in terms of the number of samples $\ell$, a better rate can be established in \eqref{eq: tmp10}. More precisely, if 
% $\sr_T(u)\in C^m{[0,1]}$ for some integer $m\geq 1$, then $C/\ell$ can be replaced with $C_m/\ell^m$ with  $C_m=\max\limits_{u \in [0,1] }^{} \sr_T(u)^{(m)}$ in $\eqref{eq: tmp10}$.
\end{enumerate}
\end{propo}
Proof of Proposition~\ref{propo: lambda_bound} is given in Section~\ref{proof:propo: lambda_bound}.
%Having a lower bound on the discrepancy between the nominal probabilities in the transformed multinomial model and the uniform multinomial distribution, 
With Proposition~\ref{propo: lambda_bound} in place, we are now ready to state the main result about the statistical power of our PCR test. We start by analyzing the power of the PCR test when it is used with the finite-sample threshold $\th^{\mathsf {finite}}_{L,\alpha}$.

\begin{thm}\label{thm: power_balls_bins}
Let $U_{\myg,L}$ be the PCR test statistic --output of Algorithm \ref{algorithm: model-xz}-- with the number of labels $L$, number of groups $\myg$, and number of counterfeits per sample $M$, where $M= KL-1$, and a score function $T$ that satisfies Assumptions \ref{assum: sr_continuity} and \ref{assum: dR-L1-integrable} with parameters $B,C$.  
%the Pearson $\chi^2-$CI test has size at most $\alpha$, i.e.
%\[
%\prob\left( U_{n,\alpha} \geq \ell+\sqrt{\frac{2\ell}{\alpha}} \right) \leq \alpha,\quad \text{ for all  } P_{X,Z,Y} \text{ s.t } X\indep Y|Z\,.
%\]
Suppose that for some $\beta>0$, the conditional dependency power $\Delta_T(\cL(\bX,\bZ,\bY))$ satisfies the following:
{
 \begin{equation}\label{eq: lower Delta}
\Delta_{T}(\cL(\bX,\bZ,\bY)) \geq \frac{32{L}^{1/4}}{\sqrt{\myg}}\left( \frac{1}{\sqrt{\alpha}} \vee \frac{1}{\beta} \right)^{1/2} + \frac{C}{L}+L \nu_K \,,
\end{equation}
}
with $\nu_K=2\left(\frac{9(C/2+2B)^2\log{K}}{\sqrt{K}}  \right)^{2/5}$, for $K$ sufficiently large such that $\nu_K<1$. Then the PCR test, used with the finite-sample threshold
$\th^{\mathsf{finite}}_{L,\alpha}$, achieves a power of at least $1-\beta$. Concretely, for all distributions $\cL(\bX,\bZ,\bY)$ satisfying \eqref{eq: lower Delta}, we have 
$ \prob\left(U_{\myg,L} \geq L+\sqrt{\frac{2L}{\alpha}} \right) \geq 1-\beta\,.$
\end{thm}
The proof of Theorem~\ref{thm: power_balls_bins} follows from Proposition~\ref{propo: lambda_bound}  and is given in Section~\ref{proof:thm: power_balls_bins}.
 
We next analyze PCR test power when it is employed with asymptotic threshold $\th^{\mathsf{asym}}_{L,\alpha}$.

{ 

\begin{thm}\label{thm: power-asym-full}
Let $U_{\myg,L}$ be the PCR test statistic-- output of Algorithm \ref{algorithm: model-xz}, with the number of labels $L$, and number of counterfeits per sample $M$, where $M= KL-1$, and a score function $T$ that satisfies Assumptions  \ref{assum: sr_continuity} and \ref{assum: dR-L1-integrable} with parameters $B,C$. In addition, suppose that
the following lower bound holds for the conditional dependency power $\Delta(\cL(\bX, \bZ, \bY))$ for a positive $\eps$:
\begin{equation}\label{eq: lower-Delta-asym}
\Delta(\cL(\bX,\bZ,\bY))\ge \eps+ \frac{C}{L}+{L \nu_K}\,,
\end{equation}
where $\nu_K=2\left(\frac{9(C/2+2B)^2\log{K}}{\sqrt{K}}  \right)^{2/5}$ for $K$ sufficiently large such that $\nu_K<1$. Then the PCR test, used with the asymptotic-sample threshold
$\th^{\mathsf{asym}}_{L,\alpha}$, achieves a full power. Concretely, for all distributions $\cL(\bX,\bZ,\bY)$ satisfying \eqref{eq: lower-Delta-asym}, we have 
\[
\lim\limits_{\myg\to \infty}^{} \prob\left(U_{\myg,L}\ge \th^{\mathsf{asym}}_{L,\alpha} \right) =1\,.
\]
\end{thm}

Proof of Theorem~\ref{thm: power-asym-full} uses the results of Proposition~\ref{propo: lambda_bound} and can be seen in Section~\ref{proof: asym-full-power}.

Theorem \ref{thm: power-asym-full} implies that for a fixed alternative with conditional dependency power satisfying \eqref{eq: lower-Delta-asym}, PCR achieves full statistical power (rejection with probability one), as the number of samples $n$ (accordingly the number of groups $\myg$) grows to infinity. 

To see a more discriminative power analysis for PCR with the asymptotic-sample threshold, and understand better the interplay between statistical power, conditional dependency power $\Delta(\cL(\bX,\bZ,\bY))$, and the significance level $\alpha$,  in the next theorem we consider a sequence of local alternatives that converge to a distribution satisfying the null hypothesis \eqref{eq: CI hypothesis}, as the sample size $\myg$ goes to infinity.  {Asymptotic power analysis—as the sample size grows to infinity—with respect to a sequence of local alternatives is very common in the statistical literature. Accordingly, we adopt the same setup used in \cite{lehmann2006testing} for the power analysis of the Pearson chi-squared test statistic (cf. Theorem 14.3.1).
}

{
\begin{thm}\label{thm: power_balls_bins_asympt}
Let $U_{\myg,L}$ be the PCR test statistic-- output of Algorithm \ref{algorithm: model-xz}, with the number of labels $L$, and number of counterfeits per sample $M$, where $M= KL-1$, and a score function $T$ that satisfies Assumptions  \ref{assum: sr_continuity} and \ref{assum: dR-L1-integrable} with parameters $B,C$. For some fixed (not growing with sample size) values {$\{h_\ell\}_{\ell\in [L]}$ with $\sum_{\ell=1}^L h_\ell=0$}, we consider a series of local alternatives sequenced by the number of groups $\myg$. More precisely,  we consider a sequence of alternatives with conditional relative density function $r_T^{(\myg)}(.)$ satisfying the following:
\begin{equation}\label{eq: alternatives}
\sum\limits_{j=K(\ell-1)}^{K\ell-1} \int\limits_{0}^{1} u^{j}(1-u)^{M-j} r_{T}^{(\myg)}(u)=\frac{1}{L}+\frac{h_\ell}{\sqrt{\myg}}\,, \quad \forall \ell \in [L]\,.
\end{equation}
In addition, suppose that the following lower bound holds for $\{h_\ell\}_{\ell\ge 1}$ values:
{
\begin{equation}\label{eq: tmp: power-asympt}
\sum\limits_{\ell=1}^{L}h_\ell^2 \ge \frac{1}{\sqrt{L}}
\cdot\left[\sqrt{3\log\frac{1}{\beta}}+\left( 3\log\frac{1}{\beta} +  2\sqrt{\log\frac{1}{\alpha}}+2\log\frac{1}{\alpha} \right)^{1/2}   \right]^2\,,
\end{equation}
for some values of $\alpha,\beta$ with $\min(\alpha,\beta)\le 1/2$.
}
 Then the PCR test deployed with the asymptotic threshold $\th_{L,\alpha}^{\mathsf{asym}}$ has asymptotic statistical power at least $1-\beta$ agains alternatives given in \eqref{eq: alternatives}. Formally, the following holds
{
 \[
 \lim\limits_{\myg\rightarrow \infty}^{}\prob\left(U_{\myg,\ell} \geq  \th^{\mathsf{asym}}_{L,\alpha} \right) \geq 1-\beta\,.
 \]
 }
 \end{thm}
 }

Proof of Theorem~\ref{thm: power_balls_bins_asympt} also uses the results of Proposition~\ref{propo: lambda_bound} and is deferred to Section~\ref{proof:thm: power_balls_bins_asympt}.
 
{
In formulation \eqref{eq: alternatives}, based on Proposition \ref{propo: lambda_bound} we know that when the conditional relative density function $r_T$ is equal to $1$, then the null hypothesis \eqref{eq: CI hypothesis} holds, specifically, when $r_T=1$, then $h_\ell$ must be zero. This implies that in the considered sequence of alternatives, as number of groups $\myg$ grows, the alternatives gets closer to the null distribution.

{\color{black}We emphasize that the results in Theorems \ref{thm: power-asym-full} and \ref{thm: power_balls_bins_asympt} (with $\th_{L,\alpha}^{\mathsf{asym}}$) hold for problem settings when the number of labels $L$ are fixed and does \textit{not} grow with the number of samples. However, the statement of Theorem~\ref{thm: power_balls_bins}  (with $\th^{\mathsf{finite}}_{L,\alpha}$) allows $L$ to scale with $n_g$.
The next remark is on PCR with the finite-sample threshold
$\th^{\mathsf{finite}}_{L,\alpha}$ and provides guidelines on the choice of the number of labels $L$, as the number of samples $n$ (and so the number of groups $n_g$) grows to infinity. }
}
%\mm{In asymptotic Pearson, number of labels does not scale with number of samples. We decided to remove the asymptotic threshold from the next section.}
\begin{rmk}\label{rmk: optimal-L}
%Let $U_{n,L}$ be the PCR test statistic-- output of Algorithm \ref{algorithm: model-xz}, with the number of labels $L$, and number of counterfeits per sample $M$, where $M= KL-1$. Then for both the asymptotic and finite thresholds, the growth rate $L\asymp n^{2/5}$ for number of labels $L$ in the PCR test is optimal .  Specifically, for fixed significance level $\alpha$ and type II error tolerance $\beta$, for $K$ sufficiently large, among all the values of number of labels $L$, the choice $L\asymp n^{2/5}$ has power of at least $1-\beta$ against the largest sets of alternatives. In addition, this set of alternatives is all the laws $\cL(X,Z,Y)$ such that  $\Delta(\cL(X,Z,Y))\gtrsim n^{-2/5}$.
{{Consider the PCR test with the finite-sample threshold
$\th^{\mathsf{finite}}_{L,\alpha}$}.} The lower bound~\eqref{eq: lower Delta}  on $\Delta_{T}(\cL(\bX,\bZ,\bY))$ is minimized for $L\asymp \myg^{2/5}$. This suggests that optimal scaling for the number of labels $L$ in the PCR test with the finite-sample threshold is $L\asymp \myg^{2/5}$ which results in a non-trivial power as long as $\Delta(\cL(\bX,\bZ,\bY))\gtrsim \myg^{-2/5}$. { In addition, in this setting having $K=O(\myg^4)$ would be sufficient to still get the optimal rate $\Delta(\cL(\bX,\bZ,\bY))\gtrsim \myg^{-2/5}$}.
% In addition, the PCR test with the optimal number of labels $L\asymp n^{2/5}$ for all can have the power of at least $1-\beta$.    
\end{rmk}
{
Table \ref{tab:L-conditions} summarizes the conditions on $L$ (number of labels) growth rate with respect to number of groups $n_g$ required for valid Type-I error control and valid/optimal Type-II error rates. 
\begin{table}[h]
\centering
\caption{Growth-rate conditions on $L$ for valid/optimal controlling Type-I and Type-II errors}
\label{tab:L-conditions}

\renewcommand{\arraystretch}{1.2}     % a bit more row height
\begin{tabular}{l p{4cm} p{4cm}}
\hline
            & \textbf{Type-I error}               & \textbf{Type-II error}              \\
\hline
$\theta^{\mathsf{finite}}_{L,\alpha}$ &
Every $L$ \newline
{(Thm.~\ref{thm: chi^2-CI-size})} &
$L = O(n_g^{2/5})$ \newline
{(optimal, Rmk.~\ref{rmk: optimal-L})} \\[2pt]

$\theta^{\mathsf{asym}}_{L,\alpha}$ &
$L = o(n_g^{2/5})$ \newline
{ (Thm.~\ref{thm: chi^2-CI-size})} &
$L = O(1)$ \newline
{(Thms.~\ref{thm: power-asym-full}, \ref{thm: power_balls_bins_asympt})} \\
\hline
\end{tabular}
\end{table}

}

%%%%%%%%%%%%%%%%%%%%%%%%%%%%
{
\subsection{Analytical assessment of the power advantage of PCR over CRT}

In this section, we present theoretical results illustrating the power advantages of our PCR test over the CRT, for certain CI testing setups. We focus on a regression scenario in which the CRT—using the marginal covariance score function $\smash{T(\mathbf X,\mathbf Y) = n^{-1}\,\mathbf X^\top \mathbf Y}$ can achieve at most $c_0\alpha$ power for an arbitrary but fixed $c_0>0$, even as both the \textit{sample size} and the \textit{number of counterfeits} grow to infinity.  The marginal covariance is a popular choice in many high-dimensional applications \citep{wu2010screen,mcmurdie2014waste}. In particular, \citet{wang2020power} analyze the power of CRT under high-dimensional linear regression with marginal covariance as the score function. 

The key insight behind this result is that, in certain settings, the normalized CRT scores  concentrate around central values. As a result, CRT achieves only trivial power since it checks deviations from the uniform distribution only in the two tails. We formalize this intuition in the next theorem.

\begin{thm}\label{coro: failure-crt}

  Consider the following model between response variable $Y$ and covariate $X$:
 
\begin{align}\label{eq:g} 
X\sim \normal(0,1)\,, \quad Y=g(X)+\eps\,,\quad \eps\sim \normal(0,1)\,,
\end{align}
where the regression function $g(x)$ is an even function. Define $\eta_g:= \left(\frac{1+\E[X^2g(X)^2]}{1+\E[g(X)^2]}\right)^{1/2}$.
Then, the followings hold:
 \begin{itemize} 
  \item[$(a)$]For any $\alpha\in (0,1/2)$,  the two-sided CRT at significance level $\alpha$ (rejecting $\alpha/2$-th upper and lower quantiles) run with marginal covariance test statistics has power smaller than $\frac{8}{\pi}(\eta_g^2+2\eta_g)$. Formally, for CRT p-value $p_{n}^{(M)}$ given in \eqref{eq: p-crt} we have
 \[
\lim\limits_{M\rightarrow \infty}^{} \lim\limits_{n\rightarrow \infty}^{}\prob\Big(  \Big|p_{n}^{(M)}  -\frac{1}{2} \Big|   \geq  \frac{1-\alpha}{2} \Big)\leq  \frac{8}{\pi}(\eta_g^2+2\eta_g) \,.
\]    
  
    \item[$(b)$] For any $\alpha\in (0,1/2-\gamma)$ with $\gamma>0$,  the one-sided CRT run with marginal covariance test statistics at significance level $\alpha$ (rejecting either $\alpha$-th upper or lower quantile) has power smaller than $\frac{\eta_g^2+2\eta_g}{2\pi\gamma^2}$. Formally, for CRT p-value $p_{n}^{(M)}$ given in \eqref{eq: p-crt} we have

\begin{align*}
\lim\limits_{M\rightarrow \infty}^{} \lim\limits_{n\rightarrow \infty}^{}\prob\left(  p_{n}^{(M)}   \geq  1-\alpha \right) \leq \frac{\eta_g^2+2\eta_g}{2\pi\gamma^2}\,,\quad
\lim\limits_{M\rightarrow \infty}^{} \lim\limits_{n\rightarrow \infty}^{}\prob\left(  p_{n}^{(M)}   \leq  \alpha \right) \leq \frac{\eta_g^2+2\eta_g}{2\pi\gamma^2}\,.
\end{align*}
\end{itemize}

\end{thm}
We defer the proof of Theorem~\ref{coro: failure-crt} to Section~\ref{proof:coro: failure-crt}. The next corollary follows from the above theorem.

\begin{coro}\label{coro: 1-sqrt}
For $g(x) = \frac{1}{\sqrt{\theta^2+x^2}}$, a simple algebraic calculation shows that $\eta_g \le \frac{5\theta}{\sqrt{2\pi}}$. Therefore, by having $\theta$ small enough (depending on $\alpha$), the power of CRT is less than $\alpha/2$. 
\end{coro}

To analytically underscore PCR’s power advantage in this setting—we compute the conditional ODC (Definition \ref{def: conditonal-odc}) for this regression setting with marginal covariance score function; the proof is presented in Section \ref{proof: propo-odc}.

\begin{propo}\label{propo: g-odc}
Consider the regression setting \eqref{eq:g} for independence testing with PCR using marginal covariance test statistics $T(\bX;\bY)=n^{-1}\bX^\sT \bY$ for groups of size $n$, i.e. $\bX,\bY\in \reals^n$. Then, the ODC function $R_{\mathsf{MC}}^{(n)}$ defined in Definition \ref{def: conditonal-odc} is given by
\[
R_{\mathsf{MC}}^{(n)}:=\prob\left(\bX^\sT \bY \le \Phi^{-1}(u) \|\bY\| \right)\,, \quad \forall u \in [0,1]\,.
\]
\end{propo}

We can empirically compute $R_T(u)$ and observe deviation from the uniform $45^{\circ}$ line as reflection for large conditional dependency power $\Delta(\mathbf X;\mathbf Y)$ as in Definition \eqref{def: conditional-dependency-power}.  We illustrate this for the regression setting outlined in the above corollary for $\smash{g(x)=\frac{1}{\sqrt{x^2+\th^2}}}$ for $\th\in\{0.01,0.1,0.5,1\}$ by plotting $R_T(u)$ from empirical simulations with 20{,}000 realizations of $(\bX,\bY)$ per $u$ for $n=10$. The estimated ODC functions $R_T(u)$, shown in Figure \ref{fig:pcr-odc}, reveal that as $\theta$ increases, the curves gets closer to the $45^\circ$ line. This behavior aligns with our expectation: a larger $\theta$ in $g_\theta(x)$ reduces the influence of $x$ on $g_\theta(x)$, accordingly 
weakens the dependence between $X$ and $Y$. 

In the next step, as the closed-forms solution for $R_{\mathsf{MC}}^{(n)}$ as given in Proposition \ref{propo: g-odc} is hard to characterize, we focus on regime when the number of samples in each group grow to infinity, and characterize the limiting dependency power.  Formally, in the next theorem we characterize $\lim_{n\to\infty}\Delta(\bX_n,\bY_n)$ for the regression setting outlined in Corollary \ref{coro: 1-sqrt}.

\begin{thm} \label{thm: one-Delta} 
Consider independence testing with PCR using marginal covariance test statistics $T(\bX;\bY)=n^{-1}\bX^\sT \bY$ for the regression setting \eqref{eq:g} for class of regression functions $g_\th(x)=\frac{1}{\sqrt{x^2+\th^2}}$ parametrized by $\th \ge0$ for groups of size $n$, i.e. $\bX,\bY\in \reals^n$.  Then, for $\eta (\th)$ given by 
\[
\eta(\th)=\left(\frac{2-\th\sqrt{2\pi}e^{\th^2/2}(1-\Phi(\th))}{1+\frac{\sqrt{2\pi}}{\th}e^{\th^2/2}(1-\Phi(\th))}\right)^{1/2}\,,
\]
the limiting conditional dependency power as a function of $\th$ can be formulated by
\[
\lim\limits_{n\to \infty}^{} \Delta(\bX,\bY)=\left|4\Phi\Big(\eta(\th) \Big(\frac{2\log \eta(\th)}{\eta(\th)^2-1} \Big)^{1/2}  \Big)- 4\Phi\Big(\Big(\frac{2\log \eta(\th)}{\eta(\th)^2-1} \Big)^{1/2}  \Big)\right|\,.
\] 
\end{thm}
In particular, we plot the limiting $\Delta(\mathbf{X},\mathbf{Y})$ as a function of $\theta$ in Figure \ref{fig:th}. It can be seen that, as $\theta$ increases, the dependency power diminishes, which is expected since for large $\theta$ the effect of $x$ on $g_\theta(x)$ becomes smaller.

We next confirm the theoretical findings in Theorem~\ref{coro: failure-crt} (trivial power of CRT), and Proposition~\ref{propo: g-odc} (non-trivial power of PCR) by a set of numerical experiments. More precisely,  we generate a data set $(\bX,\bY)$ with $n=1000$ data points according to \eqref{eq:g} with $g(x) = \frac{1}{\sqrt{10^{-6}+x^2}}$.
We run the two-sided CRT at significance level $\alpha = 0.1$ with $M = 1000$ counterfeits. The statistical power of CRT, averaged over $N = 10,000$ experiments turns out to be zero.
We also run the PCR test on the same example, with $L=5$ and different values for $K$, and with the same score function.  The number of counterfeits per each sample is therefore $M=5K-1$. In this experiment, the PCR test is considered with groups of size $4$ (with $\myg=250$) at significance level $\alpha=0.1$. We consider both of the rejection thresholds $\th^{\mathsf{asym}},\th^{\mathsf{finite}}$ for decision rule \eqref{eq:decision rule}. The PCR test achieves perfect power for different choices of $K\ge4$ (and so different numbers of counterfeits) for both of the rejection thresholds.

\begin{figure}[htbp]
  \centering
  %––––––––––––––––––––––––––––––––––––––––––––––––––––––––––––––––––––––––%
  % Subfigure (a): Delta.png
    \begin{subfigure}[b]{0.45\textwidth}
    \centering
    \includegraphics[width=0.8\textwidth]{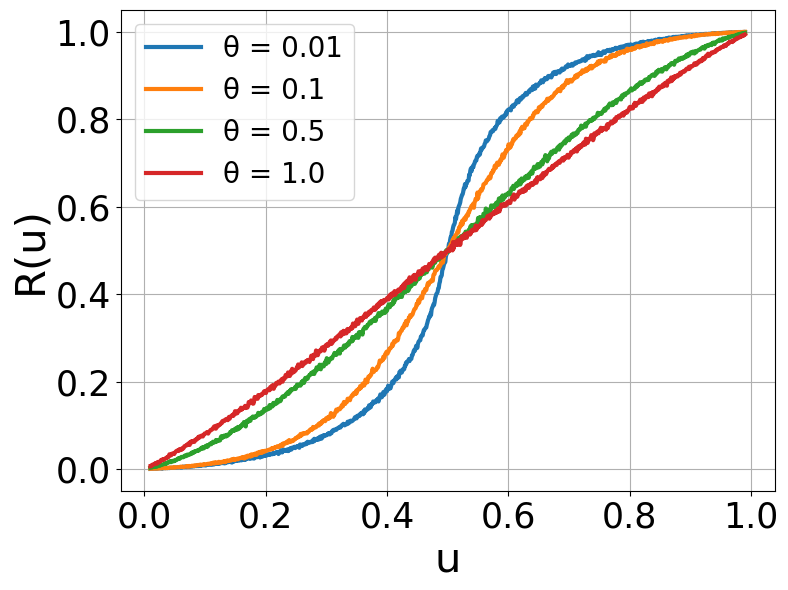}
    \caption{ODC function $R_T(u)$ for the regression setting of Corollary \ref{coro: 1-sqrt} for $n=10$ and $\th\in\{0.01,0.1,0.5,1\}$.} 
    \label{fig:pcr-odc}
  \end{subfigure}
  \hfill
  %––––––––––––––––––––––––––––––––––––––––––––––––––––––––––––––––––––––––%
  % Subfigure (b): R.png
    \begin{subfigure}[b]{0.45\textwidth}
    \centering
    \includegraphics[width=0.8\textwidth]{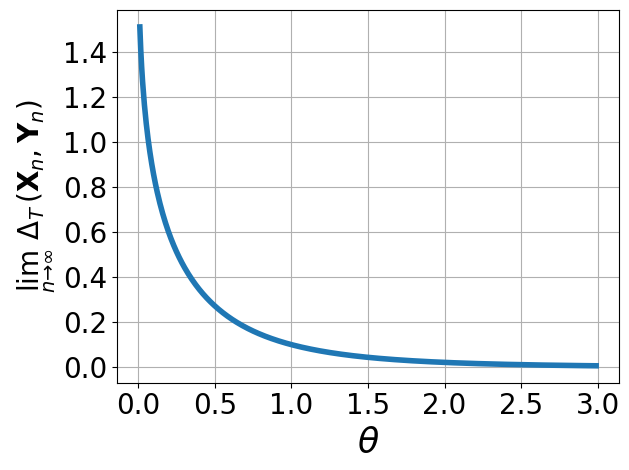}
    \caption{Precise dependency power $\lim\limits_{n\to \infty}\Delta_T(\bX,\bY)$ for the setting of Corollary \ref{coro: 1-sqrt} as a function of $\th$. }
    \label{fig:th}
  \end{subfigure}%––––––––––––––––––––––––––––––––––––––––––––––––––––––––––––––––––––––––%
  \caption{ODC function and dependency power for the regression setting of Corollary \ref{coro: 1-sqrt}.}
  \label{fig:side-by-side}
\end{figure}

 We next prove a (stronger) converse of Theorem~\ref{coro: failure-crt}, showing that no analogous statement can hold for PCR.  Informally, it states that whenever the CRT has non-trivial power, the PCR test will also have non-trivial power (it achieves any power $1-\beta$, provided a large enough sample size). \emph{Note that this result holds for any alternative hypothesis and any choice of score function.} We refer to Section~\ref{sec:converse} for its proof.

\begin{thm}\label{thm: nontrivial-PCR}
Consider an alternative hypothesis, under which the CRT achieves a not-trivial power, with a proper choice of score function. This in particular implies that the distribution of normalized rank deviates from the uniform distribution, i.e., the CRT p-value $p_n^{(M)}$ as given in \eqref{eq: p-crt} satisfies the following

\begin{equation}\label{eq: nontrivial-CRT}
\prob\Big(p_n^{(M)}\leq \alpha\Big)  \ge \alpha+ \delta\,,
\end{equation}
for some $\delta>0$. Consider the PCR test with $L$ number of labels (with $L\ge 1/\alpha$) and $n_g$ groups, each of size $n$ (so the total sample size of $nn_g$).  {Then, the PCR test asymptotically achieves full statistical power; more precisely we have
$\lim\limits_{\myg\to \infty}^{}\prob\left(U_{\myg,L}\geq \th^{\mathsf{asym}}_{L,\alpha}  \right) =1\,. $} In addition, if the gap value $\delta$ satisfies the following lower bound,
{
\begin{equation} \label{eq: lower-delta-finite}
\delta\geq \frac{32{L}^{1/4}}{\sqrt{\myg}}\left( \frac{1}{\sqrt{\alpha}} \vee \frac{1}{\beta} \right)^{1/2}\,,
\end{equation}
}
then the PCR test with the finite-sample threshold $\th^{\mathsf{finite}}_{L,\alpha}$ achieves statistical power larger than $1-\beta$, formally $\prob\left(U_{\myg,L}\geq \th^{\mathsf{finite}}_{L,\alpha}  \right) \geq 1-\beta\,.$
\end{thm}

}

\section{Parameter-free PCR test }\label{sec:pf}

The PCR test statistic described in Algorithm \ref{algorithm: model-xz} takes the parameters $K$, $L$ as input. In general, having a large $K$ (for fixed value of $L$) results in large value of $M$ (the number of counterfeits) and hence increases the statistical power of the test because we can better discern the discrepancy between the distribution of the ranks and the discrete uniform distribution. This benefit of course comes at a higher computational cost for constructing the test statistic.
The choice of $L$  (total number of labels) is however more subtle. On the one hand, a large value of $L$ implies that many of the labels occur rarely, which makes it challenging to point out significant deviations from the discrete uniform distribution (too many weak effects). On the other hand, a small value of $L$ results in a few bins over which we are comparing the test statistic with discrete uniform. In this case the test may miss sharp deviations as they are aggregated by the relatively large number of other points in the same bin. Similar observation can be made from the results of Theorem~\ref{thm: power_balls_bins} (and Theorem~\ref{thm: power_balls_bins_asympt}) where the right-hand side of~\eqref{eq: lower Delta} (and~\eqref{eq: tmp: power-asympt}) has a term decreasing in $L$ and a term increasing in $L$. Thereby, $L$ should be perceived as a tuning parameter in Algorithm \ref{algorithm: model-xz}.  

As we showed in Theorem \ref{thm: chi^2-CI-size}, any choice of $L$ results in a test with type I error control; however different choices of $L$ gives different statistical powers. %One would like to choose an $\ell$  value that gives a lower $p$-value, therefore higher number of discoveries. 
A natural approach is to run the PCR test multiple times, each time with a different value of $L$, and then `pick' the one that results in the smallest (most significant) $p$-value. However, this approach clearly violates the validity of the reported $p$-value, as we should account for the `cherry-picking'. Also, note that the obtained $p$-values (with different choices of $L$) are dependent as they are constructed from a common data set. To properly combine the $p$-values, we use the Bonferroni's method. Algorithm \ref{algorithm: model-xz-simes} describes this idea and presents a parameter-free version of Algorithm \ref{algorithm: model-xz}. The next theorem follows readily from Theorem \ref{thm: chi^2-CI-size} along with union bounding for the Bonferroni's correction. 

{
\begin{algorithmic}[t]
	\begin{algorithm}
		{{
		\SetAlgoLined
		\REQUIRE $n$ data points $(\bX_j,\bZ_j,\bY_j)\in \reals^{1\times d_x} \times \reals^{1\times d_z}\times \reals^{1\times d_y}$, significance level $\alpha\in(0,1)$, a real-valued score function $T:\reals^{\mys \times d_x}\times\reals^{\mys\times d_z}\times\reals^{\mys \times d_y} \to \reals$, $K\ge 1$ and a gird of $N$ values $\{L_1,...,L_N\}$.\\
		\ENSURE Decision on the conditional independence hypothesis \eqref{eq: CI hypothesis}.\\
	
		\For{$i\in [N]$}{
		\begin{itemize}
			\item Run Algorithm \ref{algorithm: model-xz} with $L=L_i$ labels to get test statistic $U_{\myg,L_i}$.
			\item Construct $p$-value $P_i$ using \eqref{eq: p-val} (for finite sample)  or \eqref{eq: p-val-asymptotic} (for asymptotic case).
			\end{itemize}
	}
	\begin{itemize}[leftmargin=*]
	\item Reject the null hypothesis if $P^*:= N\min\limits_{i\in[N]}^{} P_i\leq \alpha$.
	\end{itemize}

	%\item Sort $P_1,...,P_N$ and denote them by $P^{(1)}\leq P^{(2)}\leq...\leq P^{(N)}$.

		%\ENSURE {Reject the null hypothesis if $P^*:=\min\limits_{s\geq 1}^{} P_s\leq \alpha/N$. } 
		\caption{Parameter-free PCR test}\label{algorithm: model-xz-simes}
		}}
	\end{algorithm}

%Having p-value $P^*$ allows us to naturally consider the following decision rule at significance level $\alpha$.
%
%\begin{equation}\label{eq:decision rule-simes}
%	\phi\left(X,Z,Y\right)=\begin{cases} 1 & P^*\leq \alpha \quad ~(\text{reject } H_0)\,,\\
%		0&  \text{otherwise}\quad (\text{accept }H_0)\,. \
%	\end{cases}
%\end{equation}

\end{algorithmic}
}
%\vspace{-1cm}
\begin{thm}
{ Under the null hypothesis \eqref{eq: CI hypothesis}, the p-value $P^*$ constructed in Algorithm \ref{algorithm: model-xz-simes} for the finite-sample threshold is super-uniform, i.e. $\prob\left(P^*\leq t\right) \leq t$, for all $t\in [0,1]$.  In addition, for the asymptotic-sample threshold, the p-value $P^*$ is asymptotically super uniform, where we have 
$\lim_{\myg\to \infty} \prob(P^*\le t) \le t$, for all $t\in [0,1]$.}
  \end{thm}

{{\begin{rmk}
Note that the p-values $P_i$, $i\in [N]$ in Algorithm~\ref{algorithm: model-xz-simes} are in general dependent and the Bonferroni's combination is used to correct for that.
However, it will often be conservative, resulting in the test size smaller than the target level. In addition, as a practice guideline, we suggest to choose $L_i = 2^i$, for $i\in[N]$
with $N\le \log(n_g/50)$ so that the sample size $n_g$ remains significantly larger than the  number of labels $L$.  
\end{rmk}}}

\section{Robustness of the PCR test}\label{sec:robust}
In this section, we investigate the conditional independence problem when the exact conditional distribution $P_{X|Z}$ is not available; rather we use $\hP_{X|Z}(\cdot|Z)$  an estimate of $P_{X|Z}(\cdot|Z)$ for sampling the counterfeits. We would like to modify the PCR test so it still controls the type I error, when access to the exact conditional law $P_{X|Z}(\cdot|Z)$ is not feasible. To this end, the next theorem introduces a new test statistic which is based on the discrepancy between conditional laws $P_{X|Z}(\cdot|Z)$ and  $\hP_{X|Z}(\cdot|Z)$ along with the rejection thresholds for both the asymptotic setting and the finite-sample setting. We use the expected total variation metric to assess the distance between conditional laws. 

{
\begin{thm}\label{thm:robust-model X}
Let $W_\ell$, for $\ell\in [L]$, be the number of  groups with label $\ell$ as defined in Algorithm \ref{algorithm: model-xz}. For $\delta$ such that $\E_{\bZ}\left[d_{\tv}\left(P_{\bX|\bZ}(.|\bZ),\hP_{\bX|\bZ}(.|\bZ)  \right)\right]\leq \delta$,
introduce
{
\begin{equation}\label{eq:robust}
	\begin{split}
		U_{\myg,L}(\delta):=\min_{\{p_\ell\}_{\ell\in[L]}} \quad & \frac{L}{\myg(1+L\delta)}\sum\limits_{\ell=1}^{L}{\left(W_\ell-\myg p_\ell\right)^2}\\
		{\rm s.t.} \quad\quad  &p_\ell\geq 0,\quad  |p_\ell-1/{L}|\leq \delta,\quad \textrm{ for } \ell\in[L]\,,\\
		&{\rm and }\;\;\sum\limits_{\ell=1}^{L}p_\ell=1\,. \\
	\end{split}
	\end{equation}
}
Recall the thresholds $\th^{\mathsf{finite}}_{L,\alpha} $ and $\th^{\mathsf{asym}}_{L,\alpha}$ from~\eqref{eq:thresholds}. Under the null hypothesis, we have the following relations:
{
\begin{align}
&\prob\left( U_{\myg,L}(\delta) \geq  \th^{\mathsf{finite}}_{L,\alpha}   \right) \leq \alpha\,,\label{eq:robust-fin}\\
&\lim\limits_{\myg\rightarrow \infty}^{}  \prob\left( U_{\myg,L}(\delta) \geq \th^{\mathsf{asym}}_{L,\alpha} \right) \leq \alpha\,.\label{eq:robust-asym}
\end{align}
}
%\[
%V_\delta=\min \limits_{p \in \reals^\ell }^{} \sum\limits_{s=1}^{\ell}\left(W_s-{n}p_j\right)^2
%\]
 \end{thm}
 }
We refer to Section~\ref{proof:thm:robust-model X} for the proof of Theorem~\ref{thm:robust-model X}.
Note that optimization~\eqref{eq:robust} is a quadratic programming and can be solved efficiently. Also, statistic $U_{\myg,L}(\delta)$, given as the optimal value of this optimization, is a decreasing function with respect to $\delta$ and when there is no mismatch between the true and the approximate version ($\delta = 0$), we recover the primary statistic $U_{\myg,L}$ that was given by Algorithm \ref{algorithm: model-xz}.

As an immediate corollary of Theorem~\ref{thm:robust-model X} we can construct valid $p$-value for testing the conditional independence (i.e., super-uniform under the null hypothesis~\eqref{eq: CI hypothesis}), following the same recipe given by~(\ref{eq: p-val}-\ref{eq: p-val-asymptotic}), but using $U_{\myg,L}(\delta)$ instead of $U_{\myg,L}$. 
{ In the next theorem, we provide an upper bound on  type-I error inflation, for the case that the standard test statistics $U_{n,L}$ is adopted while randomizations are drawn from the estimate conditional law $\hP_{X|Z}$ 

{
\begin{thm}\label{thm: robust-inflation}
Under the null hypothesis \eqref{eq: CI hypothesis}, consider the test statistic $U_{\myg,L}$ constructed in Algorithm \ref{algorithm: model-xz} with the approximate conditional law $\hP_{X|Z}$. The followings hold:
{
\begin{align*}
\prob(U_{\myg,L}\geq \th_{L,\alpha}^{\mathsf{finite}}) &\leq \alpha + \E\left[d_{\tv}(P^n_{X|Z},\hP^n_{X|Z}) \right] \,, \\
\lim\sup\limits_{\myg\to \infty}^{}\prob\left( U_{\myg,L} \geq \th_{L,\alpha}^{\mathsf{asym}} \right) &\leq \alpha+ \lim\sup\limits_{n \to \infty}^{}  \E_{\bZ}\left[d_{\tv}(P^n_{X|Z}(.|\bZ),\hP^n_{X|Z}(.|\bZ)) \right]\,,
\end{align*}
}
where $\th_{L,\alpha}^{\mathsf{finite}},\th_{L,\alpha}^{\mathsf{asym}}$ are given by~\eqref{eq:thresholds}.
\end{thm}
}
The proof of Theorem \ref{thm: robust-inflation} is deferred to Section \ref{proof: robust-inflation}.   It is worth noting that in the model-X setup, $\hp_{X|Z}$ is often approximated via a set of unlabeled samples $\{(\tX_j,\widetilde{Z}_j)\}_{j=1:N}$. 
Specifically, when $P_{X|Z}$ belongs to a parametric family with $k$ parameters and $N\gg kn$,  the aforementioned total variation distance is of order $o_p(1)$. We refer to \citep[Section 5.1]{berrett2020conditional} for a detailed discussion of conditions under which $\E[d_{\tv}(P^n_{X|Z},\hP^n_{X|Z})]=o_p(1)$. 
}

\section{Numerical Experiments}\label{sec: numerical}

\subsection{Size, power, and robustness of PCR}
In this section, we evaluate the performance of PCR test and its extensions on synthetic datasets.  We consider groups each of size $1$ ($\myg=n$), unless otherwise is stated.

\medskip

{
\begin{figure}[t]
\begin{minipage}[t]{.45\textwidth}
\begin{center}
\begin{subfigure}[b]{0.3\textwidth}
		\centering
		\includegraphics[width=1.1\linewidth]{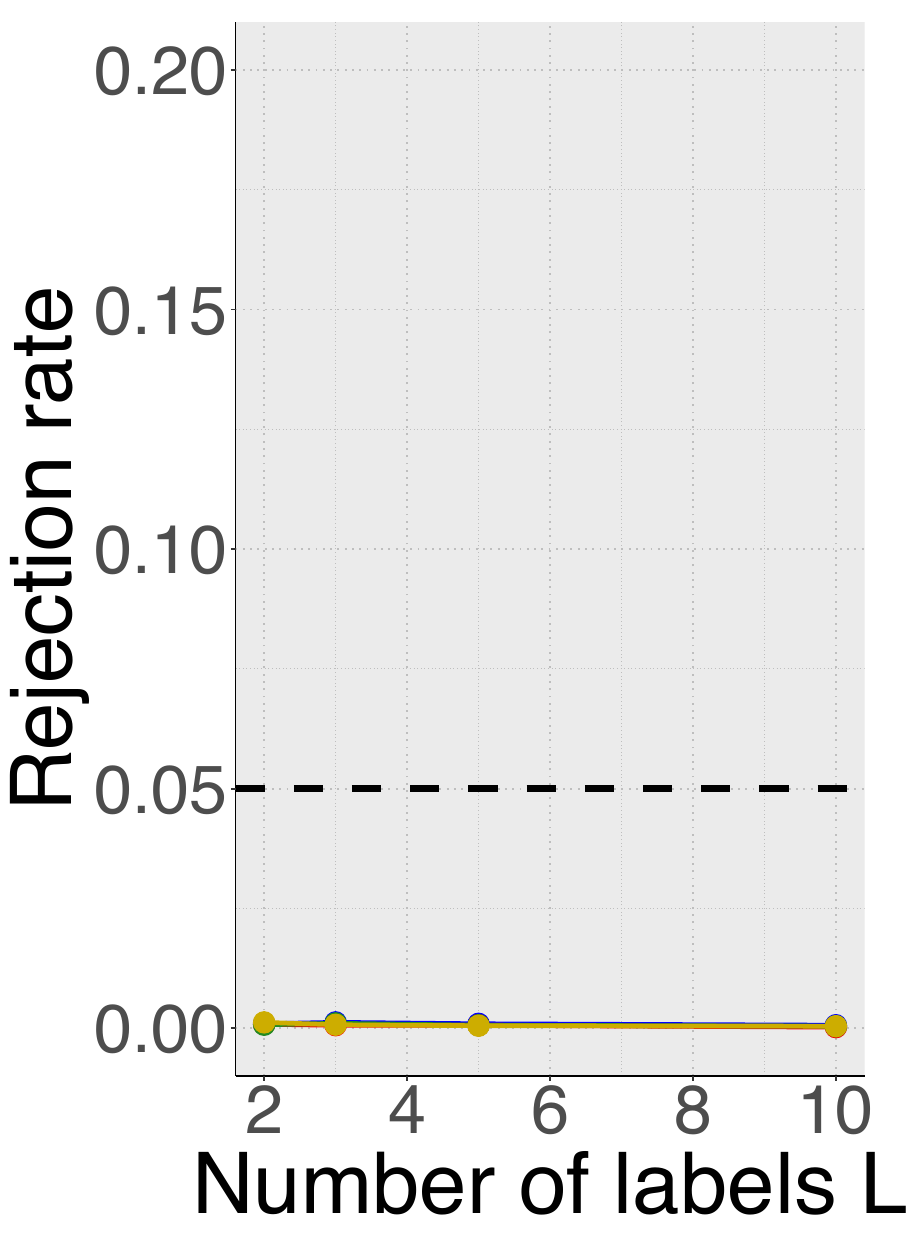}
		\caption{$\alpha=0.05$}
		\label{fig:one_sided}
	\end{subfigure}%
	\hfill
	\begin{subfigure}[b]{0.3\textwidth}
		\centering
		\includegraphics[width=1.13\linewidth]{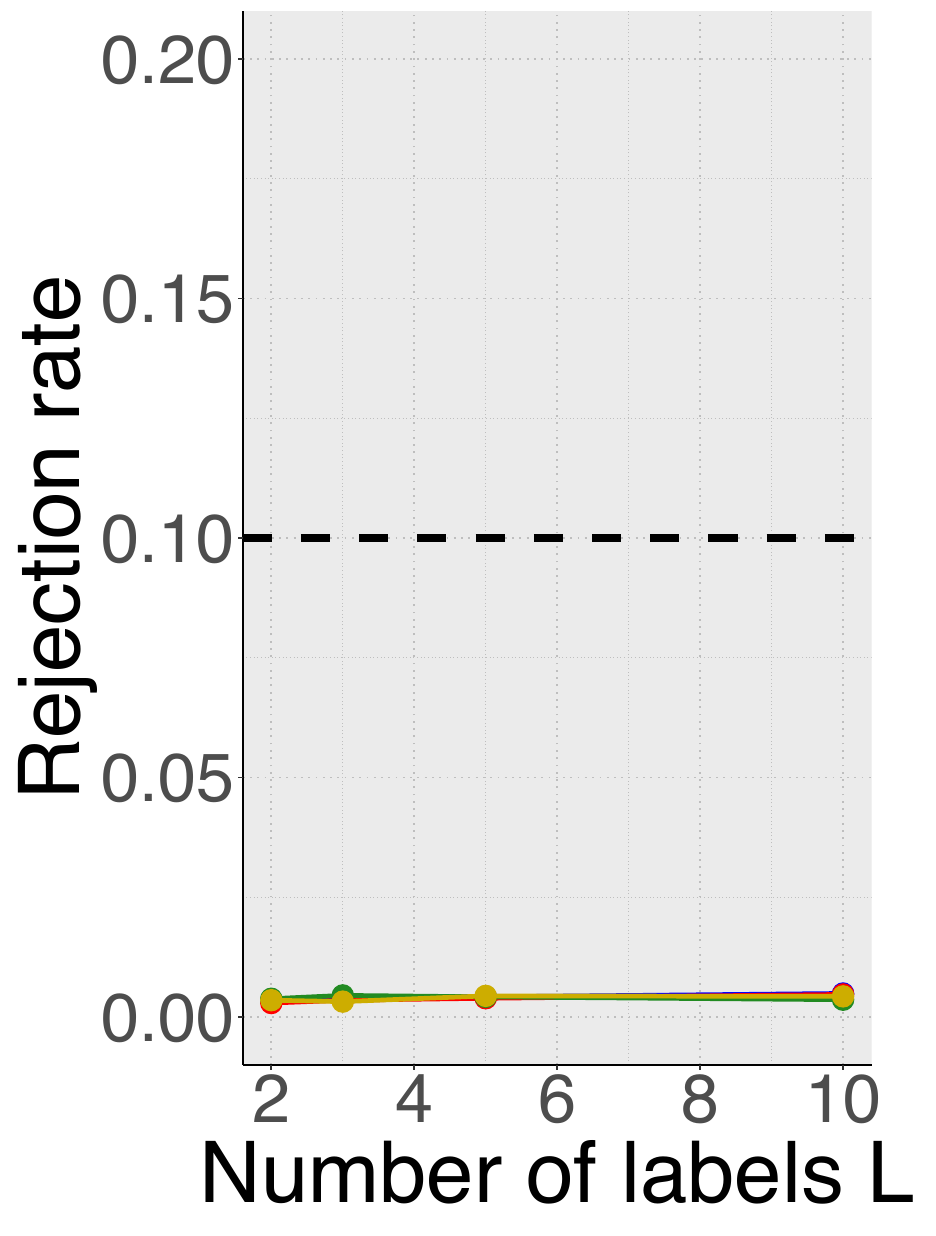}
		\caption{$\alpha=0.1$}
		\label{fig:two_sided}
	\end{subfigure}%
	\hfill
	\begin{subfigure}[b]{0.3\textwidth}
		\centering
		\includegraphics[width=1.5\linewidth]{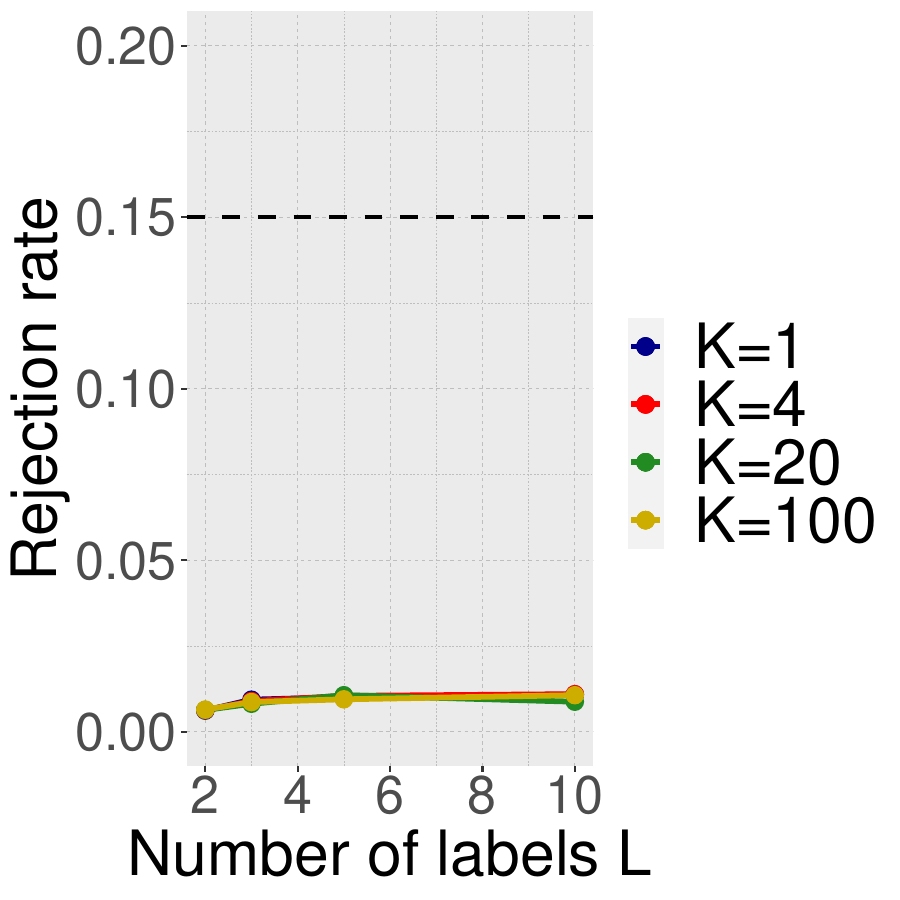}
		\caption{$\alpha=0.15$}
		\label{fig:middle}
	\end{subfigure}
	\caption{
	{{Size of PCR test with $\th^{\mathsf{finite}}_{L,\alpha}$ for dataset of size $n=100$ drawn iid from \eqref{eq:example-size}. Three significance levels $\alpha=0.05,0.1,$ and $0.15$ are considered. Reported numbers are averaged over $10,000$ trials. }}
	\vspace{0.4cm}}
	\label{fig: test-size-finite}

\end{center}
\end{minipage}% 
\hfill \begin{minipage}[t]{.45\textwidth}
\begin{center}

\begin{subfigure}[b]{0.3\textwidth}
		\centering
		\includegraphics[width=1.15\linewidth]{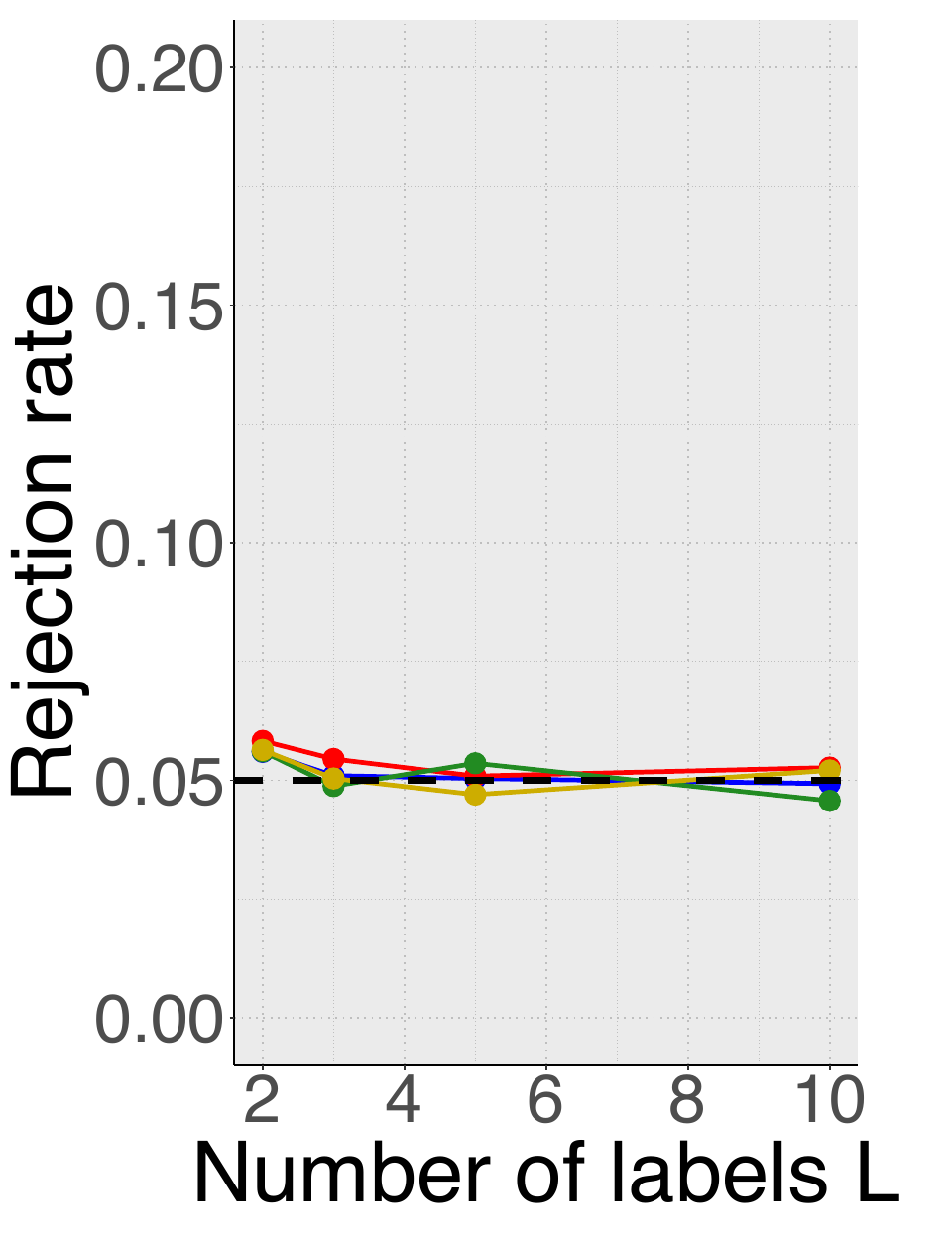}
		\caption{$\alpha=0.05$}
		\label{fig:one_sided}
	\end{subfigure}%
	\hfill
	%\hspace{1cm}
	\begin{subfigure}[b]{0.3\textwidth}
		\centering
		\includegraphics[width=1.11\linewidth]{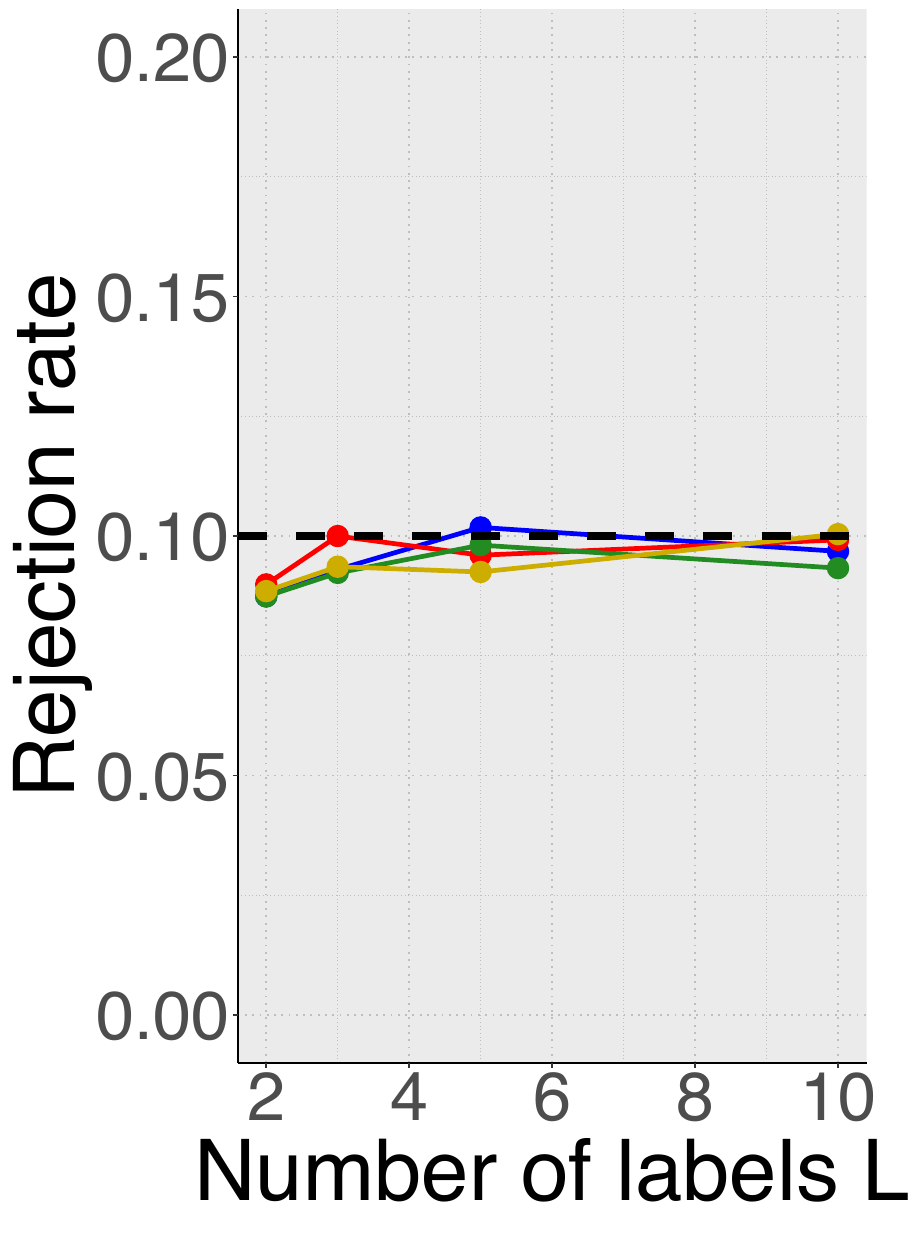}
		\caption{$\alpha=0.1$}
		\label{fig:two_sided}
	\end{subfigure}%
	\hfill
	%\hspace{0.3cm}
	\begin{subfigure}[b]{0.3\textwidth}
		\centering
		\includegraphics[width=1.5\linewidth]{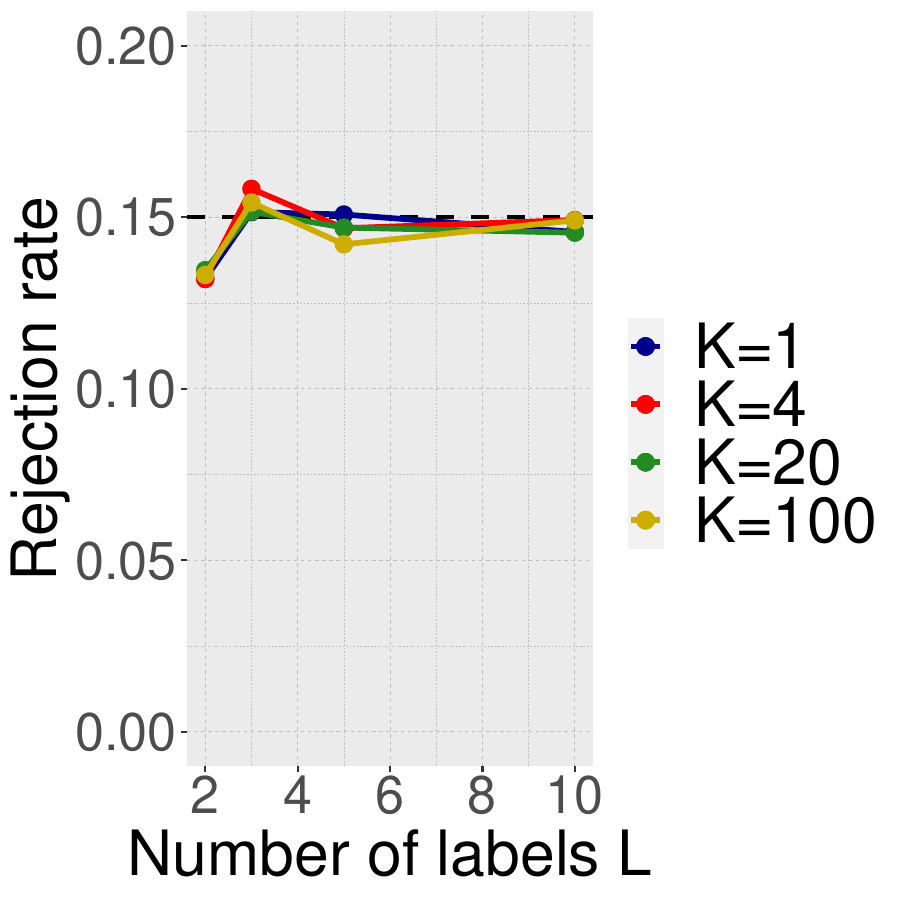}
		\caption{$\alpha=0.15$}
		\label{fig:middle}
	\end{subfigure}
	\caption{ %{\scriptsize{Size of PCR test applied on a dataset consisting of $n=100$ samples generated from model \eqref{eq:example-size}, where $X\indep Y|Z$. Three significance levels $\alpha=0.05,0.1,$ and $0.15$ are considered. Statistic $U_{n,L}$ is obtained from Algorithm \ref{algorithm: model-xz} by using the score function $T(x,z,y)=(y-x-z^\sT\mathbf{1})^2$, and the decision rule \eqref{eq:decision rule} is employed with the threshold $\th^{\mathsf{asym}}_{L,\alpha}$. Reported numbers are averaged out over $10,000$ independent realizations.}}
		{{Size of PCR test with $\th^{\mathsf{asym}}_{L,\alpha}$ for data-generating law \eqref{eq:example-size} with $n=100$. Three significance levels $\alpha=0.05,0.1,$ and $0.15$ are considered. Reported numbers are averaged over $10,000$ trials. }}
	}
	\label{fig: test-size-asym}

\end{center}

\end{minipage}

\end{figure}
}

\noindent \textbf{Size of PCR test.} We start by showing that the size of PCR test is controlled at the desired level, under various choices of input parameters $L$ and $K$. Assume $n=100$ data points $\{(X_i,Z_i,Y_i)\}_{i=1}^{n}$ are generated i.i.d. from the following model: First draw two vectors $v, u\in \reals^p$ with i.i.d. standard normal entries and $p=20$. Then,   
{
\begin{align}\label{eq:example-size}
  %v,u& \overset{\text{ i.i.d. }}{\sim} \normal(0,I_p),\quad\text{ with } p=20\,,\nonumber\\
	Z\sim \normal(0,\bI_p),\, \text{ for }Z\in \reals^p\,,\quad\quad 
	X|Z\sim \normal(v^\sT Z,1),\,\text{ for }X\in \reals\,,\quad\quad
	Y|X,Z\sim \normal\left((u^\sT Z)^2,1\right)\,.
\end{align} 
}
Clearly $X\indep Y|Z$ and the null hypothesis holds. We assume that the dependency rule $X|Z$ and the vector $v$ are known, and  therefore for every given $Z$ we can easily sample from $\normal(v^\sT Z,1)$ to construct the counterfeit variables.  Figures \ref{fig: test-size-finite} and \ref{fig: test-size-asym} exhibit the performance of the PCR test with thresholds $\th^{\mathsf{finite}}_{L,\alpha}$ and $\th^{\mathsf{asym}}_{L,\alpha}$, respectively. As expected, the $\th^{\mathsf{finite}}_{L,\alpha}$ threshold is conservative and controls the size at a level lower than $\alpha$. The $\th^{\mathsf{asym}}_{L,\alpha}$ threshold also controls the size, albeit $n$ being only $100$.

	\smallskip
	
	\noindent \textbf{Statistical Power of PCR test.} Consider a setup similar to \eqref{eq:example-size}, but with $n=1000$ data points and the conditional law 
	{
	\begin{equation}\label{eq: power-example-conditional-law}
		Y|X,Z \sim \normal  \left((u^\sT Z)^2+2 X,1 \right)\,,
	\end{equation}
	}
	Our power analysis in section \ref{sec: pwr} suggests that larger values of $M=KL-1$ would results in higher power. We fix $K=100$ and let $L$ vary in the set $L=\{2,3,\dotsc,30\}$. The significance level is fixed at $\alpha=0.1$. Figure  \ref{fig: powers} showcases the power of PCR test with both choices of rejection thresholds  $\th^{\mathsf{finite}}_{L,\alpha}$ and $\th^{\mathsf{asym}}_{L,\alpha}$. As we see, when $n$ doubles not only the power increases but also it becomes more stable with respect to the choice of $L$.

{
\begin{figure}[t]

\centering
\begin{subfigure}{0.5\textwidth}
		\centering
		\includegraphics[scale=0.35]{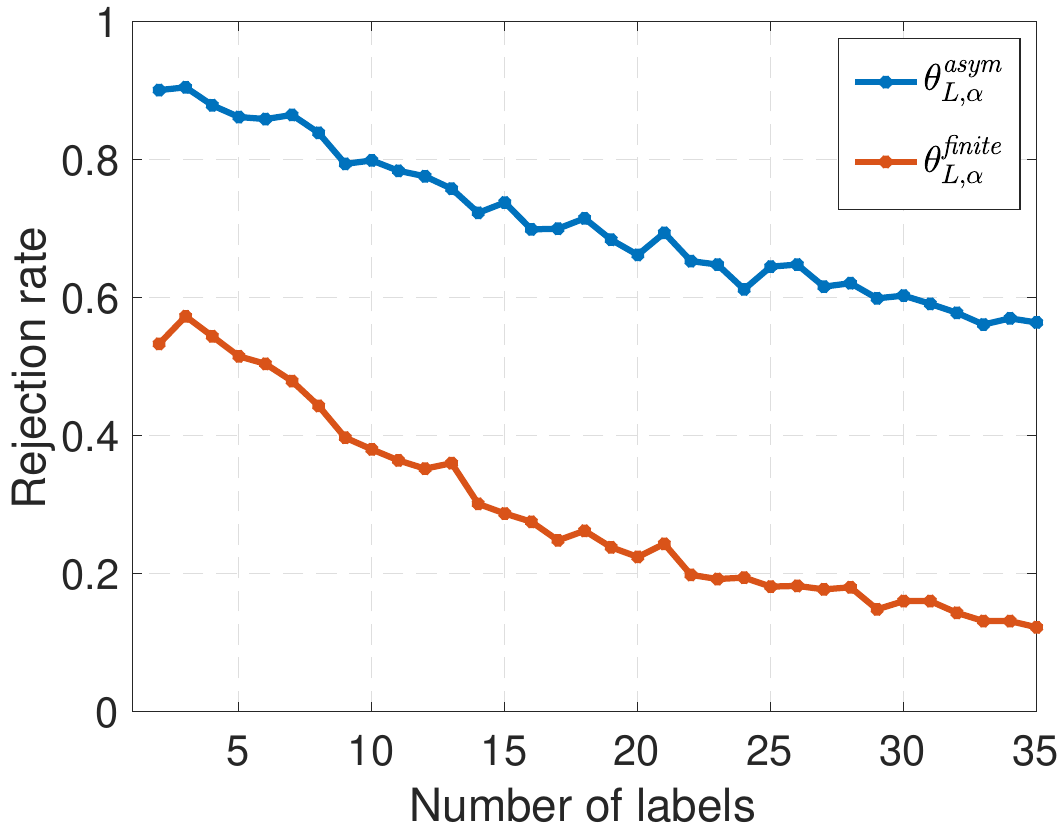}
		\caption{$n=1000$}
	\end{subfigure}%
\begin{subfigure}{0.5\textwidth}
\centering
			\includegraphics[scale=0.35]{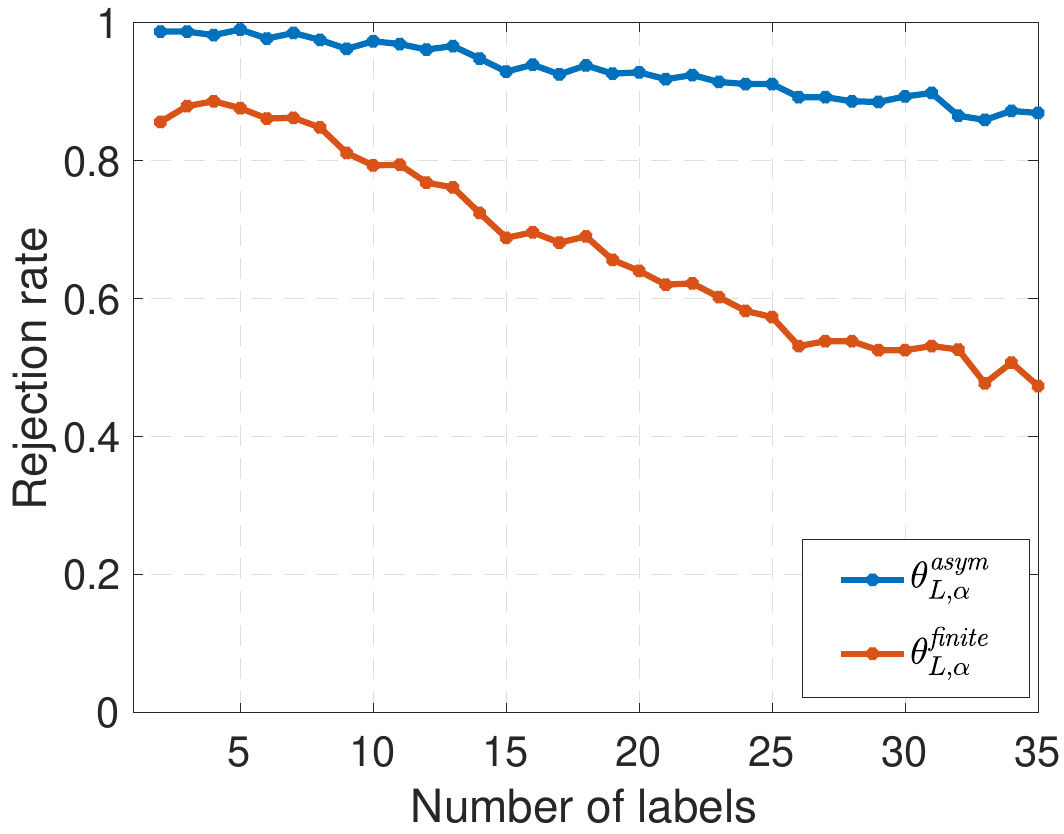}
	\caption{$n=2000$}
\end{subfigure}

\caption{  {{Power of PCR test for (left) $n=1000$  and (right) $n=2000$ data points. Data points are generated under the setup \eqref{eq:example-size} and the conditional law \eqref{eq: power-example-conditional-law}. % We consider the score function $(y-x-z^\sT\mathbf{1})^2$ and choose the significance level $\alpha=0.1$.
We consider the decision rule \eqref{eq:decision rule} with  both of the rejection thresholds $\th^{\mathsf{asym}}_{L,\alpha}$ and $\th^{\mathsf{finite}}_{L,\alpha}$. Each reported power is obtained by averaging over $1000$ trials at significance level $\alpha=0.1$. }} }\label{fig: powers}

\end{figure}
}

	\noindent
	\textbf{Parameter-free PCR test.}
	We consider a setup similar to the previous experiment \eqref{eq: power-example-conditional-law} and run the PCR test with different choices of $L\in\{2,4,8,16,32\}$. 
	We combine the obtained $p$-values using the Bonferroni's correction, as described in Algorithm~\ref{algorithm: model-xz-simes}.  With $n=1000$ data points, we get a statistical power of $0.192$ (with the finite-sample threshold), and $0.815$ (with the asymptotic threshold). Note that in this case, the power of the PCR test with different individual choices of $L$ (without combining the $p$-values) ranges in $(0.13-0.53)$, for the finite-sample threshold, and in $(0.576-0.887)$, for the asymptotic-threshold. 
	
	For $n=2000$ data points, and with the Bonferroni's correction, we get a power of $0.613$, with the finite-sample threshold, and a power of $0.972$, with the asymptotic threshold. Here, the power of the PCR test with individual choices of $L$  ranges in $(0.477-0.83)$, for the finite-sample threshold, and in $(0.8560-0.981)$, for the asymptotic threshold. 
	\smallskip
	
	\noindent
	\textbf{Robustness of the PCR test.}	In this part, we consider cases where the exact dependency law $P_{X|Z}$ is not available, and we use an estimate of it denoted by $\hP_{X|Z}$ (see Section~\ref{sec:robust} for the details and the description of the robust PCR test). 
	%To this end, we show that, using the robust test statistic, which is constructed in \eqref{eq:robust}, allows us to control Type I error.
	 %however, using the regular statistic $T$ from Algorithm \ref{algorithm: model-xz}, will significantly exceed the predetermined significance level $\alpha$.
	  %In addition, we investigate Chi-squared CI test power, when the robust test statistic is used. 
	  Consider a setup similar to \eqref{eq:example-size}, but with $n=5000$ data points and the conditional law 
	{
	\begin{equation}\label{eq:model-aj}
		Y|X,Z \sim \normal  \left((u^\sT Z)^2+a X,1 \right)\,.
	\end{equation}
	}
	When $a=0$, then the null hypothesis is true ($X\indep Y|Z$) and the rejection rate amounts to the type I error. For $a\neq 0$, the null hypothesis is false and the rejection rate amounts to the power of the test. In this experiment, we assume that the counterfeits are sampled from $\hP_{X|Z}$ with
	 %\begin{equation }\label{eq: ex-robust}
	 $\hX|Z\sim\normal(v^\sT Z,(1+\eta)^2)\,.$
	 %\end{equation}
	Note that when $\eta=0$, we get the true distribution $P_{X|Z}$ defined in \eqref{eq:example-size}. We use the Pinsker's inequality, i.e.,  {$2d^2_{\tv}(P,Q)\leq {d_{\mathsf{KL}}(P,Q)} $}, to bound the expected total variation distance $\E_{Z}\left[d_{\tv}\left(P_{X|Z}(\cdot|Z),\hP_{X|Z}(\cdot|Z)  \right)\right]$. Note that for two 1-dimensional Gaussian distributions we have 
	{
	$$d_{\mathsf{KL}}\left(\normal(\mu,\sigma^2_1), \normal(\mu,\sigma^2_2)\right) =\log \frac{\sigma_2}{\sigma_1} +\frac{\sigma_1^2}{2\sigma_2^2}-\frac{1}{2}\,,$$
	}
	which combined with Pinsker's inequality implies that
	{
	\begin{equation}\label{eq: robust-delta}
	 \E_{Z}\left[d_{\tv}\left(P_{X|Z}(\cdot|Z),\hP_{X|Z}(\cdot|Z)  \right)\right] \leq \delta:= \frac{1}{\sqrt{2}} \left(\log(1+\eta)+\frac{1}{2(1+\eta)^2}-\frac{1}{2}\right)^{1/2}. 
	 \end{equation}
	 }

The results for $a = 0$ and $a=4$ are summarized in Table~\ref{tbl:robust}. As we see the robust PCR test controls the type I error under the level $\alpha = 0.1$ for different choices of $\eta$. In addition, it achieves a high power for $a = 4$. If we use the test statistics $U_{n,L}$ (instead of $U_{n,L}(\delta)$) we observe an inflation in type I errors. Concretely, when $\eta= 0.04$ we obtain an inflated type I error of $0.595$ (with the finite-sample threshold $\th^{\mathsf{finite}}_{L,\alpha}$) and an inflated type I error of  $0.1860$ (with the asymptotic threshold $\th^{\mathsf{asym}}_{L,\alpha}$), while the target level is $\alpha = 0.1$. This highlights the importance of adjusting for the errors in estimating the model-X conditional distribution (Section~\ref{sec:robust}).

{ 
\begin{table}[]
%\begin{minipage}[t]{.45\textwidth}
\begin{center}
\scalebox{0.95}{
	\begin{tabular}[t]{|c|c|c|c|c||c|c|c|c| }\hline

       &\multicolumn{4}{c||}{$a=0$} & \multicolumn{4}{c|}{$a=4$}\\  \hline
 \diagbox{setting}{$\eta$}&\makebox{$0$}&\makebox{$0.01$}&\makebox{$0.02$}&\makebox{$0.04$}&\makebox{$0$}&\makebox{$0.01$}&\makebox{$0.02$}&\makebox{$0.04$} \\ \hline
  $U_{n,L}(\delta)$ with $\th^{\mathsf{finite}}_{L,\alpha}$  &0.008 &  0&0 & 0&1 &0.998 & 0.973& 0.63\\ \hline
    $U_{n,L}(\delta)$ with $\th^{\mathsf{asym}}_{L,\alpha}$ &0.1050 &  0.003&0 &0 &1 & 1&0.995 &0.8790 \\ \hline
\end{tabular}
}
\caption{ {{Size ($a=0$) and power ($a=4$) of the robust PCR test.  %for the setting of \eqref{eq:model-aj}  and the approximate distribution \eqref{eq: ex-robust} available for sampling the counterfeits.
 Reported numbers are obtained by averaging over $1000$ trials, with $n=5000$, $L=4$ at significance level $\alpha=0.1$. 
%The PCR  test is run with $L=4$ number of labels. For each value of the discrepancy level $\eta$, we use \eqref{eq: robust-delta} to get an upper bound $\delta$ on the expected total variation distance  $\E_{Z}\left[d_{\tv}\left(P_{X|Z}(\cdot|Z),\hP_{X|Z}(\cdot|Z)  \right)\right]$, and use it in constructing the robust statistic $U_{n,L}(\delta)$. Note that $\eta=0$ implies $\delta=0$, and therefore the robust statistic $U_{n,L}(\delta)$ matches the statistic $U_{n,L}$ in Algorithm \eqref{algorithm: model-xz}. For $a=0$ (true null hypothesis), the size is controlled at the significance level $\alpha=0.1$. 
%Reported numbers are obtained by averaging over $1000$ trials.
}} } \label{tbl:robust}
\end{center}
%\end{minipage}% 
%\hfill \begin{minipage}[t]{.45\textwidth}
%\vspace{0.5cm}

%\end{minipage}
%\caption{Two tables side by side}
%\label{fig:tables}
\end{table}
}

 \subsection{Power comparison} \label{sec: compare-crt-alternatives} 
In this section, we compare the performance of PCR with other model-X CI tests. For this end, we consider CRT, dCRT (distilled CRT) \citep{liu2020fast}, and HRT (holdout randomization test) \citep{tansey2018holdout}. We focus on the following data generating law
{
\begin{equation}\label{eq: compare-generative}
Y|X,Z\sim\normal\left(\frac{\nu}{\sqrt{X^2+c^2}}+\nu\beta X+\gamma^\sT Z,1\right)\,,
\end{equation}
}
for  $\beta\in \reals $, $\gamma\in \reals^p$, $c=0.001$, and $X,Z$ with i.i.d. standard normal entries.    We focus on two different settings: $(i)$ low-dimensional ($n>p$)
, and $(ii)$ high-dimensional ($n<p$).  For each one, we consider four different values of $\nu\in \{0,0.3,0.7,1\}$ at significance level $\alpha=0.1$, and compare statistical power of a few  model-X CI tests. For $\nu=0$ the CI holds and the rejection rates correspond to type I error which is expected to be smaller than $\alpha$.

For the low-dimensional setting, we consider $n=8000$ (total number of samples) and $p=50$. In addition, we let $\beta=0.1$, and draw $\gamma$ from $\normal(0,\bI_p)$ distribution. We use the ordinary least square (OLS) estimator to construct the score function $T(\bX,\bZ,\bY)$. Specifically,  we first regress $Y$ on $[X,Z]$ and let $\widehat{\beta}_N$, $\widehat{\gamma}_N$ denote estimate coefficients, where $N$ stands for the number of samples used in the estimation process. We also indicate the computed intercept value by $\widehat{\alpha}_N$. Next, for CRT, similar to \citep{candes2018panning}, we consider the regression coefficient of $X$ as the score function: 
$T_{\mathsf{CRT}}^{}(\bX,\bZ,\bY)=|\widehat{\beta}_n|.$
%, and $T_{\mathsf{CRT}}^{\mathsf{residual}}(\bX,\bZ,\bY)=\|\bY-\widehat{\beta}_n\bX-\bZ\widehat{\gamma}_n-\widehat{\alpha}_n\mathbf{1}\|_2^2$.

 For HRT, we split the entire samples into two equal size datasets $\cD_1,\cD_2$ with $N=4000$. We compute $\widehat{\beta}_N$ and $\widehat{\gamma}_N$ via $\cD_1$, and consider the score function $T_{\mathsf{HRT}}^{}(\bX,\bZ,\bY)=\|\bY-\widehat{\beta}_N\bX-\bZ\widehat{\gamma}_N-\widehat{\alpha}_N\mathbf{1}\|_2^2,$
 for $\bZ,\bY$ in $\cD_2$.
 
 In addition, we use the distilled CRT test statistic \citep{liu2020fast} which is given by the score function 
$T_{\mathsf{dCRT}}^{}(\bX,\bZ,\bY)= \frac{| (\bY-\bZ\widetilde{\gamma}_n-\widetilde{\alpha}_n\mathbf{1})^\sT \bX |}{\|\bX\|^2_2} \,,$
where $\widetilde{\gamma}_n$, $\widetilde{\alpha}_n$ respectively denote computed least square coefficients and the intercept value by one-time regression of $Y$ on $Z$ (full data). 

For PCR, we use the data splitting similar to HRT, and then partition $\cD_2$ into groups of size $g=5$ (with $\myg=800$) and use the score function $T_{\mathsf{PCR}}^{}(\bX,\bZ,\bY)=\|\bY-\widehat{\beta}_{N}\bX-\bZ\widehat{\gamma}_N-\alpha_N\mathbf{1}\|_2^2.$ Concerning the number of randomizations, we consider $100$ randomizations for CRT, dCRT and HRT. In addition, we run PCR with $M=99$ counterfeits and $L=5$ labels ($K=20$). 
Figure \ref{fig: OLS} exhibits average rejection rates for $200$ independent experiments. It can be seen that both versions of PCR achieve higher statistical power. Note that in this experiment all score functions belong to the same estimation family (OLS), and we used the specific score functions $T_{\mathsf{CRT}}$,  $T_{\mathsf{HRT}}$, $T_{\mathsf{dCRT}}$, which were suggested by the corresponding work. 

For the high-dimensional setting experiment, we let $n=5000$ (total number of data points), $p=6000$, and $\beta=0.2$.  For the vector $\gamma \in \reals^p$, we consider the sparsity level $s=300$ with non-zero entries drawn independently from $\normal(0,\sigma^2)$ with $\sigma=0.5$. We follow similar guidelines for score functions of HRT and dCRT as per low-dimensional experiments with the only difference that we use cross-validated lasso instead of OLS. For HRT and PCR we use the sample splitting $|\cD_1|=500$ and $|\cD_2|=4500$. In addition, for PCR, we consider the following score function, which is motivated by the distilled CRT score function in \citep{liu2020fast}, and is given by
$ T_{\mathsf{PCR}}^{\mathsf{distilled}}(\bX,\bZ,\bY)= \frac{| (\bY-\bZ\widehat{\gamma}_n-\widehat{\alpha}_n\mathbf{1})^\sT \bX |}{\|\bX\|^2_2} \,.$

Here, $\widehat{\gamma}_n$ and $\widehat{\alpha}_n$ are computed by the cross-validated lasso coefficients on $\cD_1$ (not the entire data), by regressing $Y$ on $Z$. In addition, in this score function we let $\bX,\bZ,\bY$ come from groups of size $g=5$, so $n_g=900$. We omit CRT for the high-dimensional experiment, because of the high computational complexity. The rejection rates can be seen in Figure \ref{fig: lasso}. Results are averaged over $50$ independent experiments.  It can be seen that similar to the low-dimensional experiment, PCR variants achieve higher power than other methods.

%%%%%%%%%%%%%%%%%%%%%%%%%%%%%%%%%%%%%%%%%%%%%%%%%%%%%%%%%%%%%%%%%%%%%
%%%%%%%%%%%%%%%%%%%%%%%%%%%%%%%%%%%%%%%%%%%%%%%%%%%%%%%%%%%%%%%%%%%%%

{
\begin{figure}[]
\begin{minipage}[h]{.63\textwidth}
\begin{center}
\begin{subfigure}{0.45\textwidth}
	        \centering
		\includegraphics[width=0.9 \textwidth]{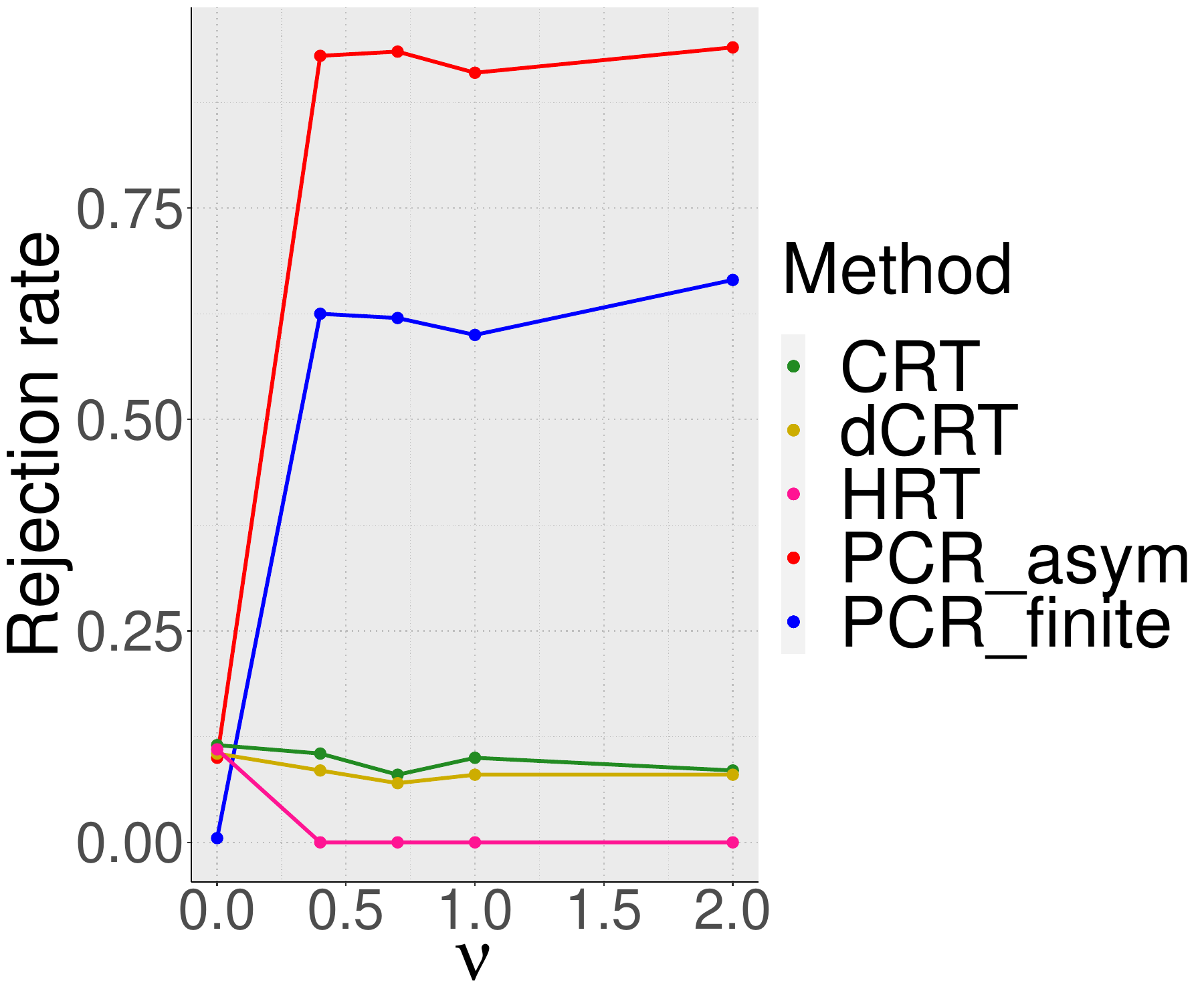}
		\caption{{Low-dim nonlinear setting} } \label{fig: OLS}
	\end{subfigure}%
\begin{subfigure}{0.45\textwidth}
\centering
			\includegraphics[width=0.9 \textwidth]{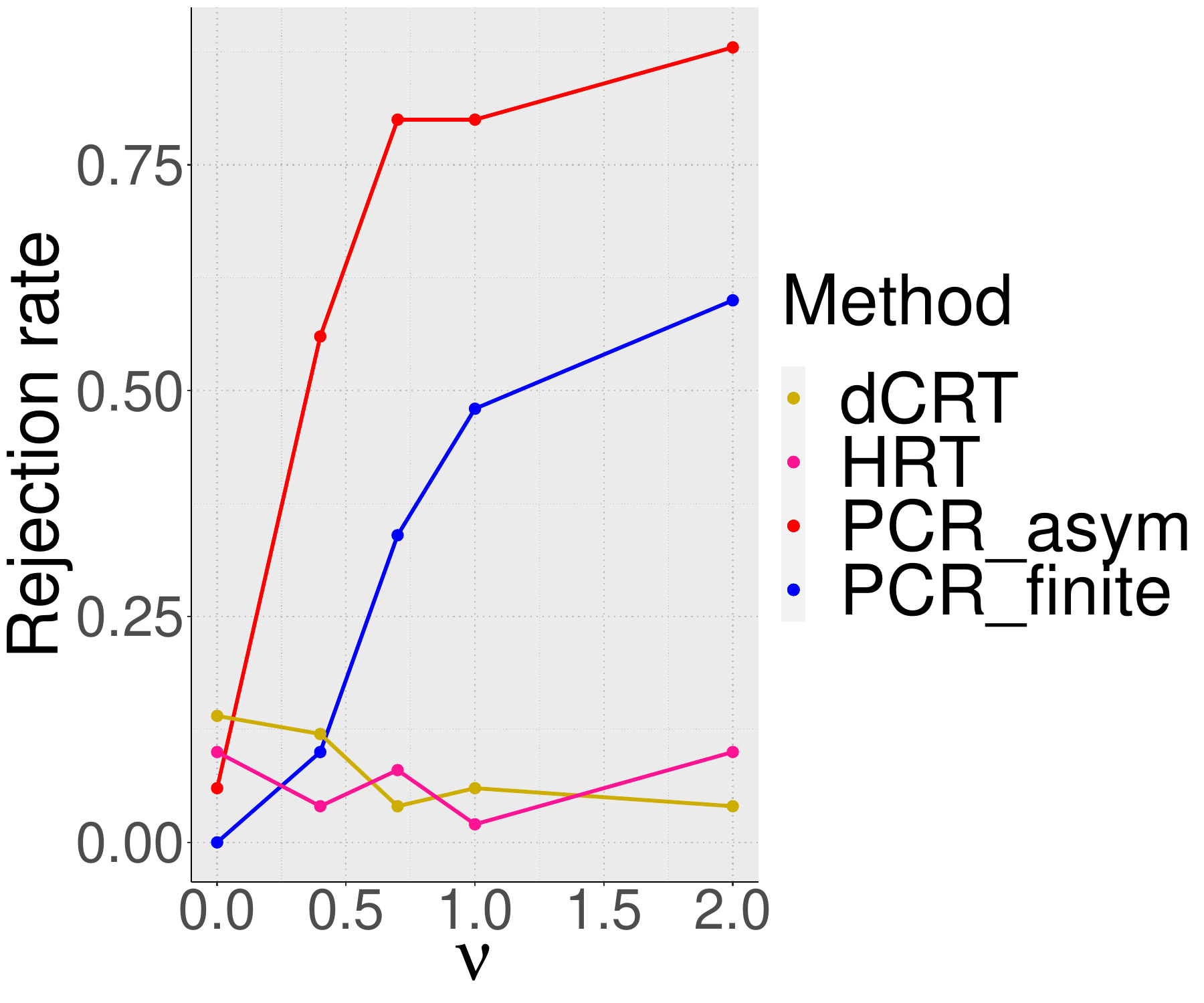}
	\caption{{High-dim nonlinear setting}}\label{fig: lasso}
\end{subfigure}
%\vspace{0.9cm}
\caption{{{Comparison between statistical power of PCR and a group of model-X CI tests for the data generating law \eqref{eq: compare-generative} for \textbf{low-dimensional} (left) and \textbf{high-dimensional} (right) settings. For the low-dimensional setting, we consider  $n=8000$, $p=50$ with the ordinary least square as the score function.  For the high-dimensional setting, we consider  $n=5000$ and $p=6000$ with the cross-validated lasso as the score function.}}  }\label{fig: OLS-lasso}
\end{center}
\vspace{-.6cm}
\end{minipage}% 
%\vspace{-1cm}
\hfill \begin{minipage}[h]{.33\textwidth}
\begin{center}
\begin{subfigure}{0.9\textwidth}
\includegraphics[height=0.7 \textwidth]{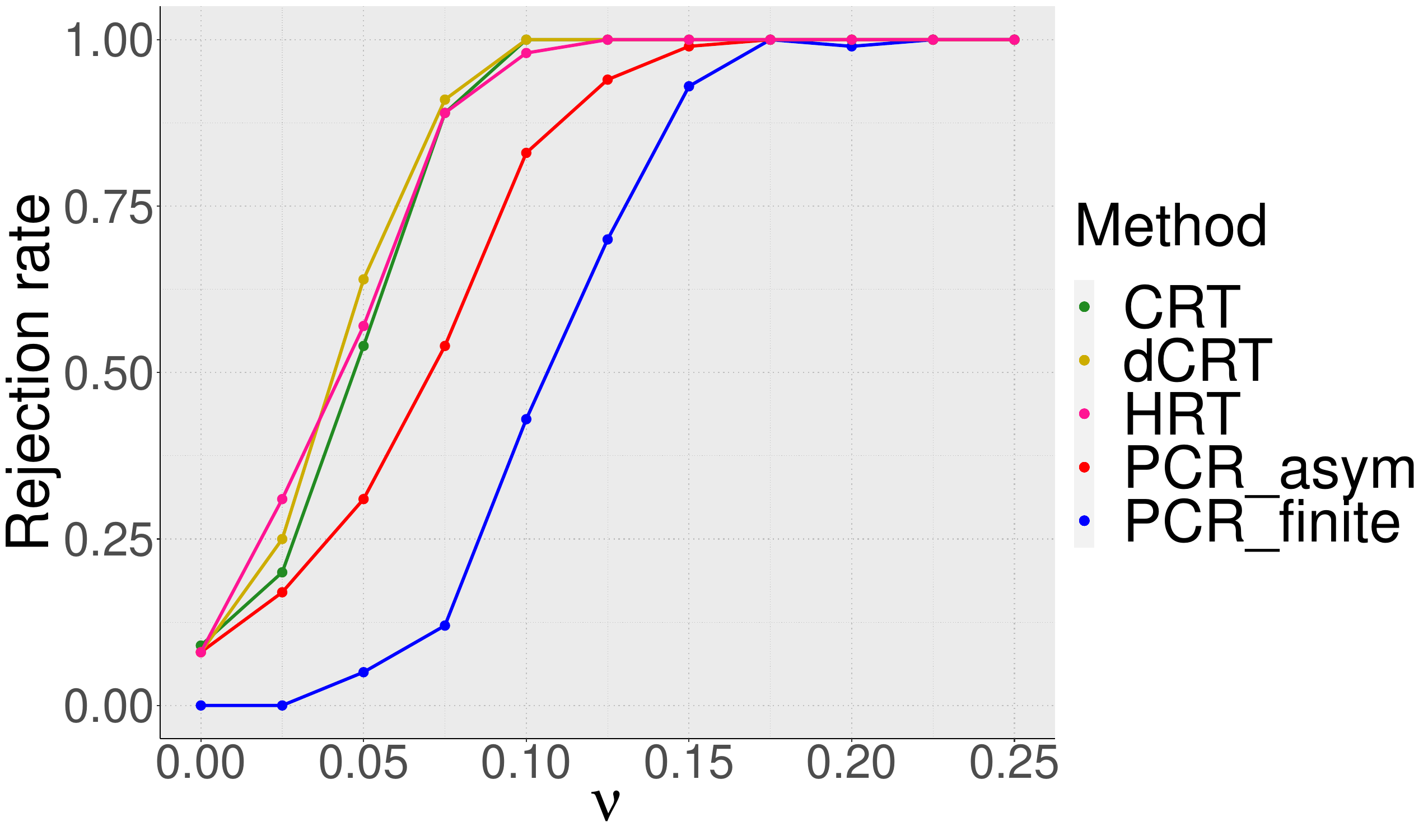}
%\vspace{0.2cm}
\caption{{Linear setting}}
\end{subfigure}
%\vspace{-0.2cm}
\caption{{Comparison between statistical power of PCR and a group of model-X CI tests. %for the data generating law \eqref{eq: compare-regular}.  
In this experiment, we consider  $n=2000$, $p=50$ with the OLS as the score function.  The results are averaged over $100$ experiments at $\alpha=0.1$.} }\label{fig: regular-OLS}
\end{center}

\end{minipage}

\end{figure}
%\vspace{2cm}
}

%\vspace{20cm}

%%%%%%%%%%%%%%%%%%%%%%%%%%%%%%%%%%%%%%%%%%%%%%%%%%%%%%%%%%%%%%%%%%%%%
%%%%%%%%%%%%%%%%%%%%%%%%%%%%%%%%%%%%%%%%%%%%%%%%%%%%%%%%%%%%%%%%%%%%%
%%%%%%%%%%%%%%%%%%%%%%%%%%%%%%%%%%%%%%%%%%%%%%%%%%%%%%%%%%%%%%%%%%%%%%

In the next experiment, we consider the standard linear regression setup (no non-linear term). Concretely, we consider
$Y|X,Z\sim\normal\left(\nu X+\gamma^\sT Z,1\right)\,.$
We let $n=2000$, $p=50$, and  $\gamma$ with i.i.d. entries drawn from $\normal(0,1)$. We pick 11 values for $\nu$ from $[0,0.25]$ and compare the performance of PCR with  CRT, dCRT, and HRT. We use the similar score functions as in the previous experiments in the low-dimensional setting. For PCR and HRT, we split data points into two disjoint groups such that $|\cD_1|=200$ and $|\cD_2|=1800$. For PCR, we consider groups of size $g=5$, so $\myg=400$ with number of labels $L=5$.  Figure \ref{fig: regular-OLS} shows the rejection rates averaged over $100$ independent experiments. As we see in this experiment, CRT, dCRT, HRT have similar performance and achieve higher power than PCR at fixed $\nu$. For $\nu\ge 0.17$, all the considered tests have perfect power. Note that in this experiment, the data is generated according to a linear model, and the score functions are based on residuals from fitting a linear regression. Given this perfect alignment, the non-uniformity of labels under the alternative occurs at the tail, making CRT-based methods better suited to capture these deviations compared to PCR, which probes the entire range for potential deviations.

%=====

\subsection{Computational advantage}\label{sec:Computational}
In this section, we investigate the computational advantages of PCR through a series of experiments. Specifically, we demonstrate that in certain cases, PCR can achieve higher statistical power than CRT, even with a smaller number of randomizations (counterfeits). This can be specifically helpful for settings where sampling is constly.  

In the first setting, we focus on the the data generating process given in \eqref{eq:example-size} with only modification being the following: 
\begin{equation}\label{eq: few-generating}
\cL(Y|X,Z)=\normal((u^\sT Z)^2+c X,1)\,.
\end{equation}
In this formulation,  $c$ reflects the dependency strength between $X$ and $Y$ conditioned on $Z$. We run both tests with the marginal covariance score function $T^{\mathsf{MC}}(\bX,\bZ,\bY)=\frac{\bX^\sT \bY}{n}$.  In addition, we consider the range of values for $c \in [0,2]$, and then compute the average rejection rates of PCR and two-sided CRT, where CRT is run with fivefold number of randomizations.  Specifically, we run PCR with $K=7,L=3$ (number of randomizations is $M=20$), whereas the number of randomizations for CRT is $B=100$. In addition, we let the number of samples be $n=3000$. Figure \ref{fig: few-MC}  plots the average rejection rates for PCR (with two rejection rules) and CRT at significance level $\alpha=0.1$. Results are averaged over $500$ experiments. It can be observed that PCR, with only one-fifth the number of randomizations of CRT and the asymptotic threshold, achieves higher statistical power. Additionally, PCR with a finite threshold exhibits comparable statistical power to CRT.

In the second set of experiments, we assess the performance of PCR, when it is executed with a reduced number of randomizations and when the score function $T$ is adapted to the training samples. We use  a similar setup to that outlined in Section \ref{sec: compare-crt-alternatives} for both high-dimensional and low-dimensional scenarios. For PCR we set $K=7$ and $L=3$ (corresponding to $M=20$ randomization), while we run CRT with $100$ randomizations.  For the low-dimensional setting, we consider the ordinary least square as the score function, and  for the high-dimensional setting we consider the cross-validated lasso as the score function. The results are presented in Figures \ref{fig: OLS-few} and \ref{fig: lasso-few} for low-dimensional and high-dimensional settings, respectively. It can be seen that for this setup as well, PCR can achieve higher statistical power than CRT, even with fewer number of randomizations.

{{We conclude this section by providing further insight on why CRT requires more counterfeits than PCR. Note that the high randomization burden of the CRT is also highlighted in \cite{li2021deploying} in the context of multiple hypothesis testing using Benjamini–Hochberg (BH) procedure with FDR control guarantees. Specifically, for \(p\) covariates with an FDR threshold \(q\) (e.g., $q=0.1$), the significance levels are of the form \(\alpha_i = \frac{i\,q}{p}\), while the attainable \(p\)-values lie in \(\{\tfrac{1}{M+1},\tfrac{2}{M+1},\dots,1\}\). This implies that \(M\) must be on the order of \(p/q\) to permit any rejections, thereby driving up the required number of randomizations. In contrast, PCR is bases on multinomial testing and can provide high-resolution \(p\)-values, even with two labels ($L=2$), provided the sample size is moderately large. Nonetheless, the resolution of p-values for CRT does not change as the sample size ($n$) changes, being purely a function of number of randomizations $M$.}}

{
\begin{figure}[t]
\begin{minipage}[h]{.33\textwidth}
\begin{center}
\begin{subfigure}{\textwidth}
\centering
\includegraphics[height=0.56\textwidth]{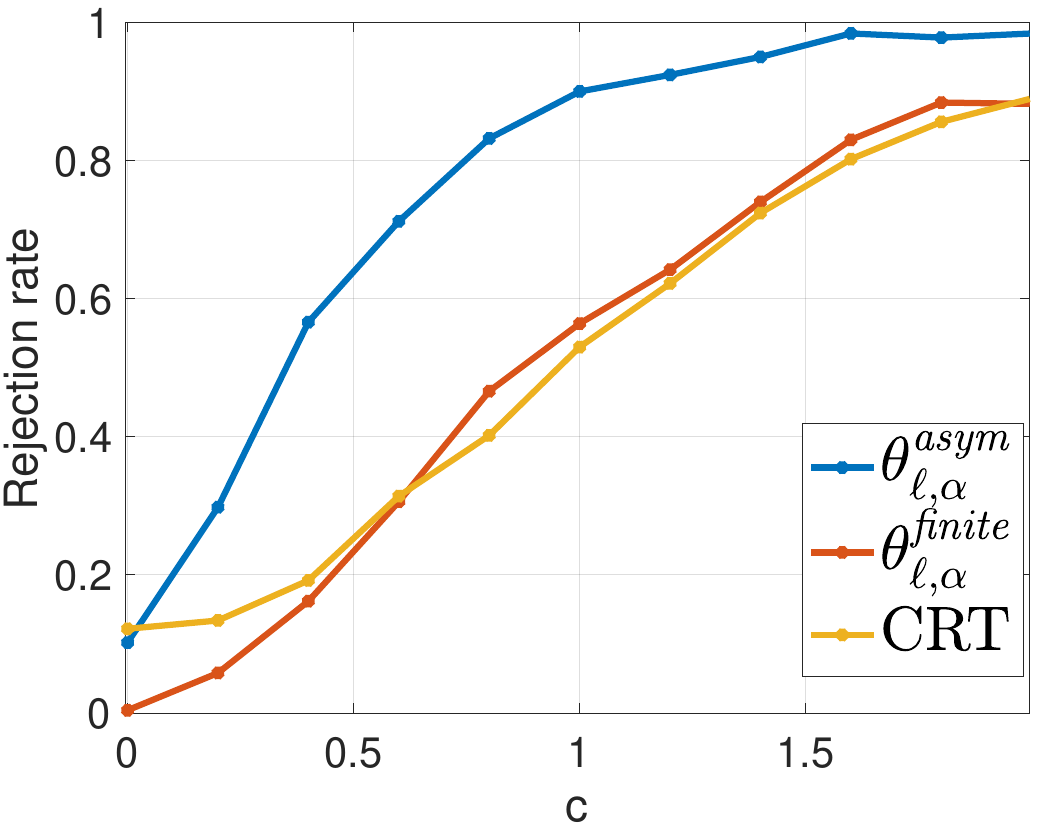}
\caption{{PCR with fewer randomization and MC score function }}
\end{subfigure}
\caption{{Average rejection rates for CRT and PCR for data generating law \eqref{eq: few-generating} and the marginal covariance score function. In this experiment, PCR is run with only one-fifth number of randomizations compared to CRT. } }\label{fig: few-MC}
\end{center}
\end{minipage} 
\hfill
\begin{minipage}[h]{.63\textwidth}
\begin{center}
\begin{subfigure}{0.42\textwidth}
	        \centering
		\includegraphics[height=0.75 \textwidth]{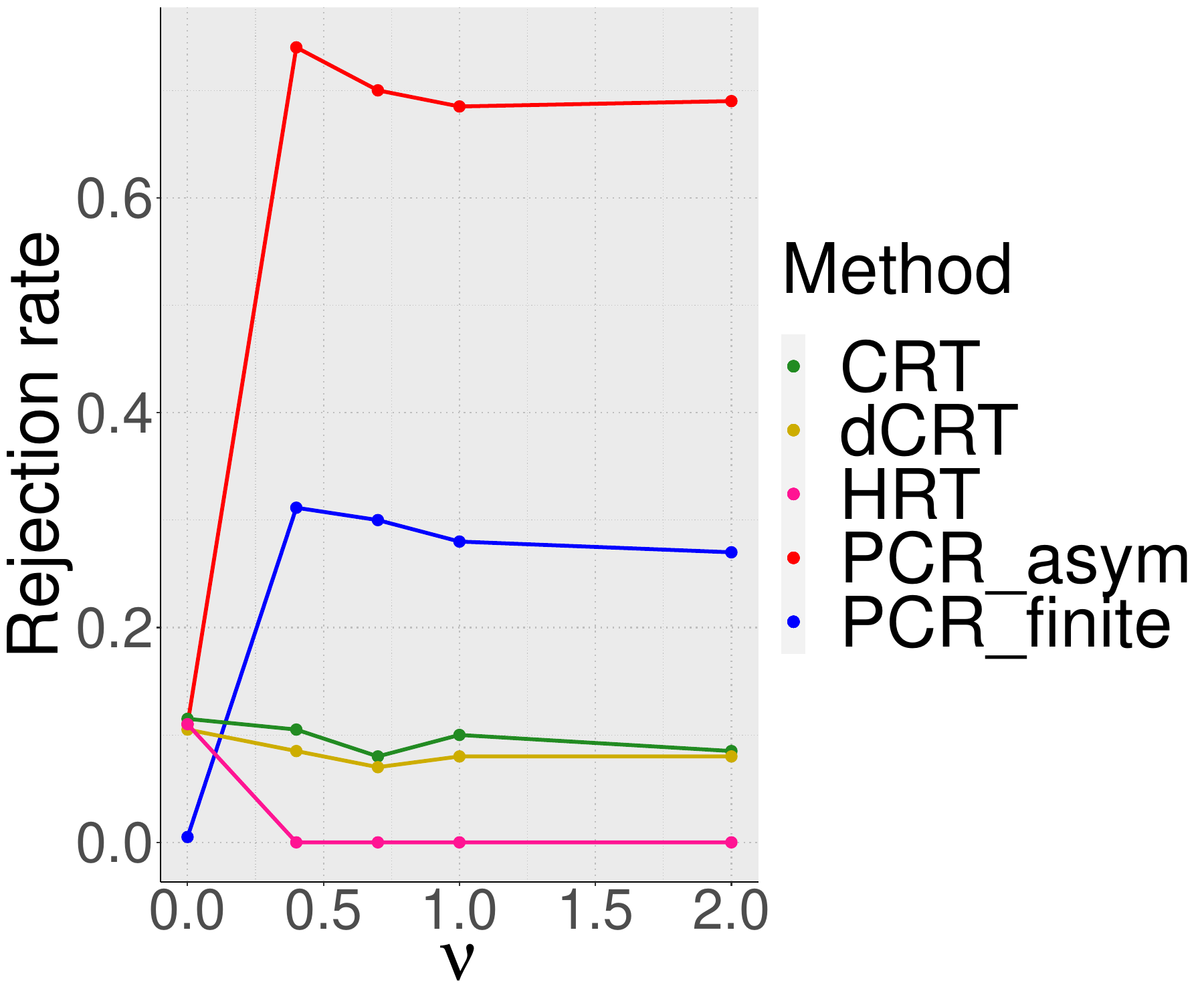}
		\caption{{PCR with fewer randomizations (low-dim setting)}}  \label{fig: OLS-few}
	\end{subfigure}%
	\hspace{0.3cm}
\begin{subfigure}{0.42\textwidth}
\centering
			\includegraphics[height=0.75 \textwidth]{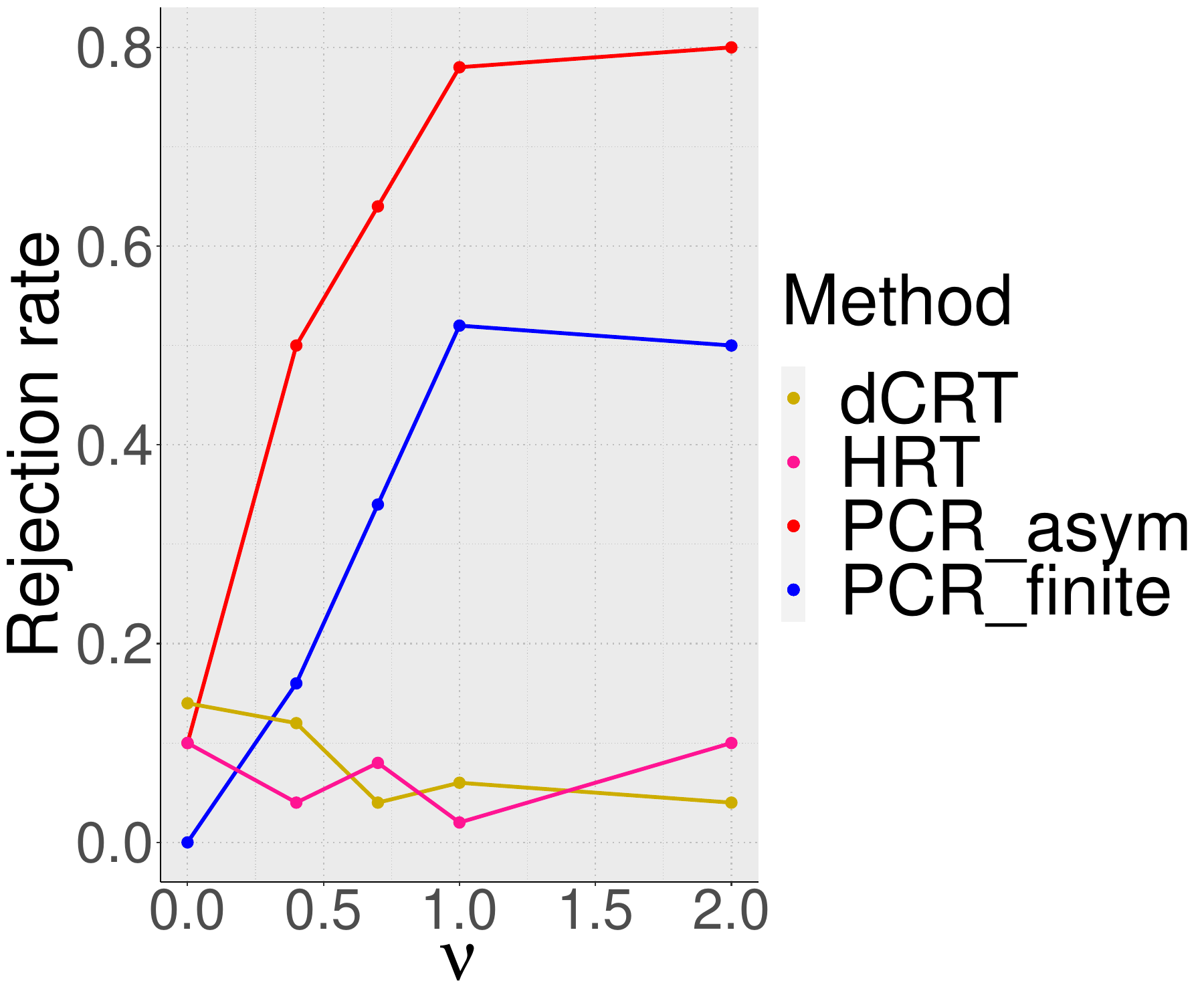}
	\caption{{PCR with fewer randomizations (high-dim setting)}}\label{fig: lasso-few}
\end{subfigure}
\caption{{{Average rejection rates for PCR and a group of model-X CI tests for the data generating law \eqref{eq: compare-generative} for \textbf{low-dimensional} (left) and \textbf{high-dimensional} (right) settings when PCR is run with only \textit{one-fifth} number of randomizations compared to CRT and the score function is fitted to the dataset.
For the low-dimensional setting, we consider the ordinary least square as the score function.  For the high-dimensional setting, we consider the cross-validated lasso as the score function.}}  }\label{fig: OLS-lasso-few}
\end{center}
%\vspace{-1.6cm}
\end{minipage}% 
%\vspace{-1cm}
%\hfill 

\end{figure}
%\vspace{2cm}
}

\subsection{Real data experiment: Capital Bikeshare dataset }\label{sec: real}

In this section, we evaluate the performance of the PCR test on real data from the Capital Bikeshare\footnote{The dataset is publicly available at \url{https://www.capitalbikeshare.com/system-data}.}. Capital Bikeshare is bike-sharing system in Washington, D.C, and releases its trips data on a quarterly basis. The data includes each trip taken, start date and time, end date and time,
start and end stations, Bike ID, and the  user type indicating whether the rider was a registered member or if it was a casual ride (one-time rental or a short pass). 
%which have taken place sometime in September, October, or November 2011 in Washington, DC.  In particular, each entry incorporates information about the trip day (Monday to Friday), the start and end time, the trip duration, the rider's membership status (i.e. having a long term membership, or it is a temporary rental ride), the bicycle ID, and the start and end station. 

In this experiment, we use our proposed PCR test to study the independence of the trip duration ($X$), and other variables, such as the user type ($Y$), and provide $p-$values for their associations. A similar data and question has been studied by \citep{berrett2020conditional} using the Conditional Permutation Test (CPT). As can be imagined, the trip duration ($X$) heavily depends on the route (length of the rout, elevation ,etc) and the time of the day at the start of the ride (due to varying traffic and the rush hours). To control for the effect of such variables, we condition on the start and end locations and the day hour $Z=(Z_{\text{start loc}}, Z_{\text{end loc}}, Z_{\text{hour}})$. 

In order to implement the PCR test, we use the conditional normal distribution $X|Z \sim \normal(\mu(Z), \sigma^2(Z))$ as an approximation of $P_{X|Z}$.  We follow the procedure of \citep{berrett2020conditional} to estimate the mean $\mu(z)$ and variance $\sigma^2(z)$. We outline the procedure here for the reader's convenience. We consider a test data, consisting of the rides taken on weekdays in Oct 2011, and a training data consisting of the rides taken on weekdays in Sep 2011 and Nov 2011. The test data is used to for testing conditional independence between factors of interest, and the training data is used to estimate the conditional mean and variance ($\mu(z)$,$\sigma^2(z)$). To have reliable estimation, we eliminate the records in the test data for which the corresponding route in the training data has less than 20 rides. After this preprocessing step, the test data includes $7,346$ samples. Finally, the conditional functions $\mu(z)$ and $\sigma^2(z)$ are estimated using a Gaussian kernel with a bandwidth of $20$ minute, on the training data. (See \citep[Appendix B]{berrett2020conditional} for further details on this part.)

\begin{table}[t]
\begin{center}
\scalebox{0.8}{
\begin{tabular}[t]{|c|c|c|}
\hline
Response $Y$ & $p$-value (finite) &$p$-value (asym)\\ \hline 
User type &   $0.0014$ & 0   \\ \hline
Date & $0.3855$ & 0.0456 \\ \hline
Week day &  0.2094 & 0.0194  \\ \hline
\end{tabular} 
}
\caption{$P$-values that are computed from the PCR test on the Capital Bikeshare dataset. The null hypothesis \eqref{eq: CI hypothesis} is considered with $X$ being the duration of the ride, and the confounder variable $Z$ encoding the start and end locations, as well as the time of day at the start of the ride. We consider three different response values $Y$: (1) User type, (2) Date of the month, (3) Weekday. The $p$-values are obtained as per~\eqref{eq: p-val} with the number of labels $L=10$, and the counterfeit ratio $K=200$.}\label{table: real-data}

\end{center}
\end{table}

We test the null hypothesis \eqref{eq: CI hypothesis} with $X$ being the duration of the ride, and three different response variables $Y$: (1) User type-- registered members have acquaintance with the routes and are likely to have lower trip durations, (2) Date of the month (continuous variable from $1-30$)--  this can be used to capture effect of factors such as weather and sunlight hours. (3) Weekday (categorical variable from Monday to Friday)-- rides on the early days of the week are likely to be more work-related.  
For score function to be used in the PCR test, we consider the squared residual from regressing $Y$ on $X$. As an example, when $Y$ is the user type, we encode it  as a binary variable $Y=\ind{\{ \text{the user is a registered member}\}}$, and fit the linear model $Y=b_0+b_1X$ to the training data to obtain the estimates $\widehat{b}_0, \widehat{b}_1$. 
%Finally, for any test data point $(x,y)$, we consider the score function $T(x,y)=(y-\widehat{b}_0-\widehat{b}_1x)^2$.

In this experiment, we use the PCR test with $L=10$ number of labels and the counterfeit ratio $K=200$, and therefore $M=1999$. Further, in order to reduce the variation between the true distribution $P_{X|Z}$ distribution and its estimate  $\normal(\widehat{\mu}(Z), \widehat{\sigma}^2(Z))$, we use PCR test with groups each of size $4$. This means that in Algorithm \eqref{algorithm: model-xz} the number of groups is $\myg=\lfloor n/4 \rfloor$. In this case, for each group of data points $\cG=(\bX,\bY)$ we use the following statistic $T(\cG)=\frac{1}{4}~\|\widehat{b}_0\mathbf{1}+\widehat{b}_1\bX-\bY\|_2^2\,,$
with $\mathbf{1}$ being the all-one vector in $\reals^4$. 

\iffalse
we use the generalization of the PCR framework as discussed at the end of Section~\ref{sec:discussion}. Concretely, we partition the $n=7346$ test data samples into $N= \lfloor n/4 \rfloor$ smaller groups, each of size $4$, and label each group based on the rank of its score, among the scores of its counterfeits, as proposed in the PCR test.  Formally, for test data vectors $X_{1:n}$, $Z_{1:n}$, we construct counterfeits $\tX_{1:n}^{(1)},\tX_{1:n}^{(2)},...,\tX_{1:n}^{(M)}$, and score each sample and its counterfeits by:
\[
T=[(\widehat{b}_0+\widehat{b}_1X_j-Y_j)^2]_{j=1:n}\,, \quad \widetilde{T}_i=[(\widehat{b}_0+\widehat{b}_1\tX_j^{(i)}-Y_j)^2]_{j=1:n}\,.
\]
We then partition the samples into $N=\lfloor n/4 \rfloor$ groups of equal size and score each group (and its counterfeits) by the average score of their four members. Finally, we follow Algorithm \eqref{algorithm: model-xz} to label the groups and construct the statistic $U_{N,\ell}$.
\fi

 We calculate the $p$-values for each of the CI tests, using \eqref{eq: p-val}. The results are outlined in Table \ref{table: real-data}. As we see among the three response variables considered in this experiment, user type has the most significant (conditional) dependence to duration of the ride.

\section{Conclusion}\label{sec:discussion}
In this work, we introduced the PCR test procedure to examine CI of two variables in the presence of a high-dimensional confounding variable, in a model-$X$ setup where the distributional information on the covariate population is available.  The proposal of the PCR test was inspired by some of the alternative distributions for which the CRT (and its variants) are powerless. The PCR test is generally more flexible in capturing the conditional dependency, and under some alternatives can result in much higher statistical power compared to the CRT. We also provided a power analysis of the PCR test  in terms of the so-called conditional dependency power of the joint law $\cL(\bX,\bZ,\bY)$, sample size $n$ and the number of labels $L$ used in constructing the PCR test statistic.  In addition, the PCR test makes a novel contribution to the CI testing problem by using the i.i.d. property of the samples to obtain high-resolution $p$-values with a very small number of conditional randomizations. This  can significantly lower the computational cost in high-dimensional variable selection problems.  
We also proposed two extensions of the PCR test: $(i)$ \emph{Parameter-free PCR test}, which consists of multiple runs of PCR test with different choices of number of labels $L$, and then using Bonferroni's method to combine the obtained $p$-values. $(ii)$ \emph{Robust PCR test}, which improves the robustness of the test against errors in estimating the conditional distribution $P_{X|Z}$.  Both of these extensions would have important practical implications. Finally, the score function in the proposed PCR test can be borrowed from many score functions developed to improve the robustness and computational complexity of CRT such as dCRT, HRT, and CPT which further improves the general performance of PCR.

\section*{Acknowledgments}
A.~Javanmard is supported in part by the Sloan fellowship in mathematics,  the NSF Award
DMS-2311024, Adobe Faculty Research Award, Amazon Faculty Research Award and an outlier research in business grant from iORB at the USC Marshall School of Business.

\newpage
\bibliographystyle{plainnat} 
\bibliography{mybib}     

\newpage

\appendix
\begin{center}
{\bf \Large{Supplementary Materials:  Pearson Chi-squared\\ Conditional Randomization Test}}
\end{center}

%\chapter{Proof of theorems and technical lemmas}
{ In this supplementary file, we present proofs of theorems and propositions. We begin with 
Section \ref{sec: preliminary} and we introduce technical preliminaries that will be employed in the subsequent proofs. Following this, we provide proofs for Sections \ref{sec:PCR}, \ref{sec: pwr}, and \ref{sec:robust}.

%=============================================================

%=============================================================

\section{Technical preliminaries}\label{sec: preliminary} 

\begin{lemma}[  Pearson's $\chi^2-$test size and power] \label{lemma: multinomial-main}
Consider a multinomial model with $L$ labels $\{1,2,...,L\}$, and $n$ number of samples. For $s\in [L]$, let $W_s$ denote the number of samples with label $s$ and $p_s$  be the occurrence probability of label $s$ in one realization of the multinomial model .  Consider the following uniformity hypothesis, at the significance level $\alpha\in (0,1)$:
\begin{equation}\label{eq: multi-unif-hypothesis}
H_0: p_\ell=\frac{1}{L},\quad \text{for } 1\leq \ell \leq L\,,
\end{equation}
with the following decision rule $\Psi_{n,L}$, which is based on the Pearson's Chi-squared statistic $U_{n,L}$:
\[
\Psi_{n,L}= \ind\left( U_{n,L}: =\frac{L}{n} \sum\limits_{s=1}^{L} \left(W_s-\frac{n}{L}\right)^2  \geq L+\sqrt{\frac{2L}{\alpha}} \right )\,.
\]
The following statements hold:
\begin{enumerate}
\item  Under the null hypothesis \eqref{eq: multi-unif-hypothesis}, $U_{n,L}\overset{d}{\Rightarrow} \chi^2_{L-1}$, as $n\rightarrow\infty$. % the Chi-squared distribution with $\ell-1$ degrees of freedom.
\item  Under the null hypothesis \eqref{eq: multi-unif-hypothesis},  the size of this test is controlled at level $\alpha$:
\[
\prob(\Psi_{n,L}=1)\leq \alpha\,.
\]
\item  If for some $\beta>0$, we have the following:

\[
\sum\limits_{s=1}^{L}\left|p_s-\frac{1}{L}\right| \geq\frac{32{L^{1/4}}}{\sqrt{n}}\left[\frac{1}{\sqrt{\alpha}}\vee\frac{1}{\beta}  \right]^{1/2}\,,
\]
then the type II error does not exceed $\beta$:
\[
\prob(\Psi_{n,L}=0)\leq \beta\,.
\]
\end{enumerate}
\end{lemma}
%\subsection{Proof of Lemma \ref{lemma: multinomial-main}}
Regarding the proof of Lemma \ref{lemma: multinomial-main}, note that the first part is a classic result on the asymptotic null distribution of the Pearson's Chi-squared test (See e.g. \citep{lehmann2006testing}, Theorem 14.3.1.)
For the proof of parts 2 and 3, we refer to \citep{balakrishnan2019hypothesis}. More specifically, \citep{balakrishnan2019hypothesis} proves similar claims for the `truncated' $\chi^2$-test statistic and for more general hypotheses regarding the nominal probabilities of the labels under multinomial models. For the special case of the uniformity testing problem~\eqref{eq: multi-unif-hypothesis}, the truncated Chi-squared statistic reduces to the classic Pearson's Chi-squared test statistic.

% The threshold $\ell+\sqrt{2\ell/\alpha}$ in the decision rule $\Psi_{n,\ell}$ is presented in \citep{balakrishnan2019hypothesis} for the truncated $\chi^2$-test statistic for a general multinomial hypothesis testing problem with $n$ samples and $\ell$ categories. The important point that we'd like to emphasize here is that the 

The next lemma is the Berry-Esseen theorem for non-identical independent random variables and its statement is borrowed from 
\citep[Section 5]{barbour2005introduction}.
\begin{lemma}(\citep[Section 5]{barbour2005introduction})\label{lemma: berry-essen}
For zero-mean independent random variables $\xi_1,...,\xi_n$ with $\sum\limits_{i=1}^{n}\E[\xi_i^2]=1$, let $W=\sum\limits_{i=1}^{n}\xi_i$. If $\sum\limits_{i=1}^{n}\E[|\xi_i^3|]\leq \gamma$, then we have
\[
\sup\limits_{-\infty\leq z\leq \infty}^{}\left|\prob\left(W \leq z  \right)-\Phi(z) \right| \leq \gamma\,.
\]
\end{lemma}

\section{Proofs of Section \ref{sec:PCR}}
\subsection{Proof of Theorem \ref{thm: chi^2-CI-size}}\label{proof:thm: chi^2-CI-size}
Because of Assumption \ref{assum: mu_mapping} (continuity of probability laws),  with probability one all the score values are distinct and so there is no ambiguity (tie) in labeling the data points.
Recall $W_\ell$ as the number of data points with label $\ell$.
By construction, the joint distribution of $(\myg W_1,\myg W_2,...,\myg W_{L})$ is a multinomial distribution with $L$ distinct values (number of labels). Denote by $p_\ell$ the probability of getting label $\ell$. 
%
% where for $1\leq s\leq \ell$, the category $s$ occurs with probability $p_s$, and .  We computed a closed-form value of $p_s$ in Proposition \ref{propo: lambda_bound}. 
Then, the statistic $U_{\myg,L}$, given by~\eqref{eq:U_nell}, is the standard Pearson's $\chi^2$ test statistic for testing the null hypothesis 
\begin{equation}\label{eq: tmp11}
H'_0: p_{\ell}=\frac{1}{L}, \text{ for } \ell\in[L]\,.
\end{equation}
The claim about $\th_{L,\alpha}^{\mathsf{finite}}$ follows from Part 2 of Lemma \ref{lemma: multinomial-main}.

{{Regarding the claim on PCR with $\th_{L,\alpha}^{\mathsf{asym}}$,  we know that by using Lemma \ref{lemma: multinomial-main} (Part 1), $U_{\myg,L} \overset{d}{\Rightarrow} \chi^{2}_{L-1}$, as $\myg\to\infty$. However, establishing uniform convergence requires additional work. 

Define $\br_j\in\reals^{L-1}$ with $r_{j\ell} = 1 - 1/L$ if groups $j$ is assigned label $\ell$ and $r_{j\ell} = - 1/L$ otherwise. Since each group is assigned exactly one label, we have dropped $r_{jL}$ from the vector $\br$ because it is determined given other entries. In addition, under the null hypothesis $\E[\br_j] = \boldsymbol{0}$.
Let $\bV_{\myg}$ be the $(L-1)\times 1$ vector defined by
\[
\bV_{\myg} = \sqrt{n_g}\left(\frac{W_1}{n_g}-\frac{1}{L},\dots,\frac{W_{L-1}}{n_g}-\frac{1}{L}\right) = \frac{1}{\sqrt{\myg}}\sum_{j=1}^{\myg} \br_j\,,
\] 
where the last step follows by definition of $\br_j$. By the multivariate CLT, $\bV_{\myg}  \stackrel{(d)}{\to} \normal(0,\bSigma)$ with
\begin{eqnarray}
\Sigma_{ij} = \begin{cases}
\frac{1}{L}- \frac{1}{L^2} & \text{if }i=j,\\
-\frac{1}{L^2} & \text{otherwise}. 
\end{cases}
\end{eqnarray}
We next invoke the following Berry-Esseen type bound from~\cite{bentkus2005lyapunov}.
\begin{thm}\label{thm:lyap}(\cite[Theorem 1.1]{bentkus2005lyapunov}) Let $\bx_1,\dotsc, \bx_n\in\reals^d$ be independent random vectors with common mean $\E(\bx_j) = \boldsymbol{0}$.
Write $\bs: = \sum_{i=1}^n \bx_j$ and assume that $\bs$ has covariance $\bR$. Let $\bZ$ be a zero mean Gaussian random vector with covariance $\bR$. Write $\beta = \beta_1+\dotsc+\beta_n$ with $\beta_j:=\|\bR^{-1/2}\bx_j\|_2^3$, and 
\[
\Delta(\mathfrak{C}) : = \sup_{A\in \mathfrak{C}} \left| \prob(\bs\in A)- \prob(\bZ\in A)\right|\,,
\]
where $\mathfrak{C}$ stands for the class of all convex subsets of $\reals^d$. There exists an absolute constant $C>0$ such that
\[
\Delta(\mathfrak{C}) \le Cd^{1/4}\beta\,.
\]
\end{thm}
 We use the above theorem for the sequence of $\br_j$, $j\in [\myg]$. Note that $\sum_{j=1}^{\myg}\br_j = \sqrt{\myg}\bV_{\myg}$
has covariance $\myg \bSigma$. In addition, 
\begin{eqnarray}
(\Sigma^{-1})_{ij} = \begin{cases}
{2}{L}, & \text{if }i=j,\\
{L}, & \text{otherwise}. 
\end{cases}
\end{eqnarray}
Therefore, 
\[
\beta_j = \myg^{-3/2}(\br_j^\sT \bSigma^{-1}\br_j)^{3/2}\,.
\]
Substituting for $\bSigma^{-1}$, we have $\br_j^\sT \bSigma^{-1}\br_j =  {L}(\|\br_j\|^2 +(\boldsymbol{1}^T\br_j)^2)  = {L} (1-\frac{1}{L})$,
where the last step follows from the definition of $\br_j$. Therefore, $\beta \le \myg^{-1/2} {L} (1-\frac{1}{L})$ and by Theorem~\ref{thm:lyap} we get
\[
\Delta(\mathfrak{C}) \le C \myg^{-1/2} (L-1)^{1/4} {L} \Big(1-\frac{1}{L}\Big) =C \myg^{-1/2} (L-1)^{5/4}\,.
\]
To complete the proof of~\eqref{eq:uniform}, we consider the convex set $A:= \{\bw: \bw^\sT(\myg\bSigma)^{-1}\bw\le \eta\}$. 
For $\bZ\sim\normal(\boldsymbol{0},\myg \bSigma)$, we have $V:= \bZ^\sT(\myg\bSigma)^{-1}\bZ \sim \chi^2_{L-1}$. Also,
\begin{align*}
\prob\Big(\sum_{j=1}^{\myg}\br_j \in A\Big) = \prob(\sqrt{\myg}\bV_{\myg}\in A)
= \prob(\bV_{\myg}^\sT \bSigma^{-1} \bV_{\myg} \le \eta)\,.
\end{align*}
It is easy to see that by definition of $\bV_{\myg}$, we have $U_{\myg,L} = \bV_{\myg}^\sT \bSigma^{-1} \bV_{\myg}$. Hence, we get
\begin{align}
\sup_{\eta} \Big|\prob(U_{\myg,L} \le \eta) - \prob(V \le \eta)\Big| \le \Delta(\mathfrak{C}) \le C \myg^{-1/2} (L-1)^{5/4}\,,
\end{align}
which completes the proof of~\eqref{eq:uniform}.}}
%%%%%%%%%%%%%%%%%%%%%%%%%%%%%%%%%%%%

%%%%%%%%%%%%%%%%%%%%%%%%%%%%%%%%%%%%%%%%%%%%%%%%%%%%%%
\section{Proofs of Section \ref{sec: pwr}}
\subsection{Proof of Remark \ref{rmk: Delta upper bound} }\label{proof:rmk: Delta upper bound}
Define 
\[
g(u,\bz,\by) : = \frac{\partial}{\partial u}F_{T|\bZ\bY}\Big( F_{T|\bZ}^{-1}\big(u; \bZ,\bY\big) ; \bZ,\bY\Big)  \,.
\]

Then by Assumption \ref{assum: dR-L1-integrable} we have $\int_0^1 \E_{(\bZ,\bY) \sim \cL(\bZ,\bY)}[|g(u,\bZ,\bY)|]<\infty$ and as an application of Fubini's theorem, we can change the order of integration and the expectation and get:
\begin{align*}
&\int_0^u \E_{(\bZ,\bY) \sim \cL(\bZ,\bY)}[g(v,\bZ,\bY)]\de v \\
&= \E_{(\bZ,\bY) \sim \cL(\bZ,\bY)}\Big[\int_0^u g(v,\bZ,\bY) \de v\Big]\\
 &= \E_{(\bZ,\bY) \sim \cL(\bZ,\bY)}\Big[F_{T|\bZ\bY}\Big( F_{T|\bZ}^{-1}\big(u; \bZ,\bY\big) ; \bZ,\bY\Big) - F_{T|\bZ\bY}\Big( F_{T|\bZ}^{-1}\big(0; \bZ,\bY\big) ; \bZ,\bY\Big)\Big]\\
 &=  \E_{(\bZ,\bY) \sim \cL(\bZ,\bY)}\Big[F_{T|\bZ\bY}\Big( F_{T|\bZ}^{-1}\big(u; \bZ,\bY\big) ; \bZ,\bY\Big)
\end{align*}
Now taking derivative of both sides with respect to $u$, we arrive at
\begin{align}\label{eq:interchange}
\E_{(\bZ,\bY) \sim \cL(\bZ,\bY)}[g(v,\bZ,\bY)] =  \sr_T(u)\,.
\end{align}
	
	We next prove part (b) of the remark. Part(a) follows readily from part (a) since under the null hypothesis $T(\bX,\bZ,\bY)$ and $T(\tbX,\bZ,\bY)$ have the same distribution.
	
	By definition of the conditional dependency power $\Delta_T(\cL(\bX,\bZ,\bY))$, cf. Definition \ref{def: conditional-dependency-power} we have
	\begin{align*}
		\Delta_T(\cL(\bX,\bZ,\bY))&= \int_0^1 |\sr_T(u)-1|\de u\\
		&\overset{(a)}{=}\int\limits_{0}^{1}\left|\E_{(\bZ,\bY) \sim \cL(\bZ,\bY)}\left[ \frac{\partial}{\partial u} F_{T|\bZ\bY}\left( F_{T|\bZ}^{-1}\big(u;\bZ,\bY\big);\bZ,\bY\right)  \right] -1\right| \de u \\
		&\overset{(b)}{\leq} \int\limits_{0}^{1} \E_{(\bZ,\bY) \sim \cL(\bZ,\bY)}\left[ \left| \frac{\partial}{\partial u} F_{T|\bZ\bY}\left( F_{T|\bZ}^{-1}\big(u;\bZ,\bY\big); \bZ,\bY\right)-1\right|\right]\de u \\
		&\overset{(c)}{=} \E_{(\bZ,\bY) \sim \cL(\bZ,\bY)} \left[ \int\limits_{0}^{1}  \left| \frac{\partial}{\partial u} F_{T|\bZ\bY}\left( F_{T|\bZ}^{-1}\big(u; \bZ,\bY\big); \bZ,\bY\right)-1\right|\right]\de u \\
		&= \E_{(\bZ,\bY) \sim \cL(\bZ,\bY)}  \left[ \int\limits_{0}^{1} \left| \frac{f_{T|\bZ\bY}\left( F_{T|\bZ}^{-1}\big(u; \bZ,\bY\big); \bZ,\bY\right)}{ f_{T|\bZ}\left( F_{T|\bZ}^{-1}\big(u; \bZ,\bY\big); \bZ,\bY\right) }-1\right|\right]\de u \\
		&= \E_{(\bZ,\bY) \sim \cL(\bZ,\bY)}  \left[ \int\limits_{-\infty}^{\infty} \left| \frac{f_{T|\bZ,\bY}\left(t;\bZ,\bY \right)}{ f_{T|\bZ}( t;\bZ,\bY) }-1\right|f_{T|\bZ}(t;\bZ,\bY)\de t\right]\\
		%&\overset{(d)}{=}\E_{(Z,Y)\sim P_{Z,Y}}\left[  \E_{X\sim P_{X|Z}}\left[\left| \frac{  f_{T|ZY}\left(T(X,Z,Y)|Z,Y\right)  }{  f_{T|Z} \left(T(X,Z,Y)|Z,Y\right)  }-1\right| \right]\right]\\
		%&=\E_{(X,Z,Y)\sim P_{X|Z}\otimes P_{Y,Z} } \left[ \left| \frac{  f_{T|ZY}\left(T(X,Z,Y)|Z,Y\right)  }{  f_{T|Z} \left(T(X,Z,Y)|Z,Y\right)  }-1\right|     \right]\\
		%&\overset{(e)}{=}\E_{(X,Z,Y)\sim P_{X|Z}\otimes P_{Y,Z} } \left[ \left| \frac{  f_1(T(X,Z,Y))  }{  f_2(T(X,Z,Y)) }-1\right|     \right]\\
		&=\E_{(\bZ,\bY) \sim \cL(\bZ,\bY)}\left[2d_{\tv}\left(  (T(\tbX,\bZ,\bY)|\bZ,\bY), (T(\bX,\bZ,\bY)|\bZ,\bY) \right)\right], 
	\end{align*}
	with $\bX\sim \cL(\bX|\bZ,\bY)$, $\tX\sim \cL(\bX|\bZ)$, and $f_{T|\bZ,\bY}$ and $f_{T|\bZ}$ representing the density functions corresponding to cdfs  $F_{T|\bZ,\bY}$ and $F_{T|\bZ}$ . Note that in (a) we used \eqref{eq:interchange}; (b) is a direct result of Jensen's inequality, and (c) follows from Assumption \ref{assum: dR-L1-integrable} in conjunction with Fubini's theorem.
	%(d) is a direct result from the definition of $f_{T|Z}(.|Z,Y)$ which is the density function of $T(X,Z,Y)$ when $X\sim P_{X|Z}(.|Z)$, and finally in $(e)$, $f_1(t), f_2(t)$ denote the probability density functions of random variable $T(X,Z,Y)$ when $X\sim P_{X|Z}$, and $X\sim P_{X|Z,Y}$, respectively; it is not hard to see that $f_1(t)=f_{T|Z}(t|z,y)f_{Z,Y}(z,y)$
	%he fact that $f_{T|ZY}(.|Z,Y)$ stands for the density function of  $T(X,Z,Y)$ when 

%%%%%%%%%%%%%%%%%%%%%%%%%%%%%%%%%%%%%%%%%%%%%%%%%%%%%%%%%%
\subsection{Proof of Proposition \ref{propo: lambda_bound}}\label{proof:propo: lambda_bound}

	%It is easy to see that each $(X_j,Z_j,Y_j)$ can be categorized precisely in one of the $m+1$ possible values $1\leq \ell\leq m+1$ and  $Y_\ell$ denotes the normalized number of $\ell-$type triples.
Based on the Pearson $\chi^2$-CI statistic $U_{\myg,L}$ construction that is described in Algorithm \ref{algorithm: model-xz}, for a group $\cG=(\bX,\bZ,\bY)$ and its $M$ constructed counterfeits $\{\tcG_i=(\tbX_{i},\bZ,\bY)\}_{i=1:M}$ we have the following rank value 
\[
R=1+\sum\limits_{j=1}^{M}\ind{ \{ T(\cG) \geq T(\tcG_j)  \}} \,.
\]
This allows us to compute the  probability of $\cG$ getting label $t\in[L]$:
\begin{align}
	\prob\left(\cG \text{ has label } t \right)&=\prob\left((t-1)K+1 \leq R\leq tK\right)\nonumber\\
	&=\sum\limits_{j=K(t-1)+1}^{Kt}\prob(R=j)\nonumber\\
	&=\sum\limits_{j=K(t-1)+1}^{Kt}\E_{\bZ\bY}[\prob(R=j|\bY=\by,\bZ=\bz)]\label{eq:Ryz}\,.
\end{align}
%Introduce
%
%\[
%\gamma(z,y):= \prob\big(\mu(X,Z,Y)\geq \mu(\tX,Z,Y)|Z=z,Y=y\big),\quad\quad \text{with } X\sim P_{X|ZY}\text{ and } \tX\sim P_{X|Z}\,. 
%\]
Note that by conditioning on $(\bZ,\bY)=(\bz,\by)$, random variables $T(\cG)$ and $T(\tcG_i)$ are independent as $\cG=(\bX,\bZ,\bY)$ and $\tcG_i= (\tbX_i,\bZ,\bY)$. To lighten the notation, we use the shorthands $T:=T(\cG)$ and $\tT_i=T(\tcG_i)$, and proceed as follows:
%and $R$ has binomial distribution with $M$ trials and success probability $\gamma(z,y)$, therefore we get
\begin{align}
	&\prob\left( R=j|\bY=\by,\bZ=\bz \right)\nonumber\\
	&= \prob\left( T \text{ is exactly larger than } j-1\text{ of } \tT_i  |\bZ=\bz,\bY=\by \right)\nonumber\\
	&\overset{(a)}{=}\int \prob\big( t \text{ is exactly larger than } j-1\text{ of } \tT_i  |\bZ=\bz,\bY=\by \big)\; \de F_{T|\bZ\bY}(t;\bz,\by)\nonumber\\
	&\overset{(b)}{=}\binom{M}{j-1} \int F_{T|\bZ}(t;\bz,\by)^{j-1}(1-F_{T|\bZ}(t;\bz,\by))^{M-j+1}\de F_{T|\bZ\bY}(t;\bz,\by)\nonumber \\
	&{=}\binom{M}{j-1}\int\limits_{0}^{1} u^{j-1}(1-u)^{M-j+1}\de F_{T|\bZ\bY}\big(F_{T|\bZ}^{-1}(u;\bz,\by);\bz,\by\big)\label{eq: tmp6} \,,
\end{align}
where (a) comes from the fact that $T|\bZ\bY$ has density $F_{T|\bZ\bY}(.)$, and (b) holds since $\tT_i|\bZ\bY$ is distributed according to $F_{T|\bZ}(.)$, independent of $T$.  Note that for a function $f(x)$, the notation $\de f(x)  =f'(x)\de x$ denotes the differential of $f(x)$.
%In step $(c)$ we use the following identity which follows from the chain rule: 
%\[
%\partt f(t) = \partu f(g^{-1}(u))\,,
%\]
%if $u = g(t)$, for any differentiable functions $g,f$ with $g$ being an invertible function.

We next plug in equation \eqref{eq: tmp6} into \eqref{eq:Ryz} to get

\begin{align}
	\prob\left(\cG\text{ has label } t \right)&=
	\sum\limits_{j=K(t-1)+1}^{Kt}\binom{M}{j-1}\E_{\bZ\bY}
	\left[ \int\limits_{0}^{1} u^{j-1}(1-u)^{M-j+1}\de F_{T|\bZ\bY}\big(F_{T|\bZ}^{-1}(u;\bZ,\bY);\bZ,\bY\big)       \right]\nonumber\\
	&\overset{(a)}{=}	\sum\limits_{j=K(t-1)+1}^{Kt}\binom{M}{j-1} \int\limits_{0}^{1} u^{j-1}(1-u)^{M-j+1}\E_{\bZ\bY}
	\left[\de F_{T|\bZ\bY}\big(F_{T|\bZ}^{-1}(u;\bZ,\bY);\bZ,\bY\big) \right]\nonumber\\
	&\overset{{(b)}}{=} \sum\limits_{j=K(t-1)+1}^{Kt}\binom{M}{j-1} \int\limits_{0}^{1} u^{j-1}(1-u)^{M-j+1}\de\E_{\bZ\bY}
	\left[F_{T|\bZ\bY}\big(F_{T|\bZ}^{-1}(u;\bZ,\bY);\bZ,\bY\big) \right]\nonumber\\
	&=\sum\limits_{j=K(t-1)+1}^{Kt}\binom{M}{j-1} \int\limits_{0}^{1} u^{j-1}(1-u)^{M-j+1}\de \sR_{T}(u)\nonumber\\
	&=\sum\limits_{j=K(t-1)}^{Kt-1}\binom{M}{j} \int\limits_{0}^{1} u^{j}(1-u)^{M-j} \sr_T(u)\de u\label{eq: tmp7}\,,
\end{align}
in (a) we used Fubini's theorem along with Assumption \ref{assum: dR-L1-integrable} and the fact that for every $0\leq u\leq 1 $ we have $|u^{j}(1-u)^{M-j}|\leq 1$.  Also, (b) is a direct result of Assumption \ref{assum: dR-L1-integrable} and dominated convergence theorem. This completes the proof of claim~\eqref{eq: tp_j}.

It is worth noting that, when $X\indep Y|Z$, we have $P_{\bX|\bZ\bY}=P_{\bX|\bZ}$ which implies $\sR_{T}(u)=u$, so the conditional relative density function $\sr_{T}(u)$ always attains the constant value $1$. In this case, we have 
\begin{align}
	p_{t}&= \sum\limits_{j=K(t-1)}^{Kt-1}\binom{M}{j} \int\limits_{0}^{1} u^{j}(1-u)^{M-j}\de u\nonumber\\
	&=\sum\limits_{j=K(t-1)}^{Kt-1}\binom{M}{j} B(j+1,M-j+1)\nonumber\\
	&=\sum\limits_{j=K(t-1)}^{Kt-1}\binom{M}{j} \frac{\Gamma(j+1)\Gamma(M-j+1)}{\Gamma(M+2)}\nonumber\\
	&=\sum\limits_{j=K(t-1)}^{Kt-1}\binom{M}{j} \frac{j!(M-j)!}{(M+1)!}\nonumber\\
	&=\sum\limits_{j=K(t-1)}^{Kt-1}\frac{1}{M+1}\nonumber\\
	&=\frac{K}{M+1}=\frac{1}{L}\label{eq: tmp8}\,,
\end{align}	
where $B(a,b)$ is the Beta function and $\Gamma(a)$ is the Gamma function.

Now, we are ready to prove Part $(i)$. First note that deriving a more explicit characterization of $p_t$ from~\eqref{eq: tmp7} is in general intractable, due to the relative density term $\sr_{T}(u)$ in the inner integral expression . However, it is useful to note that if $\sr_{T}(u)$ is a polynomial of $u$, then this probability can be easily computed by absorbing that into the integral formulation of the Beta function and then leveraging the connection between the Gamma function and Beta function for integer values.  Inspired by this observation, our strategy is to  approximate $\sr_{T}(u)$ with polynomials.  To this end, note that by Assumption \ref{assum: sr_continuity}, $\sr_{T}(u)$ is a continuous function over $[0,1]$ interval, which allows us to use the Weierstrass theorem to uniformly approximate $\sr_T(u)$ as closely as desired by polynomials.  Formally, for any $\eps>0$ there exists a polynomial $\widetilde{r}(u)$ with real coefficients such that 
\begin{equation}\label{eq: def-eps_i}
\sup\limits_{u\in [0,1]}^{} |\widetilde{r}(u)-\sr_T(u)| <\eps \,.
\end{equation}
%we have 
%\begin{equation}\label{eq: lim-eps_j}
%\lim\limits_{i\rightarrow \infty}^{}\eps_i=0\,.
%\end{equation}
In addition, from \eqref{eq: tmp7} for every $\ell\in[L]$ we have
\begin{align}
	\sum\limits_{t=1}^{\ell}p_t&= \sum\limits_{j=0}^{\ell K-1} \binom{M}{j}\int\limits_{0}^{1}u^{j}\big(1-u\big)^{M-j}\sr_{T}(u)\de u\nonumber\\
	&\geq \sum\limits_{j=0}^{\ell K-1} \binom{M}{j}\int\limits_{0}^{1}u^{j}\big(1-u\big)^{M-j}\widetilde{r}(u)\de u -\eps  \sum\limits_{j=0}^{\ell K-1} \binom{M}{j}\int\limits_{0}^{1}u^{j}\big(1-u\big)^{M-j}\de u\nonumber\\
	&= \sum\limits_{j=0}^{ \ell K-1} \binom{M}{j}\int\limits_{0}^{1}u^{j}\big(1-u\big)^{M-j}\widetilde{r}(u)\de u -\frac{\ell\eps}{L}\label{eq:lower-p_ell}\,,
\end{align}
where in the last equality we used the result in \eqref{eq: tmp8} that when $\sR_{T}(u)=u$, we have $p_{t}=1/L$. We are left with lower bounding the right-hand side summation in \eqref{eq:lower-p_ell}. Let $\widetilde{r}(u)$ be a polynomial of degree $N$ and coefficients $a_i$, i.e. $\widetilde{r}(u)=\sum\limits_{i=0}^{N}a_iu^i$. We have 
 \begin{align}
 	&\sum\limits_{j=0}^{\ell K-1} \binom{M}{j}\int\limits_{0}^{1}u^{j}\big(1-u\big)^{M-j}\widetilde{r}(u)\de u\nonumber\\
 	&=\sum\limits_{j=0}^{\ell K-1} \binom{M}{j}\int\limits_{0}^{1}u^{j}\big(1-u\big)^{M-j}\sum\limits_{i=0}^{N}a_iu^i\de u\nonumber\\
 	&=\sum\limits_{j=0}^{ \ell K-1}
 	\sum\limits_{i=0}^{N}a_i \binom{M}{j}\int\limits_{0}^{1}u^{j+i}\big(1-u\big)^{M-j}\de u\nonumber\\
 	&=\sum\limits_{j=0}^{\ell K-1}
 	\sum\limits_{i=0}^{N}a_i \binom{M}{j}B(j+i+1,M-j+1)\nonumber\\
 	&=\sum\limits_{j=0}^{\ell K-1}
 	\sum\limits_{i=0}^{N}a_i \binom{M}{j} \frac{(j+i)!(M-j)!}{(M+i+1)!\nonumber}\\
 	&=	\sum\limits_{i=0}^{N} a_i\frac{M!i!}{(M+i+1)!}\sum\limits_{j=0}^{\ell K-1}
 \binom{j+i}{i}\nonumber\\
 &=	\sum\limits_{i=0}^{N} a_i\frac{M!i!}{(M+i+1)!}
 \binom{\ell K+i}{i+1}\nonumber\\
 &=	\sum\limits_{i=0}^{N} \frac{a_i}{i+1}\prod\limits_{h=0}^{i}\frac{\ell K +h}{M+1+h}\label{eq:sum-p_ell-polynomial}\,,
 \end{align}
where in the penultimate equation, we used the Hockey-stick identity. Next, use the following simple inequality in \eqref{eq:sum-p_ell-polynomial}
 $$\frac{\ell K+h}{M+1+h}\geq \frac{\ell K}{M+1}=\frac{\ell}{L}\,,$$
to arrive at
\begin{align*}
\sum\limits_{j=0}^{\ell K-1} \binom{M}{j}\int\limits_{0}^{1}u^{j}\big(1-u\big)^{M-j}\widetilde{r}(u)\de u\nonumber\geq \sum\limits_{i=0}^{N} \frac{a_i}{i+1}\left( \frac{\ell}{L} \right)^{i+1}=\int\limits_{0}^{\frac{\ell}{L}} \widetilde{r}(u)\de u\,.
\end{align*}
Next we plug the above lower bound into \eqref{eq:lower-p_ell} to get
\[
\sum\limits_{t=1}^{\ell}p_{t} \geq \int\limits_{0}^{\frac{\ell}{L}} \widetilde{r}(u)\de u-\frac{\ell\eps}{L}\,,
\]
which along with \eqref{eq: def-eps_i} implies that
\[
\sum\limits_{t=1}^{\ell}p_{t} \geq \int\limits_{0}^{\frac{\ell}{L}} \sr_T(u)du-\frac{2\ell\eps}{L}=\sR_{T}\left(\frac{\ell}{L} \right)-\frac{2\ell\eps}{L}\,.
\]
Finally, since $\eps>0$ can be chosen arbitrarily small, by letting $\eps\to 0$ we get the desired claim of \eqref{eq:sum-p_ell-lower}.

We next proceed to Part $(ii)$. In Part $(i)$, we use a general form of the Weierstrass approximation theorem, to uniformly approximate $\sr_T$ as closely as desired, while the rate of convergence (in terms of the polynomial degree) was not needed.
For establishing an upper bound on the sum of labels probabilities,  $\sum\limits_{s=1}^{\ell}p_s$, we need to upper bound the polynomial-approximation error, and knowing the convergence rate becomes important. For this reason, we use a more refined  version of the Weierstrass approximation theorem. For the reader's convenience, we state this version in the following lemma, borrowed from~\citep{gzyl1997weierstrass}:
\begin{lemma}[ \citep{gzyl1997weierstrass}, Theorem 1]\label{lem:poly}
	Let $f$ be a $B$-bounded and  $C$-Lipschitz continuous function on $[0,1]$. Then, for every positive integer $N$, there exists a polynomial $\widetilde{f}_N$ of degree $N$ such that
	\[
\sup\limits_{u\in[0,1]}^{}|f(u)- \widetilde{f}_N(u)|\leq(C/2+2B) \sqrt{\frac{\log{N}}{N}}\,.
	\]
\end{lemma}
	Recall that by Assumption \ref{assum: sr_continuity}, $\sr_T(u)$ is $B$-bounded and $C$-Lipschitz, and therefore, by an application of Lemma~\ref{lem:poly} there exists a polynomial $\widetilde{r}_N$ of degree $N$, such that for $D=C/2+2B$ we have
	\begin{equation}\label{eq:unif-Bernstein-polynomials}
	\|\sr_{T}-\widetilde{r}_N\|_\infty \leq D\sqrt{\frac{\log{N}}{N}}\,.
	\end{equation}
	Let $\widetilde{r}_N(u)=\sum\limits_{i=0}^{N}a_iu^i$. By a similar argument used in deriving \eqref{eq:lower-p_ell} and \eqref{eq:sum-p_ell-polynomial}, we get
	\begin{align}\label{eq:dum0}
	\sum\limits_{t=1}^{\ell}p_t&\leq \sum\limits_{i=0}^{N} \frac{a_i}{i+1}\prod\limits_{h=0}^{i}\frac{\ell K +h}{M+1+h} + \frac{\ell D}{L}\sqrt{\frac{\log{N}}{N}}\,.
	\end{align}
To further simplify the right-hand side, we use the following simple algebraic manipulations. Since $h\le i\le N$ and $M+1 = LK\ge \ell K$ we have $(M+1-\ell K)(N-h)\ge0$, from which we get 
\begin{align*}
\frac{\ell K+h}{M+1+h} &\leq  \frac{\ell K+N}{M+1+N}\\
&= \frac{\ell K+N}{LK+N} =\frac{\ell}{L}\left(\frac{K+\tfrac{N}{\ell}}{K+\tfrac{N}{L}}\right)\le \frac{\ell}{L}\left(1+\frac{N}{K}\right)\,.
\end{align*}
Using this bound in~\eqref{eq:dum0}, for $h\ge 1$, we arrive at
\begin{align*}
	\sum\limits_{t=1}^{\ell}p_t &\leq\left(1+\frac{N}{K}\right)^N \sum\limits_{i=0}^{N} \frac{a_i}{i+1}\left(\frac{\ell}{L}\right)^{i+1}+ \frac{\ell D}{L}\sqrt{\frac{\log{N}}{N}}\\
	&=\left(1+\frac{N}{K}\right)^N\int\limits_{0}^{\frac{\ell}{L}}\widetilde{r}_N(u)\de u+ \frac{\ell D}{L}\sqrt{\frac{\log{N}}{N}}\\
	&\leq e^{N^2/K}\int\limits_{0}^{\frac{\ell}{L}}\widetilde{r}_N(u)\de u+ \frac{\ell D}{L}\sqrt{\frac{\log{N}}{N}}\,.
	\end{align*}
By using \eqref{eq:unif-Bernstein-polynomials} again, we obtain 
\begin{align*}
	\sum\limits_{t=1}^{\ell}p_t&\le e^{N^2/K}\int\limits_{0}^{\frac{\ell}{L}}{r}_T(u)\de u+ \frac{\ell D}{L}\sqrt{\frac{\log{N}}{N}}\left(1+e^{N^2/k} \right)\\
	&=e^{N^2/K}\;\sR_{T}\left(\frac{\ell}{L} \right)+ \frac{\ell D}{L}\sqrt{\frac{\log{N}}{N}}\left(1+e^{N^2/k} \right)\ \,.
\end{align*}

Set $N=\sqrt{K\log(1+\delta) }$ for a fixed $0<\delta<1$ and rewrite the above bound as
\[	\sum\limits_{t=1}^{\ell}p_t \leq
(1+\delta)\sR_{T}\left(\frac{\ell}{L} \right)+ \frac{3\ell D}{L}\left( \frac{\log\left(K\log(1+\delta)\right)}{2\sqrt{K\log(1+\delta)}}\right)^{1/2}\,.
\]

By using the relations  $\ell\leq L$, $\delta<1$, $\sR_T(u)\leq 1$, and $\log(1+\delta)\geq\delta/2$, for $\delta\in[0,1]$, we obtain

\[	\sum\limits_{t=1}^{\ell}p_t \leq
\sR_{T}\left(\frac{\ell}{L} \right)+ \delta +3D\left( \frac{\log{K}}{\sqrt{K\delta}}\right)^{1/2}\,.
\]
Minimizing the right-hand side over $\delta$, we get $\delta = \left(\frac{9D^2\log{K}}{\sqrt{K}}  \right)^{2/5}$ , which is smaller than one for $k$ sufficiently large. Plugging in for this value of $\delta$ we obtain 
\[
\sum\limits_{t=1}^{\ell}p_t \leq \sR_{T}\left(\frac{\ell}{L} \right)+\nu_K\,,
\]
with $\nu_K=2\left(\frac{9D^2\log{K}}{\sqrt{K}}  \right)^{2/5}$.
%%%%%%%%%%5
%%%%%%%%%%%%%

We next proceed to prove Part $(iii)$. 
%Equation \eqref{eq: lambda_n} gives us 
%\begin{align}
%\lambda_n&=n(m+1)\sum\limits_{\ell=1}^{m+1}\left( p_\ell-\frac{1}{m+1} \right)^2\nonumber\\
%&=n\left( (m+1) \sum\limits_{\ell=1}^{m+1}p_\ell^2-1\right)\label{eq:lambda_n-simplified}\,.
%\end{align}
For $t \in [L]$, let 
\begin{equation}\label{eq: tmp9}
q_t:= \sR_{T}\left( \frac{t}{L}\right)- \sR_{T}\left( \frac{t-1}{L}\right)\,,
\end{equation}
By employing the results of parts $(i)$ and $(ii)$ we have
\begin{align*}
\left| p_t-q_t \right| \le \Big|\sum_{j=1}^{t} p_j - \sR_{T}\left( \frac{t}{L}\right) - \sum_{j=1}^{t-1} p_j
+ \sR_{T}\left( \frac{t-1}{L}\right) \Big| \leq \nu_K\,.
\end{align*}
Therefore,
\begin{align}
\sum\limits_{t=1}^{L}\left|p_t-\frac{1}{L}\right|
\geq \sum\limits_{t=1}^{L}\left|q_t-\frac{1}{L}\right| -\sum\limits_{t=1}^{L}\left|p_t-q_t\right|
\geq -L\nu_K+ \sum\limits_{t=1}^{L}\left|q_t-\frac{1}{L}\right|\label{eq:lambda_n_lower_1}\,.
%\sum\limits_{\ell=1}^{m+1}p_\ell^2&=\sum\limits_{\ell=1}^{m+1}q_\ell^2+\sum\limits_{\ell=1}^{m+1}(p_\ell^2-q_\ell^2)\\
%&\geq \sum\limits_{\ell=1}^{m+1}q_\ell^2 -\nu_k \sum\limits_{\ell=1}^{m+1}(p_\ell+q_\ell)\\
%&=\sum\limits_{\ell=1}^{m+1}q_\ell^2 -2\nu_k\,,
\end{align}
%\begin{equation}\label{eq:lambda_n_lower_1}
%\sum\limits_{\ell=1}^{m+1}\left|p_\ell-\frac{1}{m+1}\right| \geq \sum \limits_{\ell=1}^{m+1}\left|q_\ell-\frac{1}{m+1} f\right|
%\end{equation}
Next, by applying the mean value theorem in the definition of $q_t$ in \eqref{eq: tmp9}, for every $t\in [L]$, there exists $\xi_t \in \left(\frac{t-1}{L},\frac{t}{L}  \right)$, such that $q_t=\sr_{T}(\xi_t)/L$. Therefore,
\begin{align*}
	\sum\limits_{t=1}^{L} \left|q_t-\frac{1}{L}\right| &=\frac{1}{L}\sum\limits_{t=1}^{L} |\sr_{T}(\xi_t)-1|\\
	&=\sum\limits_{t=1}^{L}\int\limits_{\frac{t-1}{L}}^{\frac{t}{L}}|\sr_{T}(\xi_t)-1|\de u\\
	&\geq\sum\limits_{t=1}^{L}\int\limits_{\frac{t-1}{L}}^{\frac{t}{L}}|\sr_{T}(u)-1|\de u-\sum\limits_{t=1}^{L}\int\limits_{\frac{t-1}{L}}^{\frac{t}{L}}\left|\sr_{T}(u)-\sr_{T}(\xi_t)\right|\de u\\
	&\geq \int\limits_{0}^{1}|\sr_{T}(u)-1|\de u-  \sum\limits_{t=1}^{L} \int\limits_{\frac{t-1}{L}}^{\frac{t}{L}} {C|u-\xi_t|}{} \de u\\
	&\geq \int\limits_{0}^{1}|\sr_{T}(u)-1|\de u-\sum\limits_{t=1}^{L}\frac{C}{L^2}=\int\limits_{0}^{1}|\sr_{T}(u)-1|\de u-\frac{C}{L}\,.
\end{align*}	
Using the above lower bound into \eqref{eq:lambda_n_lower_1} gives  

\[
\sum\limits_{t=1}^{L}\left|p_t-\frac{1}{L} \right|\geq 
\int\limits_{0}^{1}|\sr_{T}(u)-1|\de u-L\nu_K-\frac{C}{L}\,.
\]

%%%%%%%%%%%%%%%%%%%%%%%%%%%%%%%%%%%%%%%%%%%%%%%%
\subsection{Proof of Theorem \ref{thm: power_balls_bins}}\label{proof:thm: power_balls_bins}

The primary arguments here are similar to the initial reasonings in the proof of Theorem \ref{thm: chi^2-CI-size}, where we arrived at the point that the joint distribution of $(W_1, W_2,..., W_{L})$ is a multinomial distribution with $L$ categories, such that category $\ell\in[L]$ happens with probability $p_\ell$. Next, recall Lemma \ref{lemma: multinomial-main}, part 3, where it implies that if for some $\beta>0$, the following holds:
\begin{equation}\label{eq: tmp12}
\sum\limits_{\ell=1}^{L}\left|p_\ell-\frac{1}{L}\right| \geq\frac{32{L^{1/4}}}{\sqrt{\myg}}\left[\frac{1}{\sqrt{\alpha}}\vee\frac{1}{\beta}  \right]^{1/2}\,,
\end{equation}
then the type II error is bounded by $\beta$. On the other hand, from Proposition \ref{propo: lambda_bound} we have
\begin{align}\label{eq:dum1}
\sum\limits_{\ell=1}^{L} \left| p_\ell-\frac{1}{L}\right| \geq 
\int\limits_{0}^{1}|\sr_T(u)\de u-1|-L \nu_L-\frac{C}{L}\,.
\end{align}
Combining equations~\eqref{eq: tmp12} and \eqref{eq:dum1},  in conjunction with the definition of the conditional dependency in Definition \ref{def: conditional-dependency-power} completes the proof.

 %For the reader's convenience,  we present this theorem here: 

 %Theorem 2 of \citep{balakrishnan2019hypothesis} states that

%This theorem for uniformity testing in multinomial distributions implies that
%\[
%P_0\left( n(m+1)\sum\limits_{\ell=1}^{m+1}(Y_\ell-\frac{1}{m+1})^2 \geq 
%(m+1)+\sqrt{\frac{2}{\alpha}(m+1)}   \right) \leq \alpha\,.
%\]
%Further, multinomial distributions with probability mass values   $(p_\ell)_{\ell=1:m+1}$ that for some $\beta>0$ satisfies

%\[
%\frac{1}{4}\left(\sum\limits_{\ell=1}^{m+1}\left|p_\ell-\frac{1}{m+1}\right| \right)^2\geq\frac{1024\sqrt{d}}{n}\left[\frac{1}{\sqrt{\alpha}}\vee\frac{1}{\beta}  \right]\,,
%\]
%have the Type II error smaller than $\beta$.

%From classical results on Pearson's statistic for Hypothesis $H'_0$ (  \citep{lehmann2006testing}, Theorem 14.3.1)  we know that $T\overset{d}{\rightarrow}\chi^2_{m}(\lambda_n)$ with the same $\lambda_n$ defined in Theorem \ref{thm: power_balls_bins}.

%%%%%%%%%%%%%%%%%%%%%%%%%%%%%
{
\subsection{Proof of Theorem \ref{thm: power-asym-full}} \label{proof: asym-full-power} 
Similar to the proof of Theorem \ref{thm: power_balls_bins}, we know that $(W_1,\dots, W_L)$ has a multinomial distribution with $L$ categories where outcome $\ell\in [L]$ occurs with probability $p_\ell$. In addition, from Proposition \ref{propo: lambda_bound}, we know that the probability values $\{p_\ell\}_{\ell\ge 1}$ are connected to the conditional dependency power by the following 
\[
\sum\limits_{\ell=1}^{L}\Big|p_\ell-\frac{1}{L}\Big|\ge \Delta_T(\cL(\bX,\bZ,\bY)) -L\nu_k-\frac{C}{L} \,.
\] 
Using this with \eqref{eq: lower-Delta-asym}, we get $\sum\limits_{\ell=1}^{L}\Big|p_\ell-\frac{1}{L}\Big| \ge \eps$. This implies that there exists $\ell^*\in [L]$ such that $p_{\ell^*}\neq \frac{1}{L}$. Let $\delta=|p_{\ell^*} - \frac{1}{L}|$, so $\delta>0$.  In the next step, by an application of the strong law of large numbers for sum of independent Bernoulli random variables  we have 
\begin{equation}\label{eq: tmp-power-full}
\Big(\frac{W_{\ell^*}}{\myg}-\frac{1}{L}\Big)^2 \overset{(\mathsf{a.s})}{\to}  \Big(p_\ell^*-\frac{1}{L}\Big)^2\,.
\end{equation}
Given that $\frac{U_{\myg,L}}{\myg} \ge \left(\frac{W_{\ell^*}}{\myg}-\frac{1}{L} \right)^2$, therefore by using \eqref{eq: tmp-power-full} we arrive at
\begin{equation}\label{eq: power-full-final}
\prob\left( \frac{U_{\myg,L}}{\myg} \ge \delta^2 \right)=1\,.
\end{equation}
Finally, from \eqref{eq: power-full-final} it is straightforward to get $\lim\limits_{\myg\to \infty}\prob\left( U_{\myg,L} \ge \th^{\mathsf{asym}}_{L,\alpha}\right)=1$.  This completes the proof. 

}

%%%%%%%%%%%%%%%%%%%%%%%%%%%%%%%%%%%%%%%%%%%
\subsection{Proof of Theorem \ref{thm: power_balls_bins_asympt}} \label{proof:thm: power_balls_bins_asympt} 
{
 Similar to the first part of the proof of Theorem \ref{thm: chi^2-CI-size}, we know that  $(W_1^{(\myg)}, W_2^{(\myg)},..., W_{L}^{(\myg)})$ is a multinomial distribution with $L$ categories, such that the category $\ell\in [L]$ occurs with probability $p_\ell^{(\myg)}$. By an application of Proposition \ref{propo: lambda_bound} for $\ell \in [L]$ we have
 \[
p^{(\myg)}_\ell= \sum\limits_{j=(\ell-1)K}^{\ell K-1} \binom{M}{j}\int\limits_{0}^{1}u^{j}\big(1-u\big)^{M-j}r^{(\myg)}_{T}(u)\de u\,.
 \]

The local alternative assumption implies that $p^{(\myg)}_\ell=\frac{1}{L}+\frac{h_\ell}{\myg}$.

 We then use the following asymptotic result on the Pearson's $\chi^2$ test statistic for multinomial models (see e.g., \citep[Theorem 14.3.1]{lehmann2006testing}):
	\begin{equation}\label{eq: asympt-power-multi-pearson}
	U_{\myg,L} \overset{(d)}{\Rightarrow} \chi^2_{\lambda,L-1}\,,
	\end{equation}   
where $\chi^2_{\lambda,L-1}$ stands for the $\chi^2$ distribution with $L-1$ degrees of freedom and the non-central parameter $\lambda=L \sum\limits_{\ell=1}^{L} h_\ell^2 $. This implies that for $Q\sim \chi^2_{\lambda, L-1}$ we have

\begin{align}\label{eq: tmp-power}
\lim\limits_{\myg \rightarrow \infty}^{} \prob\left(U_{\myg ,\ell}\geq \th^{\mathsf{asym}}_{L,\alpha} \right)&= \prob \left(Q\geq \th^{\mathsf{asym}}_{L,\alpha} \right)\,.
\end{align}

 Using the lower bound on $h_\ell$ values, we obtain  $\lambda \ge A^2 L^{1/2}$, where $A$ is given by:

\begin{equation}\label{eq: tmp-A}
A=\left[\sqrt{3\log\frac{1}{\beta}}+\left( 3\log\frac{1}{\beta} +  2\sqrt{\log\frac{1}{\alpha}}+2\log\frac{1}{\alpha} \right)^{1/2} \right]\,.
\end{equation}

Thereby, by introducing $\widetilde{\lambda}= A^2 L^{1/2}$ from \eqref{eq: tmp-power} for $\widetilde{Q}\sim \chi^2_{\widetilde{\lambda},L-1}$ we have

\begin{equation}\label{eq: tmp-power-2}
\lim\limits_{\myg \rightarrow \infty}^{} \prob\left(U_{\myg ,\ell}\geq \th^{\mathsf{asym}}_{L,\alpha} \right) \ge \prob \left(\widetilde{Q}\ge \th^{\mathsf{asym}}_{L,\alpha} \right)\,.
\end{equation}

We then provide the following inequality borrowed from \citep{birge2001alternative} on tails of non-central $\chi^2$ random variables.

\begin{lemma}[\citep{birge2001alternative}, Lemma 8.1]\label{lemma: chi-tails}
Suppose that  random variable $X$ has a $\chi^2$ distribution with $m$ degrees of freedom and non-central parameter $\lambda$. Then for every $t\geq 0$  we have 
\begin{align*}
&\prob\left(X\leq m+\lambda -2\sqrt{(m+2\lambda)t}\right) \leq \exp(-t)\,,\\
&\prob\left(X\geq m+\lambda +2\sqrt{(m+2\lambda)t}+2t\right) \leq \exp(-t)\,.
\end{align*}
 \end{lemma}
As an immediate consequence of Lemma \ref{lemma: chi-tails}, we can obtain the following upper bound on the  $(1-\alpha)$-th quantile of the central $\chi^2$ distribution with $m$ degrees of freedom:
 \begin{equation}\label{eq: tmp-chi-quantile-bound}
\chi^2_{m}(1-\alpha) \leq m+2\sqrt{m\log\frac{1}{\alpha}} + 2\log\frac{1}{\alpha}\,.
 \end{equation}
By substituting $m=L-1$ in \eqref{eq: tmp-chi-quantile-bound} we get
\begin{equation}\label{eq: tmp-upper-th-asymp}
\th^{\mathsf{asym}}_{L,\alpha}\leq L-1+2\sqrt{(L-1)\log\frac{1}{\alpha}} + 2\log\frac{1}{\alpha}\,.
\end{equation}
Using  \eqref{eq: tmp-upper-th-asymp} in \eqref{eq: tmp-power-2} brings us
\begin{equation} \label{eq: tmp-power-3}
\lim\limits_{\myg\rightarrow \infty}^{} \prob\left(U_{\myg,\ell}\geq \th^{\mathsf{asym}}_{L,\alpha} \right)\geq \prob \left(\widetilde{Q}\ge L-1+2\sqrt{(L-1)\log\frac{1}{\alpha}} + 2\log\frac{1}{\alpha}  \right)\,.
\end{equation}
We next claim that 

\begin{equation}\label{eq: tmp-claim-power}
 2\sqrt{(L-1)\log\frac{1}{\alpha}} + 2\log\frac{1}{\alpha}  \leq A^2L^{1/2}-2\sqrt{(L-1+2A^2L^{1/2})\log\frac{1}{\beta}}\,.
\end{equation}
Deploying \eqref{eq: tmp-claim-power} (we provide the proof of claim \eqref{eq: tmp-claim-power} later) in \eqref{eq: tmp-power-3} yields 
\begin{equation} \label{eq: tmp-power-4}
\lim\limits_{\myg\rightarrow \infty}^{} \prob\left(U_{\myg,\ell}\geq \th^{\mathsf{asym}}_{L,\alpha} \right)\geq \prob_{} \left(\widetilde{Q}\ge L-1+ A^2L^{1/2}-2\sqrt{(L-1+2A^2L^{1/2})\log\frac{1}{\beta}} \right)\,.
\end{equation}
Next by using the first tail bound of Lemma \ref{lemma: chi-tails} (for values $m=L-1, \widetilde{\lambda}=A^2L^{1/2}$, and $t=\log\frac{1}{\beta}$) in \eqref{eq: tmp-power-4} we obtain
\[
\lim\limits_{\myg\rightarrow \infty}^{} \prob\left(U_{\myg,\ell}\geq \th^{\mathsf{asym}}_{L,\alpha} \right)\geq 1-\beta\,.
\]
This completes the proof. Finally, we are left to prove the claim \eqref{eq: tmp-claim-power}. As $L\geq 1$, we have
\begin{align*}
&\widetilde{\th}:=A^2L^{1/2}-2\sqrt{(L-1+2A^2L^{1/2})\log\frac{1}{\beta}}\geq \sqrt{L}\left(A^2-  2\sqrt{(1+2A^2)\log\frac{1}{\beta}}  \right)\,.
\end{align*}
In the next step, by using $A\geq 1$, we get
\begin{align}
\widetilde{\th}&\geq \sqrt{L}\left(A^2-  2A\sqrt{3\log\frac{1}{\beta}}  \right)\nonumber\\
&\geq \sqrt{L}\left(A - \sqrt{3\log\frac{1}{\beta}}  \right)^2-3\sqrt{L}\log\frac{1}{\beta}\nonumber\\
&\geq\sqrt{L}\left( 2\sqrt{\log\frac{1}{\alpha}}+2\log\frac{1}{\alpha}\right)\label{eq: tmp-power-claim-2} \,,
\end{align}
where the last inequality follows from the definition of $A$ in \eqref{eq: tmp-A}. We then use $L\geq 1$ in \eqref{eq: tmp-power-claim-2} to arrive at 
\[
\widetilde{\th}\geq 2\sqrt{(L-1)\log\frac{1}{\alpha}}+2\log\frac{1}{\alpha}\label{eq: tmp-power-claim} \,.
\]
This proves \eqref{eq: tmp-claim-power}.

}

%%%%%

{

\subsection{Proof of Theorem \ref{coro: failure-crt}}\label{proof:coro: failure-crt}
We start by establishing a concentration bound on the normalized rank given by~\eqref{eq: p-crt}.
\begin{propo}\label{thm: crt-failure}
Consider an even function $g$, and a dataset $(\bX,\bY)$ of $n$ i.i.d. pairs $\{(X_i,Y_i)\}_{i=1}^n$, with $X_i, Y_i\in \reals$, generated from the following regression model:
\begin{align} \label{eq: example:thm}
X&\sim \normal(0,1)\,, \nonumber \\
Y&=g(X)+\eps\,, \quad \eps\sim\normal(0,1)\,.
\end{align}
 For the marginal covariance score function $T(\bX,\bY)= n^{-1}\bX^\sT\bY$, and counterfeit datasets $\tbX_{j}\sim\normal(0,I_{n})$, recall the CRT $p$ statistic:
\begin{align}\label{eq:pMn}
p^{(M)}_n=\frac{1+\sum\limits_{j=1}^{M}\ind{ \{T(\bX,\bY)\geq T(\tbX_{j},\bY)\} } }{M+1}\,.
\end{align}
Then, the statistic $p^{(M)}_n$ concentrates around $1/2$. In particular, for any $\delta>0$ and $M>1/\delta$, we have
\[
\lim\limits_{n\rightarrow \infty}^{}\prob\left(\left|  p^{(M)}_n -1/2 \right| \geq \delta \right) \leq  \frac{1}{\left( \delta-1/M\right)^2} \left(   \frac{1}{4M} + \frac{M-1}{M} \Big(\E_{Z\sim\normal(0,1)}[\Phi^2(\eta Z)] -\frac{1}{4}\Big)  \right)\,,
\]
with $\eta=\left(\frac{1+\E[X^2g(X)^2]}{1+\E[g(X)^2]}\right)^{1/2}$.
 \end{propo}
 
The proof of this proposition is given in Section~\ref{proof:thm: crt-failure}.
We next show that the deviation of the $p$-statistic from $1/2$ can be controlled by the choice of $\eta$. 
%By recalling our initial observation, it implies that by having $c$ close enough to $0$, we can get highly concentrated values of the  CRT statistic around $1/2$. To accurately capture this behavior, we want to use this upper bound on $\eta$ in Theorem \ref{thm: crt-failure} and bound the deviation of the $p$ statistic. 
Note that for the normal distribution function $\Phi$ and the normal density $\phi$ we have %has bounded derivative $1/\sqrt{2\pi}$, for every $z$ this yields 
 \[
0\leq  \Phi(\eta z)= \Phi(0)+ \int_{0}^{\eta z} \phi(t)\de t \le \frac{1}{2}+ \frac{\eta |z|}{\sqrt{2\pi}} \,.
 \]
Consequently for $Z\sim\normal(0,1)$,
\[
\E\left[\Phi^2(\eta Z)\right] \leq \E\Big[ \Big(\frac{1}{2}+\frac{\eta |Z|}{\sqrt{2\pi}}\Big)^2\Big] = \frac{1}{4}+\frac{\eta^2}{2\pi} + \frac{\eta}{\pi}\,.
\]
%by simplifying this expression and using the half-normal distribution mean value, we finally have
Therefore,
\[
\E\left[\Phi^2(\eta Z)\right] - \frac{1}{4} \leq \frac{\eta^2+2\eta}{2\pi}\,,
\]
which along with the result of Proposition~\ref{thm: crt-failure} implies
\[
\lim\limits_{M\rightarrow \infty}^{} \lim\limits_{n\rightarrow \infty}^{}\prob\Big(  \Big|p_{n}^{(M)}  -\frac{1}{2} \Big|   \geq \delta\Big) \leq \frac{\eta^2+2\eta}{2\pi\delta^2}\,.
\]
The proof of part (a) for two-sided CRT, follows by setting $\delta=(1-\alpha)/2$ and using that $\delta >1/4$.

%Next be choosing $\delta=(1-\alpha)/2$ and using~\eqref{eq:eta-upper}, we obtain that for $\theta<\alpha^2/500$ and $\alpha\le 1/2$,
%%
%\[
%\lim\limits_{M\rightarrow \infty}^{} \lim\limits_{n\rightarrow \infty}^{}\prob\Big(  \Big|p_{n}^{(M)}  -\frac{1}{2} \Big|   \geq  \frac{1-\alpha}{2} \Big)\leq  \frac{\alpha}{2}\,,
%\]
%This completes the proof of part (a) for two-sided CRT. 

Proof of part (b) follows along the same lines. The only modification is that time we set $\delta = 1/2-\alpha\ge \gamma$, which brings us to 

\begin{align*}
\lim\limits_{M\rightarrow \infty}^{} \lim\limits_{n\rightarrow \infty}^{}\prob\left(  p_{n}^{(M)}   \geq  1-\alpha \right) \leq \frac{\alpha}{2}\,,\quad
\lim\limits_{M\rightarrow \infty}^{} \lim\limits_{n\rightarrow \infty}^{}\prob\left(  p_{n}^{(M)}   \leq  \alpha \right) \leq \frac{\alpha}{2}\,.
\end{align*}
This completes the proof of part (b) for one-sided CRT. 

%%%%%%%%%%%%%%%%%%%%%%%%%%%%%%%%%%%%%%%%%%%%%%%%%%%%%%
\subsection{Proof of Proposition \ref{thm: crt-failure}} \label{proof:thm: crt-failure}
\def\tbX{\widetilde{\bX}}

 Consider $M$ counterfeits $\tbX_1,\tbX_2,...,\tbX_M$ sampled independently from $\normal(0,\bI_n)$. For $ j\in [M]$, let $I_j= \ind{\{T(\bX,\bY)  \geq T(\tbX_j,\bY)\}} $. Let $T(\bX,\bY)=n^{-1}\bX^\sT\bY$ and
 \[
 \mu_n=\E\left[ \ind{ \{T(\bX,\bY) \geq T(\tbX_1,\bY) \}}  \right]\,,\quad \sigma_n^2=\var\left[  \ind{\{T(\bX,\bY) \geq T(\tbX_1,\bY)\} } \right]\,.
 \]

It is easy to see that $\sigma_n^2=\mu_n(1-\mu_n)$. Before proceeding further we establish a lemma which will be used in proving the result.
\begin{lemma}\label{lemma: crt-fails0}
The followings hold:
\begin{align*}
&\lim_{n\to\infty} \mu_n =  \lim\limits_{n\rightarrow \infty}^{}\prob(T(\tbX,\bY) \leq T(\bX,\bY)) =1/2 \,,\\
&\lim\limits_{n\rightarrow \infty}^{}\E \left[\prob\left( \{T(\bX,\bY)\geq T(\tbX,\bY)\}|\bX,\bY \right)^2 \right] = \E_{Z\sim\normal(0,1)}[ \Phi^2(\eta Z) ] \,,
\end{align*}
where $\eta= \frac{3+2\E[X^2g(X)^2]}{3+2\E[g(X)^2]}$.
\end{lemma}
By applying Chebyshev's inequality we get
\begin{align}
\prob\Big( \Big| \sum\limits_{j=1}^{M}\frac{I_j}{M} -1/2  \Big| \geq \delta \Big) &\leq  \prob\Big( \Big| \sum\limits_{j=1}^{M}\frac{I_j}{M} -\mu_n  \Big| \geq \delta-|\mu_n-1/2| \Big) \nonumber\\
%\prob\left( \left| \sum\limits_{j=1}^{M}\frac{I_j}{M} -\mu_n  \right| \geq \eps \right)
 &\leq \frac{1}{(\delta-|\mu_n-1/2|)^2}\cdot\E\Big[  \Big| \sum\limits_{j=1}^{M} \frac{I_j}{M}-\mu_n   \Big|^2  \Big]\,,\nonumber\\
&=\frac{1}{(\delta-|\mu_n-1/2|)^2}\cdot\E\Big[ \frac{1} {M^2} \sum\limits_{j=1}^{M} (I_j-\mu_n)^2 + \frac{1}{M^2}\sum_{i\neq j}(I_i-\mu_n)(I_j-\mu_n)) \Big]\nonumber\\
&= \frac{1}{(\delta-|\mu_n-1/2|)^2}\cdot\left( \frac{\sigma_n^2}{M}+\frac{M-1}{M}\E [ (I_1-\mu_n)(I_2-\mu_n) ] \right)\nonumber \\
&=\frac{1}{(\delta-|\mu_n-1/2|)^2}\cdot\left( \frac{1}{M}\mu_n(1-\mu_n)+\frac{M-1}{M}\E [ I_1I_2-\mu_n^2 ] \right)\,. \label{eq: tmp2} 
\end{align}

We next compute $\E[I_1I_2-\mu_n^2]$.  
\begin{align*}
\E[ I_1I_2]&=\prob\left( \{T(\bX,\bY)\geq T(\tbX_1,\bY)\} \cap \{T(\bX,\bY)\geq T(\tbX_2,\bY) \}  \right)\nonumber \\
&=\E\left[\prob\left( \{T(\bX,\bY)\geq T(\tbX_1,\bY)\} \cap \{T(\bX,\bY)\geq T(\tbX_2,\bY) \}  | \bX,\bY \right)\right]\nonumber \\
&=\E \left[\prob\left( \{T(\bX,\bY)\geq T(\tbX_1,\bY)\}|\bX,\bY \right) \prob\left(\{T(\bX,\bY)\geq T(\tbX_2,\bY) \}  | \bX,\bY \right)\right] \nonumber \\
&=\E \left[\prob\left( \{T(\bX,\bY)\geq T(\tbX,\bY)\}|\bX,\bY \right)^2 \right]\,,
\end{align*}
where we used the fact that conditioned on $\bX$ and $\bY$, score values $T(\tbX_1,\bY), T(\tbX_2,\bY)$ are independent. Therefore, by using Lemma \ref{lemma: crt-fails0}, we write
\begin{align}
\lim\limits_{n\rightarrow \infty}^{}\E[ (I_1 I_2-\mu_n^2)]&= \lim\limits_{n\rightarrow \infty}^{}\E \left[\prob\left( \{T(\bX,\bY)\geq T(\tbX,\bY)\}|\bX,\bY \right)^2 \right]-\mu_n^2 \nonumber\\
&= \lim\limits_{n\rightarrow \infty}^{}\E \left[\prob\left( \{T(\bX,\bY)\geq T(\tbX,\bY)\}|\bX,\bY \right)^2 \right]-1/4\nonumber \\
&=\E_{Z\sim\normal(0,1)}[ \Phi^2(\eta Z) ]-1/4  \,.\label{eq: tmp3}
\end{align}
%where (a) is followed by \eqref{eq: lim-mu-n}, and the last equation comes from Lemma \ref{lemma: crt-fails0}. 

To summarize, we let $S_M=\sum\limits_{j=1}^{M}  \ind{\{ T(\bX,\bY)\geq T(\tbX_j,\bY) \} }$ and use \eqref{eq: tmp3} in \eqref{eq: tmp2} along with $\lim_{n\to\infty}\mu_n = 1/2$ per Lemma \ref{lemma: crt-fails0} to obtain
\begin{equation}\label{eq: tmp4}
\lim\limits_{n\rightarrow \infty}^{}\prob\left(\left|  \frac{S_M}{M}  -\frac{1}{2} \right| \geq \delta \right) \leq  \frac{1}{4M\delta^2} + \frac{M-1}{M\delta^2} \cdot(\E_{Z\sim\normal(0,1)}[\Phi^2(\eta Z)] -1/4)\,,\quad \forall \delta>0 \,.
\end{equation}
Recalling the $p$ statistic \eqref{eq:pMn}, we have $p^{(M)}_n=\frac {1+S_M}{M+1}$. As $S_M\leq M$, we have
\[
\frac{S_M}{M} \leq \frac{S_M+1}{M+1} \leq \frac{S_M}{M}+\frac{1}{M}\,,
\]
which implies that $\left|p^{(M)}_n-\frac{1}{2}\right| \leq \left| \frac{S_M}{M}-\frac{1}{2} \right|+\frac{1}{M}$. Using this relation along with the triangle inequality in \eqref{eq: tmp4} to arrive at the following:
\begin{align*}
\prob\Big(\Big|p^{(M)}_n-\frac{1}{2}\Big|\geq \delta\Big) &\leq \prob\Big(\Big|\frac{S_M}{M}-\frac{1}{2}\Big|\geq \delta-\frac{1}{M}\Big)\\
&\leq \frac{1}{\left( \eps-1/M\right)^2} \left(   \frac{1}{4M} + \frac{M-1}{M} \cdot(\E_{Z\sim\normal(0,1)}[\Phi^2(\eta Z)] -1/4)  \right)\,.
\end{align*}
This completes the proof of Proposition \ref{thm: crt-failure}.

\subsubsection{Proof of Lemma \ref{lemma: crt-fails0}}

We start by establishing a lemma which characterizes the conditional probability that the original data score exceeds a counterfeit score.

%Define the shorthands
%\begin{align}\label{eq:mv}
%m_n(\bX,\bY):=\frac{\|\bY\|_2^2}{n}+1\,,\quad
%v_n(\bX,\bY):=\frac{4 \|\bY\|_2^2}{n^2}+\frac{2}{n}\,.
%\end{align}
%
\begin{lemma}\label{lemma:indicators-cond-exp} 
 The following holds
 \begin{equation}\label{eq: berry-2}
\left|\prob({ T(\tbX,\bY) \leq T(\bX,\bY) } | \bX,\bY ) -  \Phi\Big(  \frac{nT(\bX,\bY)}{\|\bY\|_2} \Big) \right|  \leq C_1\frac{\|\bY\|_3^3}{\|\bY\|_2^3} \,.\end{equation}
with 
%\[
 %\gamma_n(\bX,\bY) :=  \frac{1}{n^2 v_n(\bX,\bY)^{1.5}   } \left( C_0+\sum\limits_{i=1}^{n}\frac{C_1|Y_i|+C_2|Y_i|^2+C_3|Y_i|^3}{n} \right)\,,
%\]
where $C_1$ is an absolute constant.
\end{lemma}
We next show that $\E\big[\tfrac{\|\bY\|_3^3}{\|\bY\|_2^3}\big] \to 0$ as $n\to\infty$.
\begin{align*}
\E\left[\frac{\|\bY\|_3^3}{\|\bY\|_2^3}\right]&=\frac{1}{\sqrt{n}}\E\left[\frac{n^{-1}\sum_{i=1}^{n}|Y_i|^3}{\big(n^{-1}\sum_{i=1}^{n}|Y_i|^2 \big)^{3/2}} \right]\,.
\end{align*}
By recalling the strong law of large numbers, quantities $n^{-1}\sum_{i=1}^{n}|Y_i|^3$ and $n^{-1}\sum_{i=1}^{n}|Y_i|^2$ will almost surely converge to $\E[|g(x)+\eps|^3]$, and $\E[|g(x)+\eps|^2]$, respectively. This implies the almost sure convergence of $\tfrac{\|\bY\|_3^3}{\|\bY\|_2^3}$ to $0$ as $n$ grows to infinity. In the next step, by using $||\bY||_3/||\bY||_2\leq 1$ along with the dominant convergence theorem, we arrive at
%\begin{align*}
%\E\left[\gamma_n(\bX,\bY)\right] &=  \E\left[\frac{1}{\sqrt{n}\left(2+4\|\bY\|_2^2/n\right)^{1.5}   } \left( C_0+\sum\limits_{i=1}^{n}\frac{C_1|Y_i|+C_2|Y_i|^2+C_3|Y_i|^3}{n} \right) \right] \,,\\
%& \leq \frac{1}{2^{3/2}\sqrt{n}} \E\left[ \left( C_0+\sum\limits_{i=1}^{n}\frac{C_1|Y_i|+C_2|Y_i|^2+C_3|Y_i|^3}{n} \right) \right] \\
%&= \frac{1}{2^{3/2}\sqrt{n}} \left( C_0+ C_1\E[|Xg(X)+\eps|] +C_2\E[|Xg(X)+\eps|^2]+C_3\E[|Xg(X)+\eps|^3] \right) \,,
%\end{align*}
\begin{align}\label{eq:Ezero}
\lim\limits_{n\to \infty}^{} \E\left[\frac{\|\bY\|_3^3}{\|\bY\|_2^3}\right] =0.
\end{align}
In the next lemma, we characterize the distribution of the other quantity in~\eqref{eq: berry-2}.

\begin{lemma}\label{lemma:crt-failure-1}
 We have
\[
 \frac{nT(\bX,\bY)}{\|\bY\|_2 } \overset{d}{\to} \normal(0,\eta^2), \quad \text{as } n \rightarrow \infty\,.
\]
with  $\eta=\left(\frac{1+\E[X^2g(X)^2]}{1+\E[g(X)^2]}\right)^{1/2} $.
\end{lemma}

Using the result of Lemma \ref{lemma:crt-failure-1} and by an application of the Portmanteau theorem for the bounded continuous function $\Phi$ we get 
\begin{equation}\label{eq: tmp0}
\lim\limits_{n\rightarrow \infty}^{}  \E \left[ \Phi\left(   \frac{nT(\bX,\bY)}{ \|\bY\|_2 }  \right) \right]=\E_{Z\sim \normal(0,1)}[\Phi(\eta Z)]\,.
\end{equation}
Combining~\eqref{eq:Ezero} and \eqref{eq: tmp0} with \eqref{eq: berry-2} we arrive at
\[
\lim_{n\to\infty}\prob(T(\tbX,\bY)\leq T(\bX,\bY))=\E_{Z\sim \normal(0,1)}\left[\Phi\left(\eta Z\right) \right] \,.
\]

%Rewrite the main expression as the following:
%\begin{align*}
%\prob(T(\tbX,\bY)\leq T(\bX,\bY))=\E\left[ \prob\left(T(\tbX,\bY)\leq T(\bX,\bY) \right)| \bX,\bY \right]\,.
%\end{align*}
%We want to use Lemma \ref{lemma: indicators-cond-exp} to establish the result of Lemma \ref{lemma: crt-fails0}. To this end, we need to prove that $E[\gamma_n(\bX,\bY)] \rightarrow 0$, and the expectation of the inner quantity in \eqref{eq: berry-2} converges to  $ \E_{Z\sim \normal(0,1)}\left[ \Phi(\eta Z)\right] $. We start by the inner quantity, where by using Lemma \ref{ lemma: crt-failure-1} in along with the Portmanteau theorem for the bounded continuous function $\Phi$ we get 
%
%
%then use this and  \eqref{eq: tmp0} in Lemma \ref{lemma: indicators-cond-exp} to get
%\[
%\prob(T(\tbX,\bY)\leq T(\bX,\bY))=\E_{Z\sim \normal(0,1)}\left[\Phi\left(\eta Z\right) \right] \,.
%\]
In the next lemma we show that $\E_{Z\sim \normal(0,1)}\left[\Phi\left(\eta Z\right) \right] = 1/2$, which completes the proof of the first part.
\begin{lemma}\label{lem:phi-etaZ}
Let $\Phi(\cdot)$ denote the distribution of standard normal variable. Then, for any constant $\eta$ we have
\[
\E_{Z\sim \normal(0,1)}\left[\Phi\left(\eta Z\right) \right] = 1/2.
\]
\end{lemma}

The second part of the lemma follows by a similar argument. From Lemma \ref{lemma:indicators-cond-exp} we have 
\begin{align*}
\left| \prob(T(\tbX,\bY)\leq T(\bX,\bY)|\bX,\bY)^2-\Phi^2\Big(\frac{nT(\bX,\bY)}{\|\bY\|_2} \Big)  \right| \leq 2C_1\frac{\|\bY\|_3^3}{\|\bY\|_2^3}\,.
\end{align*}
Using \eqref{eq:Ezero} yields
\begin{equation}\label{eq: Esquare}
\lim_{n\to \infty} \E\left[\prob(T(\tbX,\bY)\leq T(\bX,\bY)|\bX,\bY)^2\right]=\lim_{n\to \infty}\E\left[ \Phi^2\Big(\frac{nT(\bX,\bY)}{\|\bY\|_2} \Big)  \right]
\end{equation}

 Also, by using Lemma \ref{lemma:crt-failure-1} and an application of the Portmanteau theorem for the bounded continuous function $\Phi^2$ we obtain
\[
\lim\limits_{n\rightarrow \infty}^{}  \E \left[ \Phi^2\left(    \frac{nT(\bX,\bY)}{ \|\bY\|_2 }   \right) \right]=\E_{Z\sim \normal(0,1)}[\Phi^2(\eta Z)]\,,
\]

which invoking \eqref{eq: Esquare} completes the proof of Lemma \ref{lemma: crt-fails0} second part.

\subsubsection{Proof of Lemma \ref{lemma:indicators-cond-exp}}
We focus on the distribution of $T(\tbX,\bY)|\bX,\bY$ and treat $\bX,\bY$ as deterministic values, so the only source of randomness is $\tbX$.  
To lighten the notation, we introduce
%\[
%v_n: = v_n(\bX,\bY) = \frac{2}{n} + \frac{4\|\bY\|_2^2}{n^2}\,,
%\]
\[
\xi_i=\frac{\tX_i Y_i}{\|\bY\|_2}\,,\quad\text{for } i\in [n]\,.
\]

By simple algebraic computations, we get that $\E[\xi_i|\bX,\bY]=0$ and $\sum\limits_{i=1}^{n} \E[\xi^2_i|\bX,\bY]=1$. Also, conditioned on $\bX,\bY$, random variables $\xi_i$ are independent. We next use the Berry-Essen theorem to characterize the distribution of $\sum_{i=1}^n \xi_i$. For the reader's convenience, the version of the Berry-Esseen theorem for non-identical random variables is provided in Lemma \ref{lemma: berry-essen}.  First, we need to bound the sum of third moments:
\begin{align*}
\sum\limits_{i=1}^{n}\E[|\xi_i^3|]&=\sum_{i=1}^n\frac{\E[|\tX_i|^3]|Y_i|^3}{\|\bY\|^3_2}\\
&=C_1\frac{\|\bY\|_3^3}{\|\bY\|_2^3}\,,
%&\leq \E_{z\sim \normal(0,1)}[|z|^3] \frac{1}{\|\bY\|_2}
\end{align*}
where the coefficients $C_1$ is a universal constant that can be precisely computed by using the third moment of the half-normal distribution. Note that here the expectation is with respect to $\tX_i$.
%\begin{align*}
%m_n(\bX,\bY)&:=\E[ T(\tbX,\bY)|\bX,\bY]\,,\\
%v_n(\bX,\bY)&:=\var[T(\tbX,\bY)|\bX,\bY]\,, \\
%\rho_n(\bX,\bY)&:=\E\left[\left(\left|T(\tbX,\bY)-m_n(\bX,\bY) \right|\right)^3\big|\bX,\bY\right]\,.
%\end{align*}

%By simple algebraic calculations we can get

%\[
% m_n(T(\bX,\bY)|\bX,\bY)=\frac{||\bY||_2^2}{n}+1\,,\quad   v_n(T(\bX,\bY)|\bX,\bY)=\frac{4 ||\bY||_2^2}{n^2}+\frac{2}{n}\,.
%\] 
Now, we employ the Berry–Esseen theorem \ref{lemma: berry-essen} to get:
\begin{align}\label{eq: berry-1}
\sup\limits_{z}^{}\left|\prob(  \sum\limits_{i=1}^{n} \xi_i \leq z  | \bX,\bY )- \Phi\left( z  \right) \right |  \leq C_1\frac{\|\bY\|_3^3}{\|\bY\|_2^3}\,.
\end{align}
From the definition of $\xi_i$ and recalling the definition of score $T(\tbX,\bY)$ we have
\begin{align*}
 T(\tbX,\bY)& = \frac{1}{n}\tbX^\sT\bY =\frac{1}{n}\|\bY\|_2\sum_{i=1}^n \xi_i\,.
\end{align*}

Using the above relation and choosing $z = \tfrac{nT(\bX,\bY)}{\|\bY\|_2}$ in \eqref{eq: berry-1} (note that $z$ is a measurable function of $\bX,\bY$), we get
\begin{align*}
\left|\prob({ T(\tbX,\bY) \leq T(\bX,\bY) } | \bX,\bY ) -  \Phi\Big(  \frac{nT(\bX,\bY)}{\|\bY\|_2} \Big) \right|  \leq C_1\frac{\|\bY\|_3^3}{\|\bY\|_2^3} \,.
\end{align*}

%Use the conditional expectation tower property to arrive at
%\[
%\left| \mu_n- \E\left[\Phi\left( \frac{T(\bX,\bY)- m_n(\bX,\bY)}{ v_n(\bX,\bY)^{0.5} } \right)\right] \right| \leq   \E\left[\frac{C\rho_n(\bX,\bY)}{v_n(\bX,\bY)^{1.5} \sqrt{n}}\right]
%\]

\subsubsection{Proof of Lemma \ref{lemma:crt-failure-1} }
Substituting for $T(\bX,\bY)$ we get
\begin{align}
 \frac{nT(\bX,\bY)}{ \|\bY\|_2 }&=  \frac{ \frac{1}{\sqrt{n}}\sum_{i=1}^{n}X_iY_i  }{\Big(\frac{1}{n}\sum\limits_{i=1}^{n}Y_i^2 \Big)^{1/2}} \label{eq: mu_n-inside-term}
 %&=\frac{\sqrt{n}\left(  \|\bX\|_2^2/n-2\bY^T\bX/n-1  \right)} { \left( 2+4\|\bY\|_2^2/n  \right)^{1/2} }\,, \label{eq: mu_n-inside-term}
\end{align}

By an application of the central limit theorem, the numerator converges in distribution to a normal random variable. More precisely,
 \begin{align*}
 \E[XY]&=\E[X(g(X)+\eps)]\\
 &=\E[Xg(X)]+\E[X\eps]=0\,,
 \end{align*}
where in the last relation we used the property that  $g$ is an even function and $X\sim\normal(0,1)$. In addition, $\var[(XY)]=1+\E[X^2g(X)^2]$ by simple calculation. Therefore, by CLT we have
\begin{equation} \label{eq:numerator-slutsky}
n^{-1/2} \bX^\sT\bY \overset{d}{\to}\normal(0,1+\E[X^2g(X)^2])
\end{equation}

On the other hand, from the weak law of large numbers we have that the denominator in \eqref{eq: mu_n-inside-term} converges in probability to $1+\E[g(X)^2]$. The proof is completed by using the Slutsky's theorem. %in conjunction with equations \eqref{eq:numerator-slutsky} and \eqref{eq:denominator-slutsky}.

%=========================
\subsubsection{Proof of Lemma~\ref{lem:phi-etaZ}}

{
For $Z'\sim N(0,1)$ independent from $Z$, we have $\E[\Phi(\eta Z)]=\prob(Z'\leq \eta Z)$. This can be written as $\E[\Phi(\eta Z)]=\prob(Z'-\eta Z \le 0)$. We next note that  $Z'-\eta Z \sim \normal(0,1+\eta^2)$ which implies that $\E[\Phi(\eta Z)]=\frac{1}{2}$. 
}

%Let $\Phi(\cdot)$ and $\phi(\cdot)$ respectively denote the distribution and the density function of the standard normal random variable. We write
%\begin{align}
%\frac{\partial}{\partial \eta} \E[\Phi(\eta Z)] = \E[Z \phi(\eta Z)]
%\stackrel{(a)}{=} \E[\eta \phi'(\eta Z)] \stackrel{(b)}{=} \E[-\eta^2 Z \phi(\eta Z)] = -\eta^2  \frac{\partial}{\partial \eta} \E[\Phi(\eta Z)]\,, 
%\end{align}
%where we used Stein's lemma in step $(a)$, and the identity $\phi'(t) = -t\phi(t)$ in step (b). Therefore, $\frac{\partial}{\partial \eta} \E[\Phi(\eta Z)] = 0$ and so $ \E[\Phi(\eta Z)] =  \E[\Phi(0)] = 1/2$.

%%%%%%%%%%%%%%%%%%%

\subsection{Proof of Proposition \ref{propo: g-odc}}\label{proof: propo-odc}
Following Definition \ref{def: conditonal-odc} we have
\[
F_{T}(t;\bY)=\prob_{\bX\sim \normal(0,\bI_n)}(\bX^\sT \bY\le t|\bY)=\Phi\Big(\frac{t}{\|\bY\|} \Big)\,.
\]
This results in $F_{T}^{-1}(u;\bY)=\|\bY\| \Phi^{-1}(u)$.  Plugging this in the conditional ODC we obtain
\begin{align*}
R_T(u)&=\E_{\bY}[ F_{T|\bY} (F^{-1}_{T}(u;\bY);\bY)]\\
&=\E_{\bY}[ F_{T|\bY} (\|\bY\| \Phi^{-1}(u);\bY)]\\
&=\E_{\bY}[ \prob(\bX^\sT \bY \le \Phi^{-1}(u) \|\bY\| \big |\bY)]\\
&=\prob\left(\bX^\sT \bY \le \Phi^{-1}(u) \|\bY\| \right)\,,
\end{align*}
where in the last line $(\bX,\bY)$ follows the data generating rule \eqref{eq:g}.

\subsection{Proof of Theorem \ref{thm: one-Delta}}
{
 Let $F_n$ be CDF of random variable $\frac{\bX^\sT\bY} {\|\bY\|}$, i.e.,  $F_n(t)=\prob\big(\frac{\bX^\sT\bY} {\|\bY\|}\le t\big)$, using this in ODC function given in Proposition \ref{propo: g-odc} results in
\[
R_{\mathsf{MC}}^{(n)}=F_n(\Phi^{-1}(u))\,.
\]
To compute the limiting dependency power, it requires establishing convergence for differentiation of $F_n(.)$ (density functions). The convergence of density functions is broadly studied as local limit theorems. Here we take another approach, we try to connect our problem to established CLT results in the metrics of total variation distance; this naturally shows the $L_1$ convergence of densities.  Let $f_n(.)$ be the density function of $\frac{\bX^\sT\bY} {\|\bY\|}$. We have
\begin{align}
\Delta_T(\bX,\bY)&=\int_0^1 \Big| \frac{\de}{\de u} R_{\mathsf{MC}}^{(n)}-1\Big|\de u\nonumber\\
&=\int_0^1 \Big|\frac{\de}{\de u} F_n(\Phi^{-1}(u))-1\Big|\de u\nonumber\\
&=\int_0^1 \Big| \frac{f_n(\Phi^{-1}(u))}{\varphi({\Phi^{-1}(u)})}-1\Big|\de u\nonumber\,.
\end{align}
We next use the change of variable $x=\Phi^{-1}(u)$ in the above integration to arrive at
\begin{align}\label{eq: Delta-limit-thm}
\Delta_T(\bX,\bY)&=\int_{-\infty}^{\infty} \Big| f_n(x)-\varphi(x)\Big| \de x\nonumber\\
&=2\;d_{\tv}\bigg(\cL\Big(\frac{\bX^\sT \bY}{\|\bY\|}\Big), \normal(0,1)\bigg)\,.
\end{align}

In Lemma \ref{lemma:crt-failure-1} for $\eta=\left(\frac{1+\E[X^2g(X)^2]}{1+\E[g(X)^2]}\right)^{1/2} $ we characterize the limiting distribution:
\[
V_n:=\frac{\bX^\sT\bY} {\|\bY\|} \overset{(d)}{\to} \normal(0,\eta^2)\,. 
\]
However the convergence in distribution is shown, it does not generally result in convergence in total variation distance. For this end, let $\sigma_Y^2=\E[Y^2]$, then we consider the following two dimensional random vector
 \[
 \bU_n:=\frac{1}{\sqrt{n}}\begin{bmatrix} \bX^\sT \bY\\  \|\bY\|^2-n\sigma_Y^2\end{bmatrix}\,.
 \]
 
 By applying Prokhorov’s local limit results (see \citep{petrov1956local} and Theorem 2.1 in \citep{bobkov2025renyi}), we deduce that if, for some $n \ge 1$, the random vector $\bU_n$ possesses a nonzero absolutely continuous component with respect to Lebesgue measure on $\reals^2$, then CLT convergence in total variation distance holds. In our setting, this condition is immediate, since one can explicitly compute the density of \(\bU_1\) via straightforward algebraic calculations for $\eps_1,X_1$ following standard normal distributions. Consequently, we obtain:
\begin{equation}\label{eq: local-limit-thm}
\lim_{n\to \infty} d_{\tv}\left(\cL(\bU_n), \normal(0,\bSigma)\right)=0\,.
\end{equation}
With the covariance matrix $\bSigma=\begin{bmatrix}\Sigma_{11} &\Sigma_{12}\\ \Sigma_{21} & \Sigma_{22} \end{bmatrix}$ having entries $\Sigma_{11}=\E[X^2Y^2]=1+\E[X^2g(X)^2]$, $\Sigma_{22}=\E[(Y^2-\sigma_Y^2)^2]$, $\Sigma_{12}=\Sigma_{21}=\E[XY(Y^2-\sigma_Y^2)]=0$, where the last relation follows $g$ being an even function.  For the function $h_n(s,t):=\frac{s}{\sqrt{n^{-1/2}t+\sigma_Y^2}}$, it is easy to see that $V_n=h_n(\bU_n)$. By considering $\bU\sim \normal(0,\bSigma)$, we arrive at
\begin{align*}
d_{\tv}\big(\cL(V_n),N(0,\eta^2)\big)&=d_{\tv}\big(\cL(h_n(\bU_n)), \normal(0,\eta^2)\big)\\
&\le d_{\tv}\big(\cL(h_n(\bU_n)), \cL(h_n(\bU))\big) + d_{\tv}\big( \cL(h_n(\bU)),\normal(0,\eta^2) \big)\,,
\end{align*}
where the last line follows the triangle's inequality.  We then use data processing inequality for mapping $h_n$ to get
\begin{align}\label{eq: tv-triangle}
d_{\tv}\big(\cL(V_n),N(0,\eta^2)\big)&\le d_{\tv}\big(\cL(\bU_n),\cL(\bU) \big) + d_{\tv}\big(\cL(h_n(\bU)) , \normal(0,\eta^2)\big)\,.
\end{align}
From \eqref{eq: local-limit-thm} we know that the first component in the above relation goes to zero, as $n\to \infty$. For the second component, we need to compute density function of 
$h_n(\bU)$ and use dominated convergence theorem to establish $L_1$ convergence. Note that in this case, $\bU\sim \normal(0,\bSigma)$ and $\bSigma$ is diagonal with entries $\Sigma_{11},\Sigma_{22}$. This implies that two components of $\bU$ are independent Gaussian random variables and the following conditional law holds
\[
\cL(h_n(\bU)|\bU_2)=\normal\left(0,\frac{\Sigma_{11}}{\sigma_Y^2+\bU_2 n^{-1/2}}\right)\,.
\]
It is easy to establish the dominated convergence for these densities when $n\to \infty$, and and noting that from definitions $\eta^2=\Sigma_{11}/\sigma^2_Y$. This gives us
\[
\lim_{n\to \infty}d_{\tv}\big(\cL(h_n(\bU)),N(0,\eta^2)\big)=0\,.
\]
As the two components of \eqref{eq: tv-triangle} converge to zero as $n\to \infty$, we obtain 
\[
\lim\limits_{n\to \infty}^{}d_{\tv}\big(\cL(V_n),N(0,\eta^2)\big)=0\,.
\]  
Using this in \eqref{eq: Delta-limit-thm} yields
\begin{align}\label{eq:Delta-TV}
\lim\limits_{n\to \infty}^{} \Delta(\bX,\bY)=2\;d_{\tv}\big(\normal(0,\eta^2),\normal(0,1)\big)\,. 
\end{align}
To compute this total variation distance, we note that if $p(x)$ and $q(x)$ respectively denote the density functions of $\normal(0,1),\normal(0,\eta^2)$, i.e., we have
\[
p(x)=\frac{1}{\sqrt{2\pi}}e^{-\frac{x^{2}}{2}}\,,\quad  q(x)=\frac{1}{\sqrt{2\pi}\,\eta}\,e^{-\frac{x^{2}}{2\eta^{2}}}\,.
\]
To find the TV distance we need to first find the crossing point $p(x^*)=q(x^*)$, it is easy to get that it is given by $x^*=\Big(\frac{2\eta^{2}\log\eta}{\eta^{2}-1}\Big)^{1/2}$ for $\eta\neq1$. For $\eta>1$ we have $p(x)>q(x)$ on $(-x^*,x^*)$ and $p(x)<q(x)$ outside; the inequalities reverse when $\eta<1$, the other case is trivial for $\eta=1$ we have $p(x)=q(x)$. So for $\eta>1$ we have the following
\begin{align*}
d_{\tv}\big(\normal(0,\eta^2),\normal(0,1)\big)&=\frac{1}{2} \int_{-\infty}^{\infty} |p(x)-q(x)| \de x \\ 
&=\int_{-x^*}^{x^*} (p(x)-q(x)) \de x\\
&=2\Phi(x^*)-2\Phi(x^*/\eta)\,.
\end{align*}
This brings us that for $\eta>1$ we have the following
\[
\lim\limits_{n\to \infty}^{} \Delta(\bX,\bY)=4\Phi\Big(\eta \Big(\frac{2\log \eta}{\eta^2-1} \Big)^{1/2}  \Big)- 4\Phi\Big(\Big(\frac{2\log \eta}{\eta^2-1} \Big)^{1/2}  \Big)\,.
\] 
When $\eta<1$, we can simply by symmetry use replace $\eta$ by $1/\eta$ in the above, and arrive at
\[
\lim\limits_{n\to \infty}^{} \Delta(\bX,\bY)=4\Phi\Big(\Big(\frac{2\log \eta}{\eta^2-1} \Big)^{1/2}  \Big)-4\Phi\Big(\eta \Big(\frac{2\log \eta}{\eta^2-1} \Big)^{1/2}  \Big)\,.
\] 
Put all together, we get that
\[
\lim\limits_{n\to \infty}^{} \Delta(\bX,\bY)=4\;\left|\Phi\Big(\eta \Big(\frac{2\log \eta}{\eta^2-1} \Big)^{1/2}  \Big)- \Phi\Big(\Big(\frac{2\log \eta}{\eta^2-1} \Big)^{1/2}  \Big)\right|\,.\quad \text{ for } \eta\ne 1\,.
\] 

In addition, for the regression function $g_{\th}(x)=\frac{1}{\sqrt{x^2+\th^2}}$ we have that 
$\eta^2=\frac{1+\E[X^2g(X)^2]}{1+\E[g(X)^2]}$. For $\E[g(X)^2]$ we have 
\begin{align*}
\E[g(X)^2]=\frac{2}{\sqrt{2\pi}}\int_{0}^{\infty} \frac{e^{-\frac{x^2}{2}}}{x^2+\th^2}\de x&= 
\frac{2}{\sqrt{2\pi}}\int_{0}^{\infty}\int_{0}^\infty {e^{-\frac{x^2}{2}}}  e^{-s(x^2+\th^2)}\de s \de x\\
&=\int_{0}^{\infty} (2s+1)^{-1/2} e^{-s\th^2} \de s\,.
\end{align*}
By change of variable $u=(2s+1)^{1/2}$ we get
\[
\E[g(X)^2]=e^{\th^2/2}\int_{1}^{\infty}e^{-\th^2u^2/2}\de u=\frac{\sqrt{2\pi}}{\th}e^{\th^2/2}\prob_{Z\sim N(0,1)}(Z/\th \ge1)=\frac{\sqrt{2\pi}}{\th}e^{\th^2/2}(1-\Phi(\th))\,.
\]
In addition, we have $1-\th^2 g(x)^2=x^2g(x)^2$, using the above relation gives us
\[
\E[X^2g(X)^2]=1-\th\sqrt{2\pi}e^{\th^2/2}(1-\Phi(\th))\,.
\]
Put all together, we arrive at
\[
\eta(\th)^2=\frac{2-\th\sqrt{2\pi}e^{\th^2/2}(1-\Phi(\th))}{1+\frac{\sqrt{2\pi}}{\th}e^{\th^2/2}(1-\Phi(\th))\,.}\,.
\]
}

%%%%%%%%%%%%%%%%%%%%%%%%%%%%%%%%%%%%%%%%%%%%%%%%%%%%%%%%%%
\subsection{Proof of Theorem \ref{thm: nontrivial-PCR}}\label{sec:converse}
We have $\myg$ number of groups, each of size $n$. Suppose that each group $\cG\sim \cL(\bX,\bZ,\bY)$ admits label $\ell$ with probability $p_\ell$. By construction of the PCR test
the rank of a subgroup $\cG$ is given by
\[
R=1+\sum\limits_{j=1}^{M}\ind \{T(\cG)\geq T(\tilde{G}_j) \}\,. 
\]
In particular, the probability of admitting label $1$ is $p_1=\prob(R\leq K)$, which by using $KL= M+1$ can be written as
\[
p_1=\prob\left(\frac{1}{M+1} R \leq \frac{1}{L}\right)\,.
\]
Therefore, with $L=1/\alpha$ and recalling the condition \eqref{eq: nontrivial-CRT} we get $p_1\ge \alpha+\delta$, which implies that
\begin{equation}\label{eq: lower-sum}
\sum_{\ell=1}^{L}\left|p_\ell-\frac{1}{L}\right| \geq \delta\,.
\end{equation}

We now focus on proving the asymptotic result. 
{
This implies that there exists $\ell^*\in [L]$ such that $p_{\ell^*}\neq \frac{1}{L}$. Let $\gamma=|p_{\ell^*} - \frac{1}{L}|$, so $\gamma>0$.  In the next step, by using the strong law of large numbers we get
\begin{equation}\label{eq: tmp-power-full-nontrivial}
\Big(\frac{W_{\ell^*}}{\myg}-\frac{1}{L}\Big)^2 \overset{(\mathsf{a.s})}{\to}  \Big(p_\ell^*-\frac{1}{L}\Big)^2\,.
\end{equation}
On the other hand, we know that $\frac{U_{\myg,L}}{\myg} \ge \left(\frac{W_{\ell^*}}{\myg}-\frac{1}{L} \right)^2$, therefore by using \eqref{eq: tmp-power-full-nontrivial} we arrive at
\begin{equation}\label{eq: power-full-final-nontrivial}
\prob\left( \frac{U_{\myg,L}}{\myg} \ge \gamma^2 \right)=1\,.
\end{equation}
Finally, from \eqref{eq: power-full-final-nontrivial} it is straightforward to get $\lim\limits_{\myg\to \infty}\prob\left( U_{\myg,L} \ge \th^{\mathsf{asym}}_{L,\alpha}\right)=1$.  This completes the proof. 
}

We then proceed to prove the result for the finite-sample threshold. By recalling the third part of Lemma \ref{lemma: multinomial-main} we have that if
\[
\sum_{\ell=1}^L\left|p_\ell-\frac{1}{L} \right| \geq  \frac{32{L^{1/4}}}{\sqrt{\myg}}\left[\frac{1}{\sqrt{\alpha}}\vee\frac{1}{\beta}  \right]^{1/2}\,,
\]
then the type II error with finite-sample threshold is bounded by $\beta$. Finally, combining \eqref{eq: lower-sum} and \eqref{eq: lower-delta-finite} completes the proof.

}
%%%%%%%%%%%%%%%%%%%%%%%%%%%%%%%%%%%%%%%%%%%%%%%
\section{Proofs of Section \ref{sec:robust}}
\subsection{Proof of Theorem \ref{thm:robust-model X} }\label{proof:thm:robust-model X}
	Consider a group $\cG=(\bX,\bZ,\bY)$ and its $M=KL-1$  counterfeits $\cG_i=(\tbX^{(1:M)},\bZ,\bY)$ where $\tbX^{(j)}$ is sampled from $\hP_{\bX|\bZ}(\cdot|\bZ)$, for $j\in [M]$.  Assume $\hbX$ is also drawn from $\hP_{\bX|\bZ}(\cdot|\bZ)$,  independently of $\tbX^{(1:M)},\bX,$ and $\bY$.
	We fix the values of $\bZ,\bY$, and for $\ell\in [L]$ define

	\[
	A_\ell=\left\{(\bx,\tbx^{(1)},...,\tbx^{(M)}): (\ell-1)K\leq \sum\limits_{j=1}^{M}\ind{\{T((\bx,\bZ,\bY))\geq T((\tbx^{(j)},\bZ,\bY))\}} \leq \ell K-1  \right\}\,.
	\]
	
	We have
	\begin{align}
		&\left|\prob\left( \cG \text{ has label } \ell|\bZ,\bY  \right)- \frac{1}{L} \right|\nonumber\\
		&\overset{(a)}{=}\left|\prob\left( (\bX,\tbX^{(1)},...,\tbX^{(M)}) \in A_\ell|\bZ,\bY  \right)-\frac{1}{L} \right|\nonumber\\
		%&\overset{(a)}{=}\left|\prob\left( (X,Z,Y) \text{ is type } s|Z,Y  \right)-  \prob\left( (\tX,Z,Y) \text{ is type } s|Z,Y  \right) \right|\\
		&\overset{(b)}{=}\left|\prob\left( (\bX,\tbX^{(1)},...,\tbX^{(M)}) \in A_\ell|\bZ,\bY  \right)-\prob\left( (\hbX,\tbX^{(1)},...,\tbX^{(M)}) \in A_\ell|\bZ,\bY  \right) \right|\nonumber\\
		&\overset{(c)}{\leq} d_{\tv}\left(((\bX,\tbX^{(1)},...,\tbX^{(M)})|\bZ,\bY),  ((\hbX,\tbX^{(1)},...,\tbX^{(M)})|\bZ,\bY)  \right)\nonumber\\
		&\overset{(d)}{=}d_{\tv}\left((\bX|\bZ,\bY),(\hbX|\bZ,\bY)   \right)\nonumber\\
		&\overset{(e)}{=}d_{\tv}\left((\bX|\bZ),(\hbX|\bZ)   \right)=d_{\tv}\left( P_{\bX|\bZ}(\cdot|\bZ),\hP_{\bX|\bZ}(\cdot|\bZ)  \right)\label{eq:robust-prob-upper}\,,
	\end{align}
	where $(a)$ comes from the process of labeling the data points; in (b) we used the fact that conditioned on $\bZ,\bY$ random variables $\hbX,\tbX^{(1)},...,\tbX^{(M)}$ are i.i.d., so the quantity $\sum\limits_{j=1}^{M}\ind{\{T(({\hbX},\bZ,\bY))\geq T((\tbX^{(j)},\bZ,\bY))\}}$ takes values  $\{0,1,...,M\}$, uniformly at random; (c) is a direct result from the total variation definition; in $(d)$ we used the property that conditioned on $(\bZ,\bY)$, random variables $(\bX,\tbX,\tbX^{(1)},...,\tbX^{(M)})$ are independent; $(e)$ comes from the fact that the under the null hypothesis, $\bX\indep \bY|\bZ$ and also $\hbX\indep \bY|\bZ$ by construction of $\hbX$. 
	
	In the current scenario that counterfeits are drawn from the approximate law $\hP_{\bX|\bZ}(.|\bZ)$, define $q_\ell$ to be the probability that under the null hypothesis, a regular group $\cG=(\bX,\bZ,\bY)$ has label $\ell$. Then by marginalizing out $\bZ$, we can upper bound the deviation amount of $q_\ell$ from $1/L$.
	%$$q_s=\prob\left( (X,Z,Y) \text{ is type } s \right),\quad \text{ for } s=1,2,...,\ell\,. $$
	%By marginalizing over $Z$, \eqref{eq:robust-prob-upper} gives us 
	\begin{align*}
		\left|q_\ell-\frac{1}{L}\right|	&=\left|\prob\left( \cG \text{ has label } \ell  \right)- \frac{1}{L} \right|\nonumber\\
	&= \left|\int \prob\left( \cG \text{ has label } \ell|\bZ,\bY  \right)\de P_{\bZ\bY}- \frac{1}{L} \right|\nonumber\\
		&= \left|\int \left(\prob\big( \cG \text{ has label } s|\bZ,\bY  \big)-\frac{1}{L}\right)\de P_{\bZ\bY}\right|\nonumber\\
		&\leq  \int \left|\prob\big( \cG \text{ has label } s|\bZ,\bY  \big)-\frac{1}{L}\right|\de P_{\bZ\bY}\nonumber\\
		&\overset{(a)}{\leq} \int d_{\tv}\left( P_{\bX|\bZ}(\cdot|\bZ),\hP_{\bX|\bZ}(\cdot|\bZ)  \right) \de P_{\bZ\bY}\\
		&=  \E_{\bZ}\left[ d_{\tv}\left( P_{\bX|\bZ}(\cdot|\bZ),\hP_{\bX|\bZ}(\cdot|\bZ)  \right) \right]\leq \delta\,,
	\end{align*}
	where (a) comes from \eqref{eq:robust-prob-upper}.  In summary we get 
	\begin{equation}\label{eq:robust-prob-upper2}
		\left|q_\ell-\frac{1}{L}\right|\leq \delta,\quad \text{ for } \ell=1,2,...,L\,.
	\end{equation}
	Recall $W_\ell$ as the number of data points with label $\ell$. Clearly, $(W_1,...,W_L)=\mathsf{multi}\left(\myg;q_1,...,q_L\right)$.

	We next use a result on the size of truncated $\chi^2$ test from \citep[Theorem 3.2]{balakrishnan2019hypothesis}, which implies the first inequality in the chain of inequalities below:
	 %as a result, from the multinomial hypothesis testing (\citep{balakrishnan2019hypothesis}, Theorem 2) we get that
	\begin{align*}
		\alpha&\geq \prob\left(\sum\limits_{\ell=1}^{L} \frac{(W_\ell-\myg q_\ell)^2-W_\ell}{\max\{q_\ell,\frac{1}{L} \}} \geq \myg\sqrt{ \frac{2}{\alpha} \sum\limits_{\ell=1}^{L} \left( \frac{q_\ell}{\max\{q_\ell,1/L \}} \right)^2 }    \right)\\
		&\geq \prob\left(\sum\limits_{\ell=1}^{L} \frac{(W_\ell-\myg q_\ell)^2-W_\ell}{\max\{q_\ell,\frac{1}{L} \}} \geq \myg\sqrt{ \frac{2}{\alpha}L}     \right)\\
		&= \prob \left(\sum\limits_{\ell=1}^{L} \frac{(W_\ell-\myg q_\ell)^2}{\max\{q_\ell,\frac{1}{L} \}} \geq \sum\limits_{\ell=1}^{L}\frac{W_\ell}{\max\{q_\ell,\frac{1}{L} \}} + \myg\sqrt{ \frac{2}{\alpha}L}     \right)\\\
		&\geq \prob \left( \sum\limits_{\ell=1}^{L} \frac{(W_\ell-\myg q_\ell)^2}{\max\{q_\ell,\frac{1}{L} \}} \geq L\sum\limits_{\ell=1}^{L}{W_\ell} +\myg\sqrt{ \frac{2}{\alpha}L}     \right)\\
		&=\prob \left(\sum\limits_{\ell=1}^{L} \frac{(W_\ell-nq_\ell)^2}{\max\{q_\ell,\frac{1}{L} \}} \geq \myg L+\myg\sqrt{ \frac{2}{\alpha}L}     \right)\\
		&\overset{(a)}{\geq} \prob \left( \sum\limits_{\ell=1}^{L} \frac{(W_\ell-\myg q_\ell)^2}{\frac{1}{L}+\delta } \geq \myg L+\myg\sqrt{ \frac{2}{\alpha}L}     \right)\\
		&\geq  \prob\left( \frac{L}{\myg(1+L\delta)}\sum\limits_{\ell=1}^{L}{\left(W_\ell-\myg q_\ell\right)^2}  \geq  L+\sqrt{ \frac{2}{\alpha}L}   \right)\\
		&\geq  \prob\left( U_{\myg,L}(\delta)  \geq  L+\sqrt{ \frac{2}{\alpha}L}   \right)\,,
	\end{align*}
	where $(a)$ comes from \eqref{eq:robust-prob-upper2} and the last inequality follows from the definition of $U_{\myg,L}$. This concludes the proof of claim \eqref{eq:robust-fin}.
	
	 For the claim~\eqref{eq:robust-asym}, we use the following  asymptotic result on the Pearson's $\chi^2$ test statistic for multinomial models (see e.g, \citep[Theorem 14.3.1]{lehmann2006testing}):
	\begin{equation}\label{eq: asympt-size-multi-pearson}
	\lim\limits_{\myg\rightarrow \infty}^{}\prob\left ( \sum\limits_{\ell=1}^{L}\frac{(W_\ell-\myg q_\ell)^2}{\myg q_\ell} \geq \th^{\mathsf{asym}}_{L,\alpha}  \right) \leq \alpha\,, 
	\end{equation}
where $\th^{\mathsf{asym}}_{L,\alpha} $ is the $\alpha$-th upper quantile of a Chi-squared distribution with $L-1$ degrees of freedom. By definition of $U_{\myg,L}(\delta)$, we have
\begin{align*}
\prob\left(U_{\myg,\ell}(\delta) \geq \th^{\mathsf{asym}}_{L,\alpha}    \right)&\leq 
\prob\left(\frac{L}{\myg(1+L\delta)}\sum\limits_{\ell=1}^{L} (W_\ell-\myg q_\ell)^2 \geq \th^{\mathsf{asym}}_{L,\alpha}    \right)\\
&\leq \prob\left(\sum\limits_{\ell=1}^{L} \frac{(W_\ell-\myg q_\ell)^2}{\myg q_\ell} \geq \th^{\mathsf{asym}}_{L,\alpha}    \right)\,,
\end{align*}
where in the last inequality we used \eqref{eq:robust-prob-upper2}. Finally, plug the above relation into \eqref{eq: asympt-size-multi-pearson} to get the following relation:
\[
\lim\limits_{\myg\rightarrow \infty}^{}\prob\left( U_{\myg,L}(\delta) \geq  \th^{\mathsf{asym}}_{L,\alpha}   \right) \leq \alpha\,.
\]
This concludes the proof.
%%%%%%%%%%%%%%%%%%%%%%%%%%%%%%%%%%%%%%%%%%%%%%
{
\subsection{Proof of Theorem \ref{thm: robust-inflation}}	\label{proof: robust-inflation}
	Consider a group $\cG=(\bX,\bZ,\bY)$ and its $M=KL-1$  counterfeits $\cG_i=(\tbX^{(i)},\bZ,\bY)$ where $\tbX^{(i)}$ is sampled from $\hP_{\bX|\bZ}(\cdot|\bZ)$, for $i\in [M]$.  Assume $\hbX$ is also drawn from $\hP_{\bX|\bZ}(\cdot|\bZ)$,  independently of $\tbX^{(1:M)},\bX,$ and $\bY$. From Algorithm \ref{algorithm: model-xz} we know that the test statistics $U_{\myg,L}$ is a function of $\bX,\bY,\bZ$ and $M$ sampled counterfeits $ \tbX^{(1:M)}$. For $t \geq 0$ and for fixed values of $\bZ,\bY$, we let
	
	\[
	A_t=\left\{(\bx,\tbx^{(1)},...,\tbx^{(M)})\in \reals^{n \times (M+1)}: U_{\myg,L}(\bZ,\bY,\bx,\tbx^{(1)},...,\tbx^{(M)} ) \geq t  \right\}\,.
	\]
	We have
	\begin{align}
		&\left|\prob\left(U_{\myg,L}\big( \bZ,\bY,\bX,\tbX^{(1:M)}\big)   \geq t |\bZ,\bY  \right)- \prob\left(U_{\myg,L}\big( \bZ,\bY,\hbX,\tbX^{(1:M)}\big)   \geq t |\bZ,\bY  \right)\right|\nonumber\\
		&\overset{(a)}{=}\left|\prob\left( (\bX,\tbX^{(1:M)}) \in A_t|\bZ,\bY  \right)- \prob\left((\hbX,\tbX^{(1:M)}) \in A_t|\bZ,\bY  \right) \right|\nonumber\\
		%&\overset{(a)}{=}\left|\prob\left( (X,Z,Y) \text{ is type } s|Z,Y  \right)-  \prob\left( (\tX,Z,Y) \text{ is type } s|Z,Y  \right) \right|\\
		%&\overset{(b)}{=}\left|\prob\left( (\bX,\tbX^{(1)},...,\tbX^{(M)}) \in A_\ell|\bZ,\bY  \right)-\prob\left( (\hbX,\tbX^{(1)},...,\tbX^{(M)}) \in A_t|\bZ,\bY  \right) \right|\nonumber\\
		&\overset{(b)}{\leq} d_{\tv}\left(\cL(\bX,\tbX^{(1:M)}|\bZ,\bY),  \cL(\hbX,\tbX^{(1:M)}|\bZ,\bY)  \right)\nonumber\\
		&\overset{(c)}{=}d_{\tv}\left(\cL(\bX|\bZ,\bY),\cL(\hbX|\bZ,\bY)   \right)\nonumber\\
		&\overset{(d)}{=}d_{\tv}\left((\bX|\bZ),(\hbX|\bZ)   \right)=d_{\tv}\left( P_{X|Z}^n(\cdot|\bZ),\hP_{X|Z}^n(\cdot|\bZ)  \right)\label{eq:robust-prob-upper}\,,
	\end{align}
	where $(a)$ comes from the definition of the set $A_t$; (b) is a direct result from the definition of total variation; in $(c)$ we used the property that conditioned on $(\bZ,\bY)$, random variables $(\bX,\tbX,\tbX^{(1)},...,\tbX^{(M)})$ are independent; $(d)$ comes from the fact that the under the null hypothesis, $\bX\indep \bY|\bZ$ and also $\hbX\indep \bY|\bZ$ by construction of $\hbX$. If we denote the constructed test statistics via $(\bX,\tbX^{(1:M)})$ by $U_{\myg,L}$ and the other variant used $(\hbX,\tbX^{(1:M)})$ by $\widetilde{U}_{\myg,L}$, then the above relation implies that
	\[
	\sup\limits_{t\geq 0}^{}\left| \prob(U_{\myg,L}\geq t|\bZ,\bY) -\prob(\widetilde{U}_{\myg,L}\geq t|\bZ,\bY)  \right|   \leq d_{\tv}\left( P_{X|Z}^n(\cdot|\bZ),\hP_{X|Z}^n(\cdot|\bZ)  \right)\,.
	\]
Next by marginalizing over $\bZ, \bY$  and an application of Jensen's inequality (namely $|E[V]|\le E[|V|]$ for a random variable $V$) we arrive at
	\begin{equation}\label{eq: robust-inflation}
	\sup\limits_{t\geq 0}^{}\left| \prob(U_{\myg,L}\geq t) -\prob(\widetilde{U}_{\myg,L}\geq t)  \right|  \leq  \E\left[ d_{\tv}\left( P_{X|Z}^n(\cdot|\bZ),\hP_{X|Z}^n(\cdot|\bZ)  \right) \right]\,.
	\end{equation}
	Since $\widetilde{U}_{\myg,L}$ is constructed from $\hbX,\tbX^{1:M}$, which are drawn i.i.d. from $\hP_{X|Z}$, by using Theorem \ref{thm: chi^2-CI-size} we have
	 \begin{align*}
\prob(\widetilde{U}_{\myg,L} \geq \th_{L,\alpha}^{\mathsf{finite}}  )& \leq \alpha\,, \\
\lim\sup\limits_{n\to \infty}^{}\prob(\widetilde{U}_{\myg,L} \geq \th_{L,\alpha}^{\mathsf{asym}}  )& \leq \alpha\,.
	\end{align*}
	The above bounds together with \eqref{eq: robust-inflation} complete the proof of the claim. 
}

\end{document}

%% file: notation.tex
\def\tX{\widetilde{X}}

\def\tX{\widetilde{X}}

\def\cG{{\cal G}}

\def\cD{{\cal D}}

\def\tG{\widetilde{G}}

\def\bz{\mathbf{z}}

\def\cG{\mathcal{G}}

\def\integers{{\mathbb Z}}
\def\reals{{\mathbb R}}

\def\eps{{\varepsilon}}
\def\prob{{\mathbb P}}
\def\E{{\mathbb E}}

\def\var{{\rm Var}}

\def\L0{{L_i}}

\def\de{{\rm d}}
\def\<{\langle}
\def\>{\rangle}

\def\tbX{\widetilde{\mathbf X}}

\def\F{{\sf F}}
\def\ind{{\mathbb I}}

\def\F{{\sf F}}
\def\normal{{\sf N}}

\def\tX{{\widetilde{X}}}

\def\sT{{\sf T}}

\def\v*{v_i}
\def\T*{T_i}

\def\u*{u_i}
\def\F*{F_i}

\def\hX{\hat{X}}

\def\tT{\widetilde{T}}

\def\tT{\widetilde{T}}

\def\bR{{\mtx{R}}}

\def\bw{{w}}

\def\cL{\mathcal{L}}

\def\th{\theta}

\def\hp {\widehat{p}}

\def\l1u{W}

\newcommand{\ajcomment}[1]{}

\makeatletter
\newcommand{\labitem}[2]{%
\def\@itemlabel{\text{#1}}
\item
\def\@currentlabel{#1}\label{#2}}
\makeatother

\DeclareMathAlphabet{\mathpzc}{OT1}{pzc}{m}{it}

\def\sR{\mathsf{R}}
\def\sr{\mathsf{r}}